\documentclass[twocolumn,showpacs,preprintnumbers,prb,superscriptaddress]{revtex4-1}

\usepackage{amssymb}
\usepackage{graphicx}
\usepackage{amsmath}
\usepackage{tikz}
\usetikzlibrary{matrix,calc,positioning}

\usepackage{bbm}
\usepackage{hyperref}

\makeatletter
\renewcommand{\vec}[1]{\ensuremath{\mathbf{#1}}}

\newcommand{\up}{\ensuremath{\uparrow}}
\newcommand{\dn}{\ensuremath{\downarrow}}
\newcommand{\ID}{\ensuremath{\mathcal{D}[\psi \psi^\dagger]}}

\begin{document}

\title{Multitier self-consistent \emph{GW}+EDMFT}
\author{F. Nilsson}
\email{fredrik.nilsson@teorfys.lu.se}
\affiliation{Department of Physics, Division of Mathematical Physics, Lund University, Professorsgatan 1, 223 63 Lund, Sweden}
\author{L. Boehnke}
\email{lewin.boehnke@unifr.ch}
\affiliation{Department of Physics, University of Fribourg, 1700 Fribourg, Switzerland}
\author{P. Werner}
\affiliation{Department of Physics, University of Fribourg, 1700 Fribourg, Switzerland}
\author{F. Aryasetiawan}
\affiliation{Department of Physics, Division of Mathematical Physics, Lund University, Professorsgatan 1, 223 63 Lund, Sweden}

\begin{abstract}
We discuss a parameter-free and computationally efficient \emph{ab initio} simulation approach for moderately and strongly correlated materials, the multitier self-consistent $GW$+EDMFT method. 
This scheme treats different degrees of freedom, such as high-energy and low-energy bands, or local and nonlocal interactions, within appropriate levels of approximation, and provides a fully self-consistent description of correlation and screening effects in the solid. 
The \emph{ab initio} input is provided by a one-shot $G^0W^0$ calculation, while the strong-correlation effects originating from narrow bands near the Fermi level are captured by a combined $GW$ plus extended dynamical mean-field (EDMFT) treatment. 
We present the formalism and technical details of our implementation and discuss some general properties of the effective EDMFT impurity action. In particular, we show that the retarded impurity interactions can have non-causal features, while the physical observables, such as the screened interactions of the lattice system, remain causal. 
As a first application, we present \emph{ab initio} simulation results for SrMoO$_3$, which demonstrate the existence of prominent plasmon satellites in the spectral function not obtainable within LDA+DMFT,
and provide further support for our recent re-interpretation of the satellite features in the related cubic perovskite SrVO$_3$. 
We then turn to stretched sodium as a model system to explore the performance of the multitier self-consistent $GW$+EDMFT method in situations with different degrees of correlation.  
While the results for the physical lattice spacing $a_0$ show that the scheme is not very accurate for electron-gas like systems, because nonlocal corrections beyond $GW$ are important, it does provide physically correct results in the intermediate correlation regime, and a Mott transition around a lattice spacing of $1.5a_0$. Remarkably, even though the Wannier functions in the stretched compound are less localized, and hence the bare interaction parameters are reduced, the self-consistently computed impurity interactions show the physically expected trend of an increasing interaction strength with increasing lattice spacing.  
\end{abstract}

\maketitle

\section{introduction}
Accurate descriptions of materials with strong electron-electron interactions remain one of the main challenges in modern condensed matter theory.
The $GW$ approximation, proposed in 1965 by Hedin \cite{Hedin65New} was one of the first successful attempts to go beyond density functional theory (DFT)\cite{Kohn65SelfConsistent}
for real materials.  
One of the major successes of the $GW$ approximation in its one-shot ($G^0W^0$) version is that it cures the band-gap underestimation of the local density approximation (LDA) for a wide range of semi-conductors.\cite{Aryasetiawan98GW,Schilfgaarde06Quasiparticle} 
However, even for weakly correlated metals such as sodium, the $G^0 W^0$ description yields a too small band narrowing,\cite{Lyo88Quasiparticle} and the theory does not capture the Mott physics that is crucial for understanding the properties of many strongly-correlated 3$d$ and 4$f$ materials.
Fully self-consistent $GW$ calculations are rarely performed, because they are computationally expensive and generally worsen the one-shot results, which are in better agreement with experiment.
This implies that the quasiparticle Green's function, $G^0$, includes (in a somewhat uncontrolled way) vertex corrections needed in the fully self-consistent $GW$ calculations.
Inspired by this observation, 
schemes based on quasiparticle self-consistency have been developed and shown to improve the $GW$ method for a number of different materials.\cite{Schilfgaarde06Quasiparticle,Bruneval06Effect}

For materials with open 3$d$ or 4$f$ shells the valence electrons are relatively localized around the atomic sites, and these materials therefore exhibit strong electron-electron 
interaction effects. For this class of materials, which includes e.g. many different types of high-$T_\mathrm{c}$ superconductors, the combination of density functional theory (usually based on the local density 
approximation (LDA)) and dynamical mean-field theory (LDA+DMFT)\cite{Georges96Dynamical} has been the method of choice. DMFT provides a good description of onsite correlations but neglects the intersite 
correlations.  Furthermore, in LDA+DMFT calculations, the local interactions are often treated as adjustable parameters, and the combination of a density-functional based and a 
diagrammatic scheme requires the introduction of a double-counting parameter, which is supposed to compensate the local correlation effects already contained in the LDA bandstructure. 
This parameter, which can have a substantial effect on the simulation results,\cite{Karolak10Double} is difficult if not impossible to define in a consistent manner. 
For this reason, LDA+DMFT cannot provide a true \emph{ab initio} description of materials.

Systematic procedures such as the constrained random phase approximation (cRPA) \cite{Aryasetiawan04Frequencydependent} in principle allow one to calculate the interaction parameters 
appropriate for LDA+DMFT type calculations, by taking into account the screening effect of the bands outside the low-energy subspace considered in the LDA+DMFT description. 
However, these interactions typically have substantial nonlocal contributions, and in a recent publication we showed\cite{Boehnke16When} that the resulting non-local screening has a big influence on 
the effective local interactions. This is even the case for materials such as SrVO$_3$, which were considered to be strongly correlated.\cite{Pavarini04Mott,Lechermann06Dynamical}
A proper description of the low-energy model thus requires an extended dynamical mean field theory (EDMFT) formalism,\cite{Si96Kosterlitz,Smith00Spatial,Chitra01Effective,Sun02Extended} in which both self-energy and 
polarization effects are treated in a consistent manner.  
A natural way to perform \emph{ab initio} simulations based on EDMFT is to combine this scheme with the $GW$ method.\cite{Biermann03FirstPrinciples} 
This formalism, which we will call $GW$+EDMFT in this paper,\footnote{This formalism has been introduced in \cite{Biermann03FirstPrinciples} as $GW$+DMFT. Here, we use the term $GW$+EDMFT to distinguish this approach from recent implementations that discard the screening of the impurity interaction} involves a fully self-consistent calculation of the interaction parameters, and takes into account the effects of local and nonlocal screening. As a fully diagrammatic scheme, it also does not suffer from the LDA+DMFT type double counting problems, and it is a promising formalism for the nonequilibrium simulation of strongly correlated materials.\cite{Golez17Nonequilibrium}  

While the $GW$+EDMFT method was formulated more than a decade ago,\cite{Sun02Extended,Biermann03FirstPrinciples}  challenges associated with the numerical treatment of retarded interactions have prevented its implementation for many years.\cite{IOPreview}
Recently, some non-self-consistent,\cite{Sakuma13Electronic} or partially self-consistent \cite{Tomczak14Asymmetry,Choi-First} calculations as well as self-consistent model studies\cite{Karlsson05Selfconsistent,Ayral13Screening,Huang14Extended,vanLoon16Double,Stepanov16Selfconsistent,Ayral17Influence} have been presented, 
but the fully self-consistent \emph{ab initio} scheme has been realized so far only in Ref.~\onlinecite{Boehnke16When}. 
An important issue with regard to self-consistency is to what extent the onsite vertex corrections provided by EDMFT counteract the detrimental effects of self-consistency in $GW$.
In this paper we describe the details of our multitier $GW$+EDMFT implementation and test it on materials with different degrees of correlation.

The paper is organized as follows.
In Sec. \ref{sec:method} we review the basic theory of the $GW$+EDMFT method. The method is derived from the  free-energy functional $\Psi$ and
we show that the
causality breakdown of the hybridization
function that has been reported using a Baym-Kadhanoff version of $GW$+DMFT in Ref.~\onlinecite{Lee17diatomic} for the hydrogen dimer, is related to the lack of the bosonic self-consistency
in the latter implementation and will likely not be present in the complete $\Psi$-derivable theory. 
We also discuss the multitier approach in general terms and
clarify the relation between the present approach and 
commonly used methods such as LDA+DMFT.
We then proceed to a detailed discussion of our implementation in Sec. \ref{sec:compdet}.
In Sec. \ref{sec:causality} we use an exactly solvable dimer model to analyze some general causality features of the effective impurity interaction and in Sec.~\ref{sec:results}  present and discuss the results of the full $GW$+EDMFT calculations.
First, the method is applied to the moderately correlated perovskite SrMoO$_3$, where we find that, contrary to LDA+DMFT which cannot reproduce the experimental satellite features,\cite{Wadati14Photo} this material is well described by $GW$+EDMFT.
By comparing the spectra with the effective impurity interaction and fully screened interaction we can distinguish Hubbard bands from plasmonic satellites and thereby deduce the nature of the satellite features. 
By systematically changing the occupation we also use SrMoO$_3$ as a model to investigate the causality of the effective impurity interaction and relate these results to our more general causality-considerations in Sec. \ref{sec:causality}. 

We then focus on sodium as a model system to investigate the performance of the method in different regimes. By successively increasing the lattice constant we investigate the effects of self-consistency and long-range screening for situations with different degrees of local correlations.
We show that the method performs well in the moderately to strongly correlated regime and captures the Mott-Hubbard metal to insulator transition at some critical value of the streching.
Furthermore we show that the self-consistency is essential to capture the correct trend in the impurity interaction.
Section~\ref{sec:conclusions} summarizes the most important findings. 

\section{Method}
\label{sec:method}
\subsection{GW-approximation}
By expanding the self-energy in terms of the screened Coulomb interaction Hedin derived the following set of exact coupled differential equations
defining the self-energy $\Sigma$ and the Green's function $G$ in terms of the polarization $\Pi$ and vertex function $\Gamma$ \cite{Hedin65New}, 
\begin{align}
 &\Sigma(1,2) = -\int d(34) G(1,3^+)W(1,4)\Gamma(3,2,4), \label{eq:HedinSigma} \\
 &G(1,2) = G_0(1,2) + \int d(34) G_0(1,3)\Sigma(3,4)G(4,2), \label{eq:HedinG} \\
 &\Gamma(1,2,3) = \delta(1-2)\delta(2-3), \nonumber \\
 &\hspace{5mm}+ \int d(4567) \frac{\partial \Sigma(1,2)}{\partial G(4,5)}G(4,6)G(7,5)\Gamma(6,7,3), \label{eq:HedinVertex} \\
 &\Pi(1,2) = \int d(34) G(1,3) \Gamma(3,4,2) G(4,1^+), \label{eq:HedinP} \\
 &W(1,2) = v(1,2) + \int d(34)v(1,3)\Pi(3,4)W(4,2). \label{eq:HedinW}
\end{align}
The above equations are for finite temperature with the notation $1=(\tau_1,\vec{r}_1)$, etc., and the spin has been omitted for simplicity.
By approximating the vertex function by the first term, $\Gamma(1,2,3) \approx \delta(1-2)\delta(2-3)$, we arrive at the well
known $GW$ approximation for the self-energy,
\begin{align}
\Sigma^{GW} = -G(1,2)W(1,2),
\end{align}
and the random-phase approximation for the polarization function,
\begin{align}
\Pi(1,2) = G(1,2)G(2,1^+).
\label{eq:PRPA}
\end{align}

With a suitable choice of basis functions the integral equations \eqref{eq:HedinG} and \eqref{eq:HedinW}-\eqref{eq:PRPA} can be mapped to matrix equations which can be treated in computer codes with standard linear algebra libraries.

\subsection{Extended Dynamical Mean-Field Theory}
\label{sec:edmft}
By introducing a localized basis set {$w_{n\vec{R}}(\vec{r})$}, where $n$ is an orbital index and $\vec{R}$ a site index, it is possible to separate the correlations into local (onsite) correlations and nonlocal (offsite) correlations.
The key assumption in EDMFT is that all correlations are local. 
In practice this means that the selfenergy and polarization can be described using only the local basis functions.
While the selfenergy and the Green's functions are one-particle quantities that can be expanded directly in the local one-particle basis, the polarization,
which is a two-particle quantity, requires a local product basis 
\begin{align}
\Psi_{\alpha\vec{R}}(\vec{r}) = w_{i\vec{R}}^*(\vec{r})w_{j\vec{R}}(\vec{r}),
\label{eqn:locprodbas}
\end{align} 
where, $\alpha=(i,j)$. 
For realistic materials the localized basis is typically chosen as linear muffin tin orbital (LMTO) basis functions\cite{Aryasetiawan98GW} or maximally localized Wannier functions (MLWFs)\cite{Marzari97Maximally,Mostofi08wannier90} that are derived from the LDA band structure.

In EDMFT the full lattice problem is mapped to an impurity problem with a 
dynamical bare propagator $\mathcal{G}$ and interaction $\mathcal{U}$ \cite{IOPreview}. These so-called Weiss-fields are determined self-consistently by 
requiring that the local part of the lattice Green's function ($G_{\mathrm{loc}}$) and fully screened interaction $W_{\mathrm{loc}}$ (as defined
by the projection onto the local one- and two- particle basis) should be equal to their impurity counterparts, $G_{\mathrm{imp}}$ and $W_{\mathrm{imp}}$, respectively, 
\begin{align}
G_{\mathrm{imp}}=G_{\mathrm{loc}}, \nonumber \\
W_{\mathrm{imp}}=W_{\mathrm{loc}} \label{eq:GWEDMFT-SCcond}.
\end{align}
In Refs.~\onlinecite{Hugel16Bosonic,Golez17Nonequilibrium} these self-consistency conditions were formally derived by constructing the free energy functional of the
impurity $\Gamma'$ and considering the variation of $\Gamma - \Gamma'$, where $\Gamma$ is the lattice system free energy functional.

The impurity action is given by
\begin{align}
S=&\int_0^\beta d\tau d\tau' \sum_{ab\sigma} c^\dagger_{a\sigma}(\tau)[\delta(\tau-\tau')\partial_\tau-\mathcal{G}^{-1}_{ab\sigma}(\tau-\tau')]c_{b\sigma}(\tau') \nonumber \\
  &+\frac{1}{2}\int_0^\beta d\tau d\tau' \sum_{\sigma\sigma'}\sum_{abcd}\mathcal{U}_{abdc}(\tau-\tau') \nonumber\\
  &\phantom{\frac{1}{2}\int_0^\beta d\tau d\tau' }
  \times c^\dagger_{a\sigma}(\tau)c_{b\sigma}(\tau)c^\dagger_{c\sigma'}(\tau')c_{d\sigma'}(\tau')
\end{align}
and the EDMFT self-consistency cycle takes the following form:
\begin{enumerate}
\item Start with an inital guess for $\Sigma^\mathrm{imp}$ and $\Pi^\mathrm{imp}$.
\item Use these for the local quantities $\Sigma^\mathrm{loc}=\Sigma^\mathrm{imp}$ and $\Pi^\mathrm{loc}=\Pi^\mathrm{imp}$ (EDMFT approximations) \label{nbr:EDMFTcondition}.
\item Use $\Sigma_\vec{k}=\Sigma^\mathrm{loc}$ and $\Pi_\vec{q}=\Pi^\mathrm{loc}$.
\item Calculate $G^\mathrm{loc}=\sum_\vec{k}\left(G^{(0)}_\vec{k}{}^{-1}-\Sigma_\vec{k}\right)^{-1}$ and $W^\mathrm{loc}=\sum_\vec{q}v_\vec{q}\left(\mathbbm{1}-\Pi_\vec{q} v_\vec{q}\right)^{-1}$,
\item Use $G^\mathrm{imp}=G^\mathrm{loc}$ and $W^\mathrm{imp}=W^\mathrm{loc}$ (EDMFT self-consistency conditions). \label{nbr:edmftsccond}
\item Calculate the fermionic Weiss field 
\begin{align}
\mathcal{G}=\left(\Sigma^\mathrm{imp}+G^\mathrm{imp}{}^{-1}\right)^{-1}
\label{eq:weissfields1}
\end{align}
and the effective impurity interaction 
\begin{align}
\mathcal{U}=W^\mathrm{imp}\left(\mathbbm{1}+\Pi^\mathrm{imp} W^\mathrm{imp}\right)^{-1}.
\label{eq:weissfields2}
\end{align}
\label{nbr:weissfields}
\item Numerically solve the impurity problem to obtain $G^\mathrm{imp}$ and the impurity charge susceptibility $\chi^\mathrm{imp}$.
\item Use the current $\mathcal{G}$ and $\mathcal{U}$ to calculate $\Sigma^\mathrm{imp}=\mathcal{G}^{-1}-G^\mathrm{imp}{}^{-1}$ and $\Pi^\mathrm{imp}=\chi^\mathrm{imp}\left(\mathcal{U}\chi^\mathrm{imp}-\mathbbm{1}\right)^{-1}$. The fully screened interaction  $W^\mathrm{imp}=\mathcal{U}-\mathcal{U}\chi^\mathrm{imp}\mathcal{U}$ only enters the calculations through the self-consistency condition in step~\ref{nbr:edmftsccond}.
\item Go back to step \ref{nbr:EDMFTcondition}.
\end{enumerate}

In view of pure model calculations, but also with simulations of cubic $t_{2g}$ materials in mind, it is instructive to explicitly determine the product basis representation of the Kanamori interaction\cite{Kanamori63Electron}
\begin{align}
  \label{eqn:kanamoriH}
  H_\mathrm{K}=&U\sum_an_{a\up}n_{a\dn}+\frac{1}{2}\sum_{a\neq b}\sum_{\sigma\sigma'}(U'-J\delta_{\sigma\sigma'})n_{a\sigma}n_{b\sigma'}\nonumber\\
&-\sum_{ab}J(\underbrace{c^\dagger_{a\up}c_{a\dn}c^\dagger_{b\dn}c_{b\up}}_\mathrm{spin-flip}+\underbrace{c^\dagger_{b\up}c^\dagger_{b\dn}c_{a\up}c_{a\dn}}_\mathrm{pair-hopping}). 
\end{align}

It is straightforward to show that this is the special case of the general rank-4 tensor representation of the interaction
\begin{equation}
  \label{eqn:tensorU}
  H=\frac{1}{2}\sum_{\sigma\sigma'}\sum_{abcd}U_{abdc}(c^\dagger_{a\sigma}c_{b\sigma}c^\dagger_{c\sigma'}c_{d\sigma'})
\end{equation}
with the choice $U=U_\mathrm{K}$,
\begin{equation}
  \label{eqn:kanamoriU}
  U_\mathrm{K}={\small
    \begin{tikzpicture}[baseline={($(current bounding box.west)-(0,1.05em)$)}] 
      \matrix (UK) [matrix of math nodes, row sep=1em, column sep=1em, nodes={anchor=base,inner sep=1pt}, left delimiter=(, right delimiter=)] { 
        U\phantom{'}&U'&U'&\phantom{U'}&\phantom{U'}&\phantom{U'}&\phantom{U'}&\phantom{U'}&\phantom{U'}\\
        U'&U&U'&\phantom{J}&\phantom{J}&\phantom{J}&\phantom{J}&\phantom{J}&\phantom{J}\\
        U'&U'&U&\phantom{J}&\phantom{J}&\phantom{J}&\phantom{J}&\phantom{J}&\phantom{J}\\
        \phantom{U'}&\phantom{U'}&\phantom{U'}&J&J&\phantom{J}&\phantom{J}&\phantom{J}&\phantom{J}\\
        \phantom{U'}&\phantom{U'}&\phantom{U'}&J&J&\phantom{J}&\phantom{J}&\phantom{J}&\phantom{J}\\
        \phantom{U'}&\phantom{U'}&\phantom{U'}&\phantom{U'}&\phantom{U'}&J&J&\phantom{J}&\phantom{J}\\
        \phantom{U'}&\phantom{U'}&\phantom{U'}&\phantom{U'}&\phantom{U'}&J&J&\phantom{J}&\phantom{J}\\
        \phantom{U'}&\phantom{U'}&\phantom{U'}&\phantom{U'}&\phantom{U'}&\phantom{U'}&\phantom{U'}&J&J\\
        \phantom{U'}&\phantom{U'}&\phantom{U'}&\phantom{U'}&\phantom{U'}&\phantom{U'}&\phantom{U'}&J&J\\
      }; 
      \draw[dashed]($0.5*(UK-1-3.north east)+0.5*(UK-1-4.north west)$) -- ($0.5*(UK-9-3.south east)+0.5*(UK-9-4.south west)$);
      \draw[dashed]($0.5*(UK-1-5.north east)+0.5*(UK-1-6.north west)$) -- ($0.5*(UK-9-5.south east)+0.5*(UK-9-6.south west)$);
      \draw[dashed]($0.5*(UK-1-7.north east)+0.5*(UK-1-8.north west)$) -- ($0.5*(UK-9-7.south east)+0.5*(UK-9-8.south west)$);
      \draw[dashed]($0.5*(UK-3-1.south west)+0.5*(UK-4-1.north west)$) -- ($0.5*(UK-3-9.south east)+0.5*(UK-4-9.north east)$);
      \draw[dashed]($0.5*(UK-5-1.south west)+0.5*(UK-6-1.north west)$) -- ($0.5*(UK-5-9.south east)+0.5*(UK-6-9.north east)$);
      \draw[dashed]($0.5*(UK-7-1.south west)+0.5*(UK-8-1.north west)$) -- ($0.5*(UK-7-9.south east)+0.5*(UK-8-9.north east)$);
      \node[above=6pt of UK-1-1] (top-1) {11};
      \node[above=6pt of UK-1-2] (top-2) {22};
      \node[above=6pt of UK-1-3] (top-3) {33};
      \node[above=6pt of UK-1-4] (top-4) {21};
      \node[above=6pt of UK-1-5] (top-5) {12};
      \node[above=6pt of UK-1-6] (top-6) {32};
      \node[above=6pt of UK-1-7] (top-7) {23};
      \node[above=6pt of UK-1-8] (top-8) {13};
      \node[above=6pt of UK-1-9] (top-9) {31};
      \node[left=12pt of UK-1-1] (left-1) {11};
      \node[left=12pt of UK-2-1] (left-2) {22};
      \node[left=12pt of UK-3-1] (left-3) {33};
      \node[left=12pt of UK-4-1] (left-4) {12};
      \node[left=12pt of UK-5-1] (left-5) {21};
      \node[left=12pt of UK-6-1] (left-6) {23};
      \node[left=12pt of UK-7-1] (left-7) {32};
      \node[left=12pt of UK-8-1] (left-8) {31};
      \node[left=12pt of UK-9-1] (left-9) {13};
    \end{tikzpicture}}.
\end{equation}
(At the borders or the matrix, we show the indices of the product basis according to Eq.~\eqref{eqn:locprodbas}.)

We first note that \eqref{eqn:kanamoriU} will have a determinant of zero and is hence not invertible.
While this forbids the use the inverted forms of the bosonic Dyson equation, e.g. $\mathcal{U}^{-1}=\Pi+W_\mathrm{loc}^{-1}$, 
this does not pose a problem because all steps of the EDMFT self-consistency cycle can be formulated without, as shown above.

\subsection{$GW$+EDMFT}

\subsubsection{Self-energy and polarization}

While (E)DMFT treats the strong local correlations in a nonperturbative fashion to all orders, the non-local correlations are omitted. 
Due to the separability of the vertex function in Eq.~\eqref{eq:HedinVertex} into the sum of the trivial vertex function used in $GW$ and the non-trivial vertex correction, the Hedin equations for the self-energy (Eq.~\eqref{eq:HedinSigma}) and the polarization (Eq.~\eqref{eq:HedinP}) also separate into the $GW$ contribution and the contribution from the vertex corrections:

\begin{align}
&\Sigma = \Sigma^{GW} + \Sigma^\mathrm{vc},\\
&\Pi = \Pi^{GG} + \Pi^\mathrm{vc}.
\label{eq:gw+vc}
\end{align}

Within the combined $GW$+EDMFT scheme $\Sigma^\mathrm{vc}$ and $\Pi^\mathrm{vc}$ are approximated by their impurity counterparts.
In this scheme the double counting is well defined and is simply the local projection of the $GW$ self-energy and polarization, respectively, 
\begin{align}
&\Sigma^{GW+\mathrm{EDMFT}}(\vec{k}) = \Sigma^{GW}(\vec{k}) + \Sigma^{\mathrm{EDMFT}}  \nonumber \\ 
&\hspace{30mm}-\sum_\vec{k}\Sigma^{GW}(\vec{k}),\label{eq:SigmaGW+EDMFT}\\
&\Pi^{GW+\mathrm{EDMFT}}(\vec{k}) = \Pi^{GG}(\vec{k}) + \Pi^{\mathrm{EDMFT}}  \nonumber \\ 
&\hspace{30mm}-\sum_\vec{k} \Pi^{GG}(\vec{k}).
\label{eq:PRPA2}
\end{align}
The self-consistency cycle is equivalent to the EDMFT self-consistency cycle (see Sec. \ref{sec:edmft}) but with $\Sigma(\vec k)$ and $\Pi(\vec k)$ in step~3 replaced by the corresponding quantities defined in Eqs.~\eqref{eq:SigmaGW+EDMFT}-\eqref{eq:PRPA2}. 

\subsubsection{Functional derivation}
In Ref.~\onlinecite{Biermann03FirstPrinciples} it was shown that the $GW$+EDMFT formalism can be rigorously derived as an approximation to the free-energy functional $\Psi$ \cite{Almbladh99Variational,Chitra01Effective}. 
In this section we briefly review the derivation following Refs.~\onlinecite{Biermann03FirstPrinciples,Chitra01Effective,Ayral13Screening,Rohringer17Diagrammatic}
and also make a connection with the Baym-Kadanoff formulation of $GW$+DMFT in, e.g., Ref~\onlinecite{Lee17diatomic}.

The partition function in the grand canonical ensemble for interacting electrons moving in the crystal potential $V_c$ is given by
\begin{align}
Z  = \int \ID \exp (-S[\psi,\psi^\dagger]),
\end{align}
where $S$ is the action
\begin{align}
&S[\psi,\psi^\dagger]=\int dx \psi^\dagger(x)\left( \partial_\tau - \frac{\nabla^2}{2m} + V_c(x)\right)\psi(x)  \nonumber \\
&-\frac{1}{2} \int dx dx' \psi^\dagger(x)\psi^\dagger(x')V(x-x')\psi(x')\psi(x).
\end{align}
The electron-electron interaction term is decoupled using a Hubbard-Stratonovic transformation which yields an additional
bosonic field $\phi$,
\begin{align}
&S[\phi,\psi,\psi^\dagger]=\int dx \psi^\dagger(x)\left( \partial_\tau - \frac{\nabla^2}{2m} + V_H(x)\right)\psi(x)  \nonumber \\
&-\frac{1}{2} \int dx dx' \phi(x)V^{-1}(x-x')\phi(x') \nonumber \\ 
&- i\alpha \int dx \phi(x)\psi^\dagger(x)\psi(x) .
\end{align}
Here $V_H$ includes both the crystal and the Hartree potential and $\alpha$ is a coupling constant that is set to 1 for the physical case.
Now we introduce additional source fields that couple to the fermionic and bosonic propagators, which yields the total action
\begin{align}
  &S[\phi,\psi,\psi^\dagger,J_f,J_b] = S[\phi,\psi,\psi^\dagger] \nonumber\\
                                     &\hspace{20mm}- \int dx dx' J_f(x,x') \psi^\dagger(x)\psi(x') \nonumber \\ 
                                     &\hspace{20mm}- \frac{1}{2} \int dx dx' J_b(x,x')\phi(x)\phi(x') .
\end{align}

\begin{figure}
  \centering
  \includegraphics{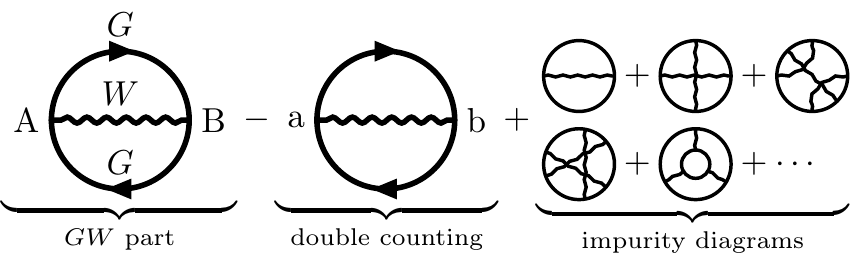}
  \caption{Diagrammatic representation of the terms in the $\Psi$ functional for $GW$+EDMFT. Impurity indices are omitted for readability. They are all lowercase to indicate that they belong to the more correlated space $s$ in the notation presented in Sec.~\ref{sec:connecting}. Combinatorial factors were omitted for clarity.}
  \label{fig:psigwedmft}
\end{figure}

The free energy of the system in the presence of the external fields is defined as 
\begin{align}
\Omega(J_f,J_b) = -\mathrm{ln}(Z(J_f,J_b)).
\end{align}
By defining the fermionic Green's function $G(x,y)=-\langle T\psi(x)\psi^\dagger(y)\rangle = \frac{\partial \Omega}{\partial J_f}$ and the corresponding bosonic propagator
$W(x,y)=\langle T\phi(x)\phi(y)\rangle = \frac{2\partial \Omega}{\partial J_b}$ and 
performing a double Legendre transform of $\Omega$ we obtain the free energy functional $\Gamma$: 
\begin{align}
&\Gamma\left[ G, W \right] = \Omega(J_f,J_b) - J_fG-\frac{J_b}{2}W = \nonumber \\
&\hspace{5mm}\mathrm{Tr[ln}(G)\mathrm{]} - \mathrm{Tr}[(G_H^{-1} - G^{-1})G ] - \frac{\mathrm{Tr[ln}(W)]}{2} \nonumber \\
 &\hspace{5mm}+\frac{\mathrm{Tr}[(V^{-1}-W^{-1})W]}{2} + \Psi[G,W],
\end{align}
where $G_H$ is the Hartree Green's function of the solid and $\Psi$ contains all further contributions,
\begin{align}
\Psi(G,W) = \int d\alpha \int dx \langle \phi(x)\psi^\dagger(x)\psi(x) \rangle\;.
\end{align}
Physically, $\Psi$ includes all two-particle irreducible diagrams constructed with the electron-boson vertex.
Setting the source terms to zero and requiring stationarity of $\Gamma$ yields the Dyson equations
\begin{align}
&G^{-1} = G_H^{-1} - \frac{\partial \Psi}{\partial G},\\
&W^{-1} = V^{-1} + 2\frac{\partial \Psi}{\partial W},
\end{align}
from which we can identify the self-energy and the polarization as
\begin{align}
\Sigma = \frac{\partial \Psi}{\partial G}, \label{eq:fd1} \\
\Pi = -2\frac{\partial \Psi}{\partial W}. \label{eq:fd2}
\end{align}
  A clear advantage of the functional formalism is  
  that the derived methods satisfy 
  conservation laws in the Baym-Kadanoff sense. However, when the self-consistency is restricted to a subspace of the full Hilbert space, the conservation laws
  may not be fulfilled anymore.
 The $GW$ approximation corresponds to the lowest order approximation to $\Psi$ in $W$:
 \begin{align}
  \Psi^{GW} = -\frac{1}{2}\mathrm{Tr}GWG.
 \end{align}
The EDMFT functional is defined by making a local approximation to $G$ and $W$.
It should be noted that by local in this context we mean an onsite approximation in the localized basis defined in Section~\ref{sec:edmft},
and the local product basis (Eq.~\eqref{eqn:locprodbas}) for the bosonic quantities. 
This yields the full $GW$+EDMFT functional:
\begin{align}
\Psi^{GW+\mathrm{EDMFT}}(G,W) =& \underbrace{\mathrm{Tr}GWG}_{\Psi^{GW}} + \underbrace{\Psi(G^{\mathrm{loc}},W^{\mathrm{loc}})}_{\Psi^{\mathrm{EDMFT}}} \nonumber \\
-&  \underbrace{\mathrm{Tr}G^{\mathrm{loc}}W^{\mathrm{loc}}G^{\mathrm{loc}}}_{\Psi^{DC}},
\label{eq:GW+EDMFTfunctional}
\end{align}
where the last term constitutes the doublecounting term between the $GW$  and EDMFT-functionals.
Figure~\ref{fig:psigwedmft} gives a visual representation of these terms of the diagrams.
\\
\subsubsection{Double counting}
The double-counting terms for the polarization and selfenergy are obtained by evaluating the functional derivatives in Eqs.\ \eqref{eq:fd1}-\eqref{eq:fd2} for $\Psi^{DC}$ defined in Eq.\ \eqref{eq:GW+EDMFTfunctional}:
\begin{align}
&\Sigma_{kl}^{\mathrm{DC}}(\tau) = G_{mn}^{\mathrm{loc}}(\tau) W_{mknl}^{\mathrm{loc}}(\tau), \label{eq:SigmaDC} \\
&\Pi_{mm'nn'}^{\mathrm{DC}}(\tau) = G_{mn}^{\mathrm{loc}}(\tau)G_{n'm'}^{\mathrm{loc}}(-\tau) \label{eq:PolDC}.
\end{align}
A summation over repeated indices is assumed and the matrix elements are taken in the localized one- and two-particle basis functions defined in Sec.~\ref{sec:edmft}, which for $W$ corresponds to
\begin{align}
 W_{ijkl} = \int d\vec{r}d\vec{r}'w_{i}^*(\vec{r})w_{j}(\vec{r})W(\vec{r},\vec{r}')w_{k}(\vec{r}')w^*_{l}(\vec{r}').
\end{align}

For the case where the orbital subspaces for the EDMFT and $GW$ calculations are the same, $\Sigma^{\mathrm{DC}}$ reduces to the local projection ($\vec{k}$-sum) of the full $GW$ selfenergy.
If the orbital subspace for the EDMFT calculation is smaller than that of the $GW$ calculation the difference between
$\Sigma^{\mathrm{DC}}$ and the local projection of the full $GW$ self-energy is
\begin{equation}
 \left( \Sigma^{\mathrm{loc}}_{GW} - \Sigma^{\mathrm{DC}}\right)_{ik} = -\sum_{j,l\in r} G^{\mathrm{loc}}_{jl}W^{\mathrm{loc}}_{ijkl},
 \label{eq:Sigmadiff}
\end{equation}
where $r$ is the part of the full Hilbert space that is not included in the EDMFT subspace. 

In Ref.\ \onlinecite{Lee17diatomic} a different formulation of $GW$+DMFT based on the Baym-Kadanoff functional was suggested and evaluated for the hydrogen dimer.
This formulation does not include the bosonic self-consistency of the full formulation presented above and therefore the double-counting between the $GW$ and the DMFT parts is different:
\begin{align}
 &\Sigma^{\mathrm{DC-BK}}_{ik} = -G_{jl}^{\mathrm{loc}}(\tau) W^{GW\mathrm{-imp}}_{ijkl}(\tau), \\
 & W^{GW\mathrm{-imp}} = \left(1-U\Pi^{\mathrm{loc}}\right)^{-1}U,\\
 & \Pi_{mm'nn'}^{\mathrm{loc}}(\tau) = G_{mn}^{\mathrm{loc}}(\tau)G_{n'm'}^{\mathrm{loc}}(-\tau),
 \end{align}
where $U$ is the impurity interaction, which in this case is fixed to the local model interaction, and all sums are restricted to the DMFT orbital subspace.
In Ref.~\onlinecite{Lee17diatomic} it was shown that this formulation yields non-causal hybridization functions in the strongly correlated regime due to the non-causality of the difference between the local projection of the $GW$ selfenergy and the double-counting term
\begin{align*}
 \Sigma^{\mathrm{loc}}_{GW} - \Sigma^{\mathrm{DC-BK}}\;.
\end{align*}
An alternative double-counting was then introduced to replace this derived double-counting with a causal one.

The full $\Psi$-derivable formulation of $GW$+EDMFT on the other hand is not expected to experience the same causality problem, 
since in that case the double-counting that follows naturally from the derivation coincides with the `causal' double-counting introduced in Ref.~\onlinecite{Lee17diatomic}
when the orbital subspaces for the EDMFT and $GW$ calculations are the same. Furthermore, if the EDMFT subspace is smaller than the $GW$ subspace
the difference $\left( \Sigma^{\mathrm{loc}}_{GW} - \Sigma^{\mathrm{DC}}\right)$ reduces to the expression in Eq.~\eqref{eq:Sigmadiff}, which is causal by construction.

\subsection{Multitier Self-Consistent GW+EDMFT}
\label{sec:MTSCGWEDMFT}
\begin{figure}
\includegraphics{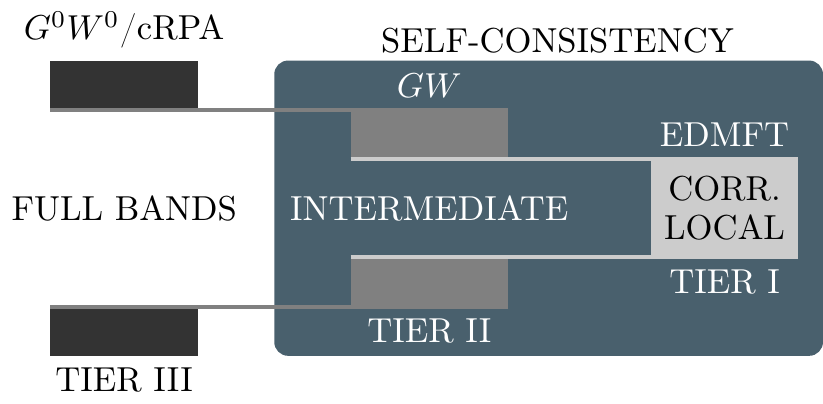}
\caption{Schematic figure of the multitier $GW$+EDMFT scheme showing the different approximations on the different tiers. (Adapted from Ref.~\onlinecite{Boehnke16When}.)}
\label{fig:GWEDMFTtiers}
\end{figure}

In the present multitier $GW$+EDMFT the full Hilbert space is divided into three subspaces and each subspace is treated with an appropriate level of approximation.
The aim of our approach is to accurately describe systems that have both strong local correlation and nonlocal correlation effects,
typically 3$d$- or $4d$-compounds such as transition-metal oxides, transition metals, and high $T_\mathrm{c}$ cuprate superconductors, and to do so at a reasonable computational cost. 
A common feature of many of these compounds is strongly correlated partially filled 3$d$-states mixed with less correlated extended $s$ or $p$ states.
While local vertex contributions from the DMFT-type impurity problem are needed for the $d$ states, the $s$ and $p$ states are typically well described within the $GW$ approximation or even LDA.
However, due to the nontrivial mixing with the correlated $d$ states the $s$ and $p$ states need to be included in the self-consistency cycle with an appropriate double counting term.
These considerations suggest adopting the following three-step procedure:
\begin{itemize}
 \item TIER III: Perform a one-shot $G^0W^0$ calculation in the full Hilbert space. Choose a basis on an intermediate subspace, typically 3-8 bands, and calculate the effective interaction $U(\omega)$ for this subspace using the constrained random phase approximation. 
 $\Sigma^{G^0W^0}$ is kept in TIER III only.
 \item TIER II: Within the intermediate subspace the self-energy and polarization are calculated self-consistently using a fully self-consistent $GW$ implementation
 \item TIER I: Choose a correlated subspace, smaller or equal to the intermediate subspace, for which local vertex contributions are calculated using an EDMFT-type impurity problem.
\end{itemize}
Self-consistent calculations are performed in the intermediate subspace and local
vertex contributions from EDMFT are included inside the correlated subspace at each
step in the self-consitency loop. The multitier $GW$+EDMFT scheme is illustrated in Fig.~\ref{fig:GWEDMFTtiers}.

Within the multitier approach all double countings are well defined. The full Green's function is given by
\begin{align}
G_{\vec{k}}^{-1}=&\overbrace{\underbrace{\mathrm{i}\omega _{n}+\mu -\varepsilon _{\vec{k}}^{\mathrm{LDA}}+V_{\mathrm{XC},\vec{k}}}_{G_{\mathrm{Hartree},\vec{k}}^{0}{}^{-1}}\underbrace{-\left(\Sigma_{\vec{k}}^{G^{0}W^{0}}-\Sigma_{\vec{k}}^{G^{0}W^{0}}\big|_{I}\right)}_{-\Sigma _{\mathrm{r},\vec{k}}\big|_{I}}}^{\text{TIER III},\; G_{\vec{k}}^{0}{}^{-1}}\notag\\
&\underbrace{-\left(\Sigma _{\vec{k}}^{GW}\big|_{I}-\Sigma ^{GW}\big|_{C,\mathrm{loc}}\right)}_{\text{TIER II}}\underbrace{-\Sigma ^{\mathrm{EDMFT}}\big|_{C,\mathrm{loc}}}_{\text{TIER I}}\;,\label{eqn:fullG}
\end{align}
and the corresponding equation for the bosonic propagators by
\begin{align}
W_\vec{q}^{-1}=&\overbrace{v_\vec{q}^{-1}\underbrace{-\left(\Pi^{G^0G^0}_\vec{q}-\Pi^{G^0G^0}_\vec{q}\big|_I\right)}_{-\Pi_{\mathrm{r},\vec{q}}}}^{\text{TIER III},\; U_{\vec{q}}^{-1}}\notag\\
&\underbrace{-\left(\Pi_\vec{q}^{GG}\big|_I-\Pi^{GG}\big|_{C,\mathrm{loc}}\right)}_{\text{TIER II}}\underbrace{-\Pi^\mathrm{EDMFT}\big|_{C,\mathrm{loc}}}_{\text{TIER I}}.\label{eqn:fullW}
\end{align}

The explicit $\vec{q}$-dependence of the polarization and self-energy, which couples back to the impurity problem in the self-consistency cycle, makes the $GW$+EDMFT solution dependent on the details of the bare lattice problem (e. g. the geometry of the lattice or the $\vec{q}$-dependence of the bare interaction) to a much greater extent than what one typically finds in approximations like DMFT, where the observables primarily depend on integrated quantities like the bare local density of states.

\subsection{Connecting the tiers}
\label{sec:connecting}
\begin{table*}
  \begin{tabular}{c||c|c||c|c||c||c}
    &  \multicolumn{2}{c||}{non-sc}  & \multicolumn{3}{c||}{selfconsistent}\\
    Colloquial name & \multicolumn{2}{c||}{TIER~III} & \multicolumn{2}{c||}{TIER~II} & TIER~I  & References / Comments\\
    & $\Sigma$ & $\Pi$ & $\Sigma$ & $\Pi$ & used quantities\\
    \hline
    model DMFT            &   |       &    |     & local &  |    & $\Sigma$ & Ref.~\onlinecite{Georges96Dynamical} \\

    model EDMFT           &   |       &    |     & local & local & $\Sigma$, $\Pi$ & Ref.~\onlinecite{Chitra01Effective} \\

    model D$\Gamma$A       &  |       &     |    & SDE\footnote{Schwinger-Dyson equation}/BSE\footnote{Bethe-Salpeter equation} &  |  & $\Sigma$, $\gamma$\footnote{$\gamma$ denotes the fully irreducible vertex of the impurity problem}  & Ref.~\onlinecite{Toschi07Dynamical} (non-self-consistent)\\

    model $GW$+EDMFT      &   |       &    |     & $GW$+imp  & $GG$+imp  & $\Sigma$, $\Pi$ & Refs.~\onlinecite{Sun02Extended,Ayral13Screening,Huang14Extended,Ayral17Influence}\\

    model TRILEX       &  |       &     |     &   $GW\Lambda$+imp & $GG\Lambda$+imp & $\Sigma$, $\Pi$, $\Lambda$\footnote{$\Lambda$ denotes the three-leg vertex} & Refs.~\onlinecite{Ayral15Mott,Ayral16Motta}\\

    $G^0W^0$              &  $G^0W^0$ & $G^0G^0$  &  $\emptyset$   &  $\emptyset$ & $\emptyset$ & often referred to as $GW$. Ref.~\onlinecite{Aryasetiawan98GW} \\

    $GW$                  &   |       &    |     &  $GW$ & $GG$  &  $\emptyset$  & rarely used. Refs.~\onlinecite{Holm98Fully,GarciaSelf-Consistent,Koval14Fully}\\

    $QP$sc$GW$            &   |       &    |     &  $G^{QP}W^{QP}$ & $G^{QP}G^{QP}$  &  $\emptyset$  &  Refs.~\onlinecite{Schilfgaarde06Quasiparticle,Bruneval06Effect,Gatti13Dynamical,Koval14Fully}\\

    LDA+DMFT              & $E^\mathrm{LDA}_\mathrm{xc}$ & | & $E^\mathrm{LDA}_\mathrm{xc}$+imp & | & $\Sigma$ & Ref.~\onlinecite{Kotliar06Electronic} \\

    LDA+DMFT+$U(\omega)$  & $E^\mathrm{LDA}_\mathrm{xc}$ & $G^0G^0$ & $E^\mathrm{LDA}_\mathrm{xc}$+imp & | & $\Sigma$ & Ref.~\onlinecite{Werner12Satellites,Sakuma13Electronic,Werner15Dynamical} \\

    SEx+DMFT              & $E^\mathrm{SEx}_\mathrm{xc}$ & $G^0G^0$ & $E^\mathrm{SEx}_\mathrm{xc}$+imp & | & $\Sigma$ & Ref.~\onlinecite{vanRoekeghem14Dynamical} (BFA\footnote{BFA stands for Bose Factorization Approach\cite{Casula12Dynamical}} for impurity) \\

    $GW$+DMFT+$U(\omega)$ & $G^0W^0$ & $G^0G^0$ & $G^0W^0$+imp & | & $\Sigma$ & Ref.~\onlinecite{Tomczak14Asymmetry} (BFA for impurity) \\

    \textit{\emph{ab initio}} D$\Gamma$A  &  $E^\mathrm{LDA}_\mathrm{xc}$  &     |    & $E^\mathrm{LDA}_\mathrm{xc}$+SDE/BSE &  |  & $\Sigma$, $\gamma$ & Refs.~\onlinecite{Toschi11Abinitio,Galler17Abinitio} (non-self-consistent)\\
    $GW$+EDMFT            & $G^0W^0$ & $G^0G^0$ & $GW$+imp & $GG$+imp & $\Sigma$, $\Pi$ & Refs.~\onlinecite{Biermann03FirstPrinciples,Boehnke16When}\\

  \end{tabular}
  \caption{Relation of existing formalisms to the multitier scheme. Embedding the TIER~I values into the TIER~II approximation is done as described in Sec.~\ref{sec:connecting}. 
  Formalisms without an approximation to the polarization $\Pi$ in TIER~II do not update the impurity interaction. The TIER~I column lists the quantities that have to be measured 
  in the impurity problem and enter the self-consistency on TIER~II. 
It is implicit that the corresponding doublecountings between the tiers are removed, which makes it necessary to add them back on TIER~II in the case of $E_\mathrm{xc}^\mathrm{LDA}$ and 
$E_\mathrm{SEx}^\mathrm{LDA}$, where the full LDA/SEx band values are used in the self-consistency. $\emptyset$ indicates that the corresponding tier is empty.
    \label{tab:context}}
\end{table*}

The multitier approach allows one to systematically remove the least relevant degrees of freedom from the description and to replace them by an effective medium into which the more relevant degrees of freedom are embedded. 
The general recipe for this procedure does not depend on the actual separation into `less relevant' and `more relevant' spaces, the only prerequisite is that there exists a physically motivated approximation for the matrix-elements of the self-energy and polarization outside of the `more relevant' space. 
To make this more explicit, let the available degrees of freedom like lattice site and orbital be $S$ and $s\subset S$ denote some small subspace thereof.
At this point there are no restrictions for these subspaces.
They may represent the local part, a limitation to a certain number of (correlated) bands or any combination thereof.

In the following, an uppercase character represents an index in $S$, a lowercase character an index in $s$.
The same character in upper- and lowercase represents the same index.
For example $A_{CD}+B_{cd}$ would be a shorthand notation for
\begin{align}
\begin{cases} \left(A_{CD}+B_{cd}\right)_{cd}= A_{cd}+B_{cd} \\
              \left(A_{CD}+B_{cd}\right)_{CD\not\in s} = A_{CD}
\end{cases}
\end{align}
With this notation, the usual Dyson equation can be written as
\begin{align}
  \label{eqn:Dyson}
  G_{AB}=&G^0_{AB}+G^0_{AC}\Sigma_{CD}G_{DB}\\
=&G^0_{AB}+G^0_{AC}(\Sigma^\mathrm{r}_{CD}+\Sigma^\mathrm{d}_{cd})G_{DB}\label{eqn:b}\\
=&\widetilde G^0_{AB}+\widetilde G^0_{Ac}\Sigma^\mathrm{d}_{cd}G_{dB},\label{eqn:a}
\end{align}
with
\begin{align}
  \widetilde G^0_{AB}=&G^0_{AB}+G^0_{AC}\Sigma^\mathrm{r}_{CD}\widetilde G^0_{DB}.\label{eqn:c}
\end{align}
The only restriction of the partitioning of $\Sigma_{CD}$ into its two constituents is that $\Sigma^\mathrm{d}$ must be non-zero on $s$ only.
We do not impose any conditions on $\Sigma^\mathrm{r}$; specifically, it can be zero or non-zero on $s$.
That this is an exact rewriting can be understood from the expression 
\begin{align}
  G^{-1}=\underbrace{G_0^{-1}-\Sigma^\mathrm{r}}_{=\widetilde G_0^{-1}}-\Sigma^\mathrm{d}.
\end{align}
If the aim is to evaluate Eq.~\eqref{eqn:a} on $s$, it can be rewritten to be completely contained within that space, 
\begin{align}
  G_{ab}=&\widetilde G^0_{ab}+\widetilde G^0_{ac}\Sigma^\mathrm{d}_{cd}G_{db}\label{eqn:d}\;,
\end{align}
thus providing the bare propagator $\widetilde G^0_{ab}$ on the $s$ space, which is effectively retarded through the channels in $S$ that are not contained in $s$ ($S\setminus s$).
It is easy to show that this reduces to the DMFT formalism when $S$ is chosen as the whole lattice, $s$ a single site of that and $\Sigma^\text{r}$ to be the periodic continuation of $\Sigma^\text{d}$.

Equations~\eqref{eqn:Dyson} through~\eqref{eqn:d} are equally valid for $W$, $U$, and $\Pi$ instead of $G$, $G^0$, and $\Sigma$, which leads to a description of the the effectively retarded $s$-space interaction.
Care has to be taken in that case of the rank~4 structure of the involved quantities, e.g., using the handling described 
in Sec.~\ref{sec:matrixformulation}. 

With this formalism in place, cRPA can be identified as ``$S$: all bands, $s$: a limited subset of those, and $\Pi^r$ the $G^0G^0$ polarization bubble outside $s$.'' 
EDMFT, in addition to the DMFT construction discussed above, includes the analog considerations for the effectively retarded local interaction.

The general formulation of the multitier approach is a versatile tool, and the implementation with $G^0W^0$@TIER~III + $GW$@TIER~II + QMC@TIER~I used in this paper and before\cite{Boehnke16When} harvests its full potential with the currently available tools.
This is just a snapshot of the development however and as better tools become available as solvers on the respective tiers, they can be systematically incorporated.

Table~\ref{tab:context} lists a number of formalisms that can be understood within this framework. Some complimentary approaches, like dual fermion/dual boson (DF/DB) based formalisms\cite{Rubtsov08Dual,Rubtsov12Dual,Stepanov16Local} and QUADRILEX\cite{Ayral16Mottb} are related, but cannot directly be cast into the multitier form. DF/DB are derived from a reformulation of the lattice action, which yields a different form of the Dyson equations with a non-local bare propagator and while QUADRILEX has a two-particle self-consistency cycle that modifies the impurity interaction, this is accomplished without introducing an explicit polarization in bosonic variables.

\section{Computational details}
\label{sec:compdet}
\subsection{TIER III-II}
In TIER III the one-shot $GW$ selfenergy as well as the effective interaction on the intermediate subspace ($U(\omega)$) are calculated within cRPA using the all electron FLAPW code \textsc{SPEX} \cite{Friedrich10Efficient,fleur}.
The intermediate subspace for the self-consistency loop on TIER II and I is defined using MLWFs as implemented in the \textsc{Wannier90} library.\cite{Marzari97Maximally,Mostofi08wannier90,Freimuth08Maximally,Sakuma13Symmetryadapted}
TIER II is treated using a custom finite-temperature Matsubara axis self-consistent $GW$ implementation on the intermediate subspace.

\subsubsection{TIER II: Matrix Formulation at Finite Temperature}
\label{sec:matrixformulation}
As single-particle basis for TIER II we use MLWFs, 
\begin{equation}
 \phi_{n\vec{q}}(\vec{r}) = \sum_{\vec{R}}e^{i\vec{q}\cdot\vec{R}}w_{n\vec{R}}(\vec{r}),
 \label{eq:wann}
\end{equation}
where $w_{n\vec{R}}(\vec{r})$ is a Wannier function centered in the unit cell specified by $\vec{R}$ and $n$ is an orbital index.

The two-particle quantities $W$ and $\Pi$ require a product basis.
In this work we use a restricted product basis of the form
\begin{equation}
 \Psi_{\alpha \vec{q}}(\vec{r}) = \sum_{\vec{R}}e^{i\vec{q}\cdot\vec{R}}w_{i\vec{R}}^{*}(\vec{r})w_{j\vec{R}}^{}(\vec{r}). 
 \label{eq:pb}
\end{equation}
where $\alpha=\{i,j\}$. Note that in general $\langle \Psi_{\alpha \vec{q}}(\vec{r}) | \Psi_{\beta \vec{q}}(\vec{r}) \rangle \neq \delta_{\alpha \beta}$.
Formally, it is possible to work in this nonorthogonal basis in the following way:
Define the matrix elements of the screened interaction as $W_{\alpha\beta}$ from \emph{downfolding} of the two-particle object $W(\vec{r},\vec{r}')$ by
\begin{equation}
  \label{eqn:downfolded}
  W_{\alpha\beta}=\int d\vec{r}d\vec{r}'\Psi_\alpha(\vec{r})W(\vec{r},\vec{r}')\Psi_\beta^\ast(\vec{r}')\;.
\end{equation}
Further define the matrix elements of the polarization $\Pi_{\alpha\beta}$ implicitly by
\begin{equation}
  \label{eqn:upfoldable}
  \Pi(\vec{r},\vec{r}')=\sum_{\alpha\beta}\Psi^\ast_\alpha(\vec{r})\Pi_{\alpha\beta}\Psi_\beta(\vec{r'}).
\end{equation}
Hence, $\Pi_{\alpha\beta}$ is an object that is \emph{upfoldable} to a real-space representation $\Pi(\vec{r},\vec{r}')$. 
The nonorthogonality of the product basis $\Psi_{\alpha}(\vec{r})$ ($\vec{k}$-index dropped for simplicity) implies that
\begin{equation}
\Pi_{\alpha \beta}\neq \int d\vec{r}d\vec{r}'\Psi_\alpha(\vec{r})\Pi(\vec{r},\vec{r}')\Psi_\beta^\ast(\vec{r}').
\end{equation}
However, this does not 
prevent us from mapping
 equations of the type
\begin{equation}
  \label{eqn:Winr}
  W(\vec{r},\vec{r}')=U(\vec{r},\vec{r}')+\int d\vec{r_1}d\vec{r_2}U(\vec{r},\vec{r}_1)\Pi(\vec{r}_1,\vec{r}_2)W(\vec{r}_2,\vec{r}')
\end{equation}
to matrix operations in the product basis notation by applying Eq.~\eqref{eqn:downfolded}:
\begin{align}
&\int\underbrace{d\vec{r}d\vec{r}'\Psi_\alpha(\vec{r})W(\vec{r},\vec{r}')\Psi^\ast_\beta(\vec{r}')}_{W_{\alpha\beta}}=U_{\alpha\beta}+\nonumber \\
&\int d\vec{R}\lefteqn{\underbrace{\phantom{\Psi_\alpha(\vec{r})U(\vec{r},\vec{r}_1)\Psi^\ast_{\alpha'}(\vec{r}_1)}}_{U_{\alpha\alpha'}}}\Psi_\alpha(\vec{r})U(\vec{r},\vec{r}_1)\overbrace{\Psi^\ast_{\alpha'}(\vec{r}_1)\Pi_{\alpha'\beta'}\lefteqn{\underbrace{\phantom{\Psi_{\beta'}(\vec{r}_2)W(\vec{r}_2,\vec{r}')\Psi^\ast_\beta(\vec{r}')}}_{W_{\beta'\beta}}}\Psi_{\beta'}(\vec{r}_2)}^{\Pi(\vec{r}_1,\vec{r}_2)}W(\vec{r}_2,\vec{r}')\Psi^\ast_\beta(\vec{r}'),
\end{align}
where $d\vec{R}=d\vec{r}_1d\vec{r}_2d\vec{r}d\vec{r}'$.
As a rule of thumb, convolutions alternating between upfoldable and downfolded quantities map to matrix operations.
This mapping to matrix operations also implies a corresponding definition for the inversion of these quantities.

With these definitions and using the Wannier basis in Eq.~\eqref{eq:wann} for the single-particle quantities
and the product basis in Eq.~\eqref{eq:pb} for the two-particle quantities the integral equations \eqref{eq:HedinG} and \eqref{eq:HedinW}-\eqref{eq:PRPA} can be mapped to
the following matrix equations (summation over repeated indices assumed):
\begin{align}
&\Sigma_{ik}(\vec{q},\tau) = -\sum_{\vec{k}} G_{jl}(\vec{k},\tau) W_{ijkl}(\vec{q}-\vec{k},\tau), \label{eq:GWSigma} \\
&G_{kl}(\vec{q},i\nu_m) = G^{0}_{kl}(\vec{q},i\nu_m) \nonumber\\
&\hspace{15mm}+ G^{0}_{km}(\vec{q},i\omega) \Sigma_{mn}(\vec{q},i\nu_m) G_{nl}(\vec{q},i\nu_m), \label{eq:GWdyson} \\
&\Pi_{mm'nn'}(\vec{q},\tau) = \sum_{\vec{k}}G_{mn}(\vec{k},\tau)G_{n'm'}(\vec{k}-\vec{q},-\tau), \label{eq:GWPol} \\
&W_{\alpha \beta}(\vec{q},i\omega_n) = U_{\alpha \beta}(\vec{q},i\omega_n) \nonumber\\
&\hspace{15mm}+ U_{\alpha \gamma}(\vec{q},i\omega_n) \Pi_{\gamma \eta}(\vec{q},i\omega_n) W_{\eta \beta}(\vec{q},i\omega_n). \label{eq:GWWdyson}
\end{align}

In TIERs~II~and~I it has to be ensured that a sufficiently large number of Matsubara frequencies and a fine enough imaginary time discretization are used.
The substantial spectral weight at large energies that TIER~II inherits in its bare propagators from TIER~III requires a much larger number than for e.g.\ LDA+DMFT calculations.
For SrMoO$_3$ and SrVO$_3$ discussed in later sections, we used at least 1500~Matsubara frequencies.

\subsubsection{Gamma point handling}

Both the bare Coulomb interaction and the screened Coulomb interaction diverge as $1/\vec{k}^2$ for $k \rightarrow 0$.
Because of this the $\Gamma$-point needs to be handled with special care.
In the $GW$ calculation using the FLAPW code \textsc{SPEX} in TIER III the treatment of the $\Gamma$-point is greatly simplified by making use of the analytic forms of both the bare Coulomb interaction and the bare Green's function.
By a basis change to the Coulomb eigenbasis the divergence is restricted to a single eigenvalue and can be treated separately using so-called kp-perturbation theory.\cite{Friedrich10Efficient} 
Furthermore, the value of the divergent term of the Coulomb interaction is 
redefined as the integral of the corresponding term in a small region around the $\Gamma$-point, whose size is determined by the $\vec{k}$-mesh:
\begin{align}
 V_{div}(\vec{k}=\Gamma) \to \frac{V}{8\pi^3}\int_{BZ}\frac{1}{k^2}d^3k - \sum_{k\neq 0} \frac{1}{k^2}.
\label{eq:div1}
\end{align}
With this definition the integral of $V$ over the Brillouin zone will be correctly reproduced by the $\vec{k}$-point sampling. 
By making use of the analytic form of the bare polarization a similar expression can be derived
for $U(\omega)$ \cite{Friedrich10Efficient}:

\begin{align}
U_{\mathrm{div}}(\vec{k}\rightarrow \Gamma,\omega) \to \frac{V}{8\pi^3}\int_{BZ}\frac{c(\omega)}{k^2}d^3k - \sum_{k\neq 0} \frac{c(\omega)}{k^2},
\label{eq:div2}
\end{align}
where the constant $c(\omega)$ is derived from the head element of the polarization matrix, $\langle E_0|\Pi |E_0 \rangle$, with $|E^0\rangle$ being the eigenfunction that corresponds to the divergent eigenvalue of the Coulomb interaction.

In TIER II we perform a self-consistent $GW$ calculation on the intermediate subspace starting from a downfolded frequency-dependent interaction $U(i\omega_n)$ and retarded bare propagator $G^0(i\nu_m)$ calculated in TIER III, and therefore we lack analytic expressions for both the Green's function and the interaction.
This forces us to resort to cruder approximations for the $\Gamma$-point than in TIER III.

When $U(i\omega_n)$ is projected onto the intermediate subspace the divergent contribution will no longer be contained in a single matrix element,
but will in general give a contribution to all matrix elements of $U(i\omega_n,\vec{k}=\Gamma)$.
Furthermore, since the intermediate subspace is much smaller than the space spanned by the complete product basis in TIER III it is not possible to isolate the divergent term by a simple rotation.

The dielectric function $\epsilon(\vec{k})$, on the other hand, is a smooth function of $\vec{k}$ and will generally have an extremal point for $\vec{k}=\Gamma$ and can therefore be approximated as a constant function of $\vec{k}$ in a small region around the $\Gamma$-point.
Hence, if the \vec{k}-mesh is chosen sufficiently dense, the inverse dielectric function may be approximated as
\begin{align} 
\epsilon^{-1}_{\alpha \beta}(i\omega_n,\vec{k}<\vec{k}_{\mathrm{cutoff}}) \approx \bar{\epsilon}^{-1}_{\alpha \beta}(i\omega_n), 
\end{align}
where $\bar{\epsilon}^{-1}(i\omega_n)$ is the average value of $\epsilon^{-1}(i\omega_n,\vec{k})$ at the boundaries defined by $\vec{k}_{\mathrm{cutoff}}$. 
Using this approximation for the inverse dielectric function the fully screened interaction can readily be calculated as
\begin{align} 
\bar{W}_{\alpha \beta}(i\omega_n,\vec{k}=\Gamma) = \sum_{\gamma} \bar{\epsilon}^{-1}_{\alpha \gamma}(i\omega_n,\vec{k}=\Gamma) \bar{U}_{\gamma \beta}(i\omega_n,\vec{k}=\Gamma),
\end{align}
where the notation $\bar{U}$ is used to emphasize that the divergent term of the interaction is redefined according to Eq.~(\ref{eq:div2}).

\begin{figure}
  \centering
  \includegraphics{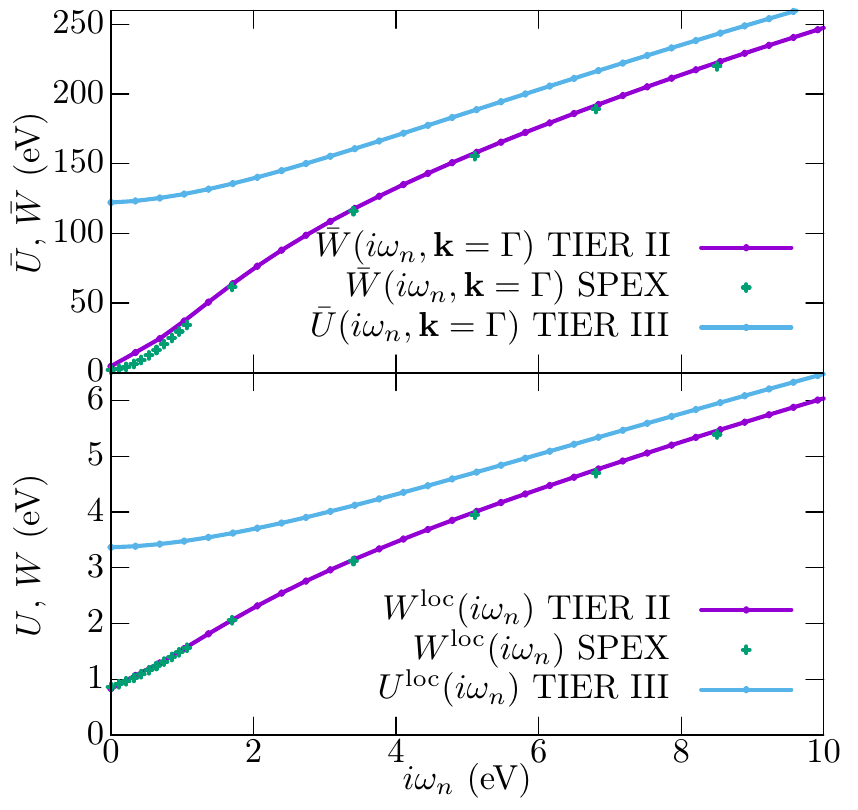}
  \caption{
  Diagonal matrix elements of the screened interaction of SrVO$_3$ obtained in the first iteration ($W^0$) using the custom $GW$ code in TIER II at inverse temperature $\beta=15\frac{1}{\mathrm{eV}}$ compared to the corresponding quantity calculated in SPEX at zero temperature.
  Also shown for comparison is the bare interaction $U$ from SPEX. 
   All calculations were done using an $8\times8\times8$ \vec{k}-point mesh. 
   }
  \label{fig:Wgamma}
\end{figure}

In Fig.~\ref{fig:Wgamma} we compare the fully screened interaction calculated in the custom $GW$ implementation for TIER II with the results from \textsc{SPEX} for the benchmark material SrVO$_3$.
One can see that the local part of the screened interaction agrees very well with the results from \textsc{SPEX}. Also the $\Gamma$-point contribution shows remarkable agreement in spite of the rather crude approximations in
TIER II.

\subsubsection{The Hartree contribution}
During the self-consistent calculations the density within the intermediate subspace, and thus the Hartree energy, will change.
In Eq.~\eqref{eqn:fullG} the Hartree contribution was hidden in the self-energies. 
In this section we derive the Hartree contribution for a downfolded model and discuss the technical details of the current implementation.

The Hartree  potential is given by
\begin{align}
 V^H(r) = \int \rho(r')v(r-r') d^3r' .
\end{align}
We now take the matrix elements of the above expression and use
\begin{align}
\rho(r)=& -\sum_\sigma G^\sigma(r,r,\tau=\beta^-) \\
      =& -\sum_{\sigma i j \vec{k}} \phi_{i\vec{k}}(r)\phi^{*}_{j\vec{k}} (r) G_{ij}^\sigma(\vec{k},\tau=\beta^-),
\end{align}
 \begin{align}
  &\langle \phi_{i\vec{k}} | V^H(r) | \phi_{j\vec{k}} \rangle = \iint d^3rd^3r' \rho(r')\phi^{*}_{i\vec{k}}(r) V(r-r') \phi_{j\vec{k}}(r) \nonumber \\
&=-\sum_{\sigma l m q}  G_{lm}^\sigma(q,\tau=\beta^-)  \nonumber \\ 
  &\times \iint d^3rd^3r' \phi^{*}_{i\vec{k}}(r) \phi_{j\vec{k}}(r)  V(r-r') \phi_{l\vec{q}}(r') \phi^{*}_{m\vec{q}} (r') .
\label{Vhartree}
\end{align}
Inserting the definition of the basis functions  [Eq.~\eqref{eq:wann}] into the above equation yields 
\begin{align}
&\langle \phi_{i\vec{k}} | V^H(r) | \phi_{j\vec{k}} \rangle =\nonumber \\
 &-\sum_{\sigma l m q}  G_{lm}^\sigma(q,\tau=\beta^-) \sum_ {\vec{R}_1 \vec{R}_2 \vec{R}_3 \vec{R}_4} e^{i\vec{k}\cdot(\vec{R_2}-\vec{R_1})}e^{i\vec{q}\cdot(\vec{R_3}-\vec{R_4})} \nonumber \\
&\times \iint d^3rd^3r' w^{*}_{i\vec{R}_1}(r) w_{j\vec{R}_2}(r)  V(r-r') w_{l\vec{R}_3}(r') w^{*}_{m\vec{R}_4} (r').
\label{eq:Hartree_full}
\end{align}

As long as the Wannier functions are sufficiently localized this term will be dominated by the $\vec{R}_1=\vec{R}_2$ and $\vec{R}_3=\vec{R}_4$ contribution, in which case the Hartree term reduces to
\begin{align}
\langle \phi_{i\vec{k}} | V^H(r) | \phi_{j\vec{k}} \rangle =& -\sum_{\sigma l m q}  G_{lm}^\sigma(q,\tau=\beta^-) V_{ijlm}(\vec{q}=0) \nonumber \\
=&\sum_{l m}  n_{lm} V_{ijlm}(\vec{q}=0).
\label{eq:Hartree_approx}
\end{align}

Since a Hartree contribution is already contained in the LDA Hamiltonian in TIER III we only need to consider the correction to the Hartree term in TIER II
due to the change in the density in the intermediate subspace ($\Delta V^\mathrm{H}_{ij}$).
Since the change in the density integrates to zero the divergent term arising from $\mathbf{q}=0$ in the Coulomb potential is eliminated,
\begin{align}
\Delta V^\mathrm{H}_{ij} =\sum_{l m}  (n_{lm}-n^\mathrm{LDA}_{lm}) \tilde{V}_{ijlm}(\vec{q}=\Gamma).
\label{eq:DeltaHartree}
\end{align}
The additional tilde on the interaction matrix is used to emphasize that the constant (divergent) eigenfunction of the interaction has been removed before projecting onto the Wannier basis.

\begin{figure*}
\begin{center}
\includegraphics[width=0.6\textwidth]{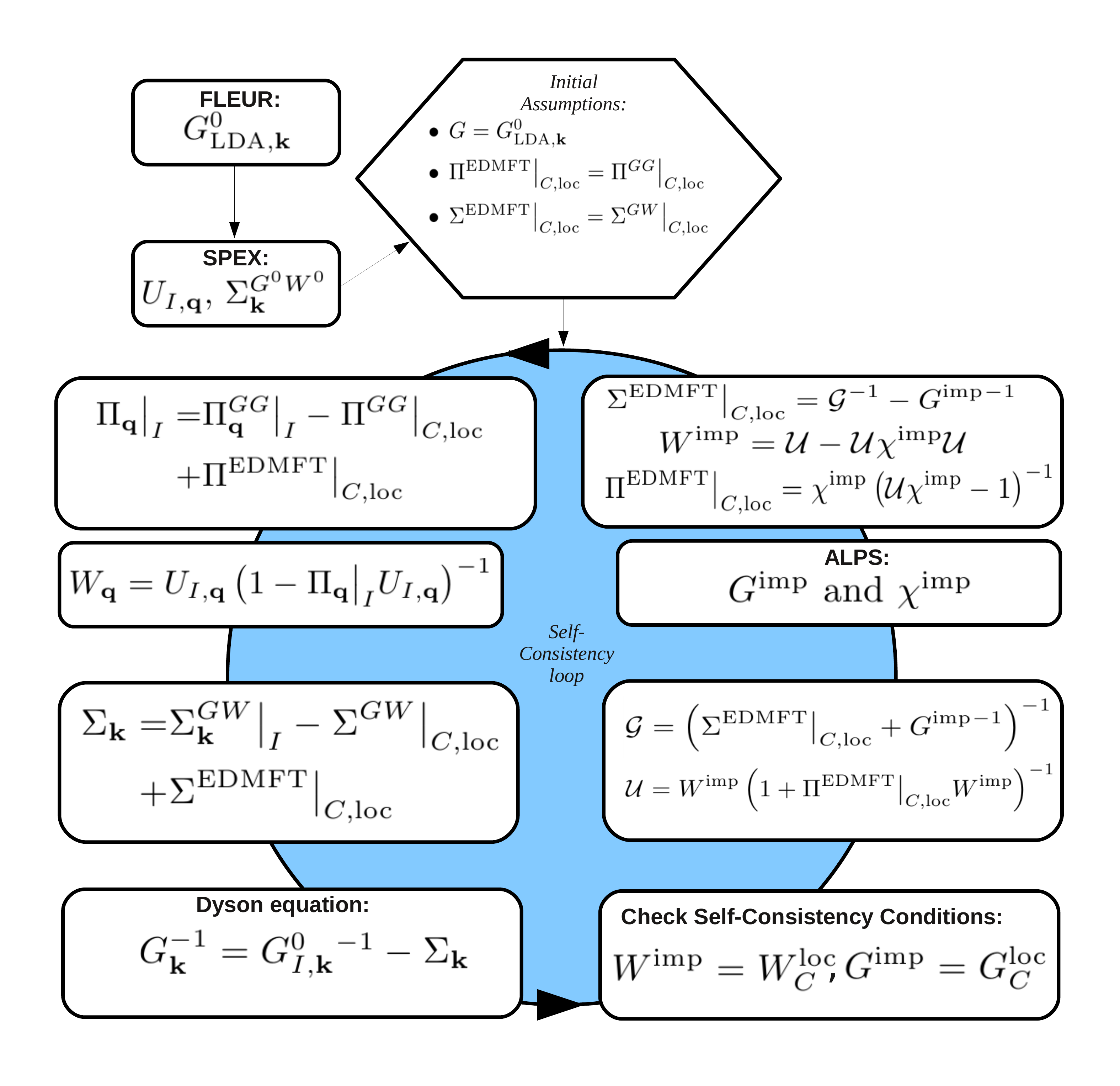}
\end{center}
\caption{Schematic illustration of the different steps in the multitier self-consistent $GW$+EDMFT simulation.}
\label{fig:SCcycle}
\end{figure*}

\subsection{TIER I}
The relevant low-energy local problem in the presence of effective bare propagators and interactions screened through all the degrees of freedom that were systematically removed in TIERs~I and~II is primarily due to the retarded effective interaction nontrivial to solve and some approximations cannot be avoided at this point.
In our calculations we employ the CT-Hyb~\cite{Werner06ContinuousTime,Werner07Efficient,Gull11Continuoustime} quantum Monte-Carlo algorithm as implemented in \textsc{Alps}.\cite{Bauer11Alps,Alps,Hafermann13Efficient} This implementation features the treatment of dynamic screening of the interaction by a numerically cheap reweighting of the Monte-Carlo configurations, as introduced in Ref~\onlinecite{Werner10Dynamical}. 
This approach handles the retardation effects exactly, 
as long as the screening affects only the monopole part of the interaction.
To prevent systematic errors from the fitting of high frequency tails we make use of the Legendre polynomial based compact representation of the Green's function.\cite{Boehnke11Orthogonal}
This is of particular importance in multitier $GW$+EDMFT, where structure can be expected up to very high frequencies.
A fitting procedure is at risk of losing those features.

The ALPS solver,\cite{Hafermann13Efficient} while efficient, restricts the impurity interactions to density-density like terms (``segment picture'') and it allows only the measurements of density-density like contributions to the impurity charge susceptibility.

The first restriction has nontrivial consequences, since the $\mathcal{U}_{abab}$ terms give rise to the $-J\delta{\sigma\sigma'}n_{a\sigma}n_{b\sigma'}$ contribution to \eqref{eqn:kanamoriH} due to the equal-spin term in Eq.~\eqref{eqn:tensorU}, which is of density-density form and should be retained, but also give rise to the spin-flip contribution for the opposite-spin term in \eqref{eqn:tensorU}, which can not be dealt with in the implementation of Ref.~\onlinecite{Hafermann13Efficient}. Thus, there will be a discrepancy between the $\mathcal{U}$-tensor used in the self-consistency and the Hamiltonian that ultimately enters the impurity problem.

The second restriction, although more severe at first glance, turns out to be of minor importance. The orbital-resolved impurity charge susceptibility $\chi^\text{imp}$ enters the EDMFT scheme only through the formula 
\begin{equation}
  \label{eqn:Wimp}
  W^\mathrm{imp}=\mathcal{U}-\mathcal{U}\chi^\mathrm{imp}\mathcal{U}\;,
\end{equation}
from which it becomes apparent that limiting $\chi^\mathrm{imp}$ to $\chi^\mathrm{imp}_{iijj}$ type contributions merely means that only the $U$ and $U'$ will be screened while $J$ remains  unscreened, which is a physically reasonable approximation (it is much harder to screen $l\neq 0$ components of a charge distribution). To partially remedy this restriction, the non-density-density components of the local polarization are in our calculations taken into account on the $GG$-bubble level.

Additionally, the solver used in the current implementation allows for an orbital diagonal hybridization only, which is however the case by construction for all systems discussed in this paper. It inhibits us however from applying the formalism to a cluster of atoms as the local $\mathcal{C}$ space on TIER~I. 
Recent developments aim to remove some of these restrictions~\cite{Steiner15Double} by introducing a segment-picture-based expansion in the hybridization and Hund's coupling parameter $J$. This approach can efficiently handle the density-density components of retarded $U$, $U'$, and $J$ interactions, while retarded spin-flip terms produce a severe sign-problem. Other choices for impurity solvers that can deal with the retarded interaction and bare propagator include the non- and one-crossing approximations (NCA/OCA)\cite{Grewe81Diagrammatic,Haule02Pseudogap,Golez15Dynamics,Werner08Phonon} 
that are viable options in the large-$U$ limit.
However, since the self-energies produced by these approximate solvers do not contain all the local diagrams that have been removed in TIER~II, but only a partial summation, numerical artifacts can be expected.\cite{Golez17Nonequilibrium} 

\subsection{Self-consistency cycle}
In Fig.~\ref{fig:SCcycle} the full flow of the calculations is illustrated.
First a DFT calculation is performed using the FLAPW-code \textsc{FLEUR}.\cite{fleur} Then, a low-energy model is defined using MLWFs\cite{Marzari97Maximally,Mostofi08wannier90,Freimuth08Maximally,Sakuma13Symmetryadapted}
and the model interaction as well as the $G^0W^0$ selfenergy is computed with the \textsc{SPEX} code\cite{Friedrich10Efficient,fleur}. 
This defines the bare propagators for the intermediate subspace [$G_{I,\vec{k}}^{0}{}^{-1}$ and $U_{I,\vec{q}}^{-1}$ in Eqs.~\eqref{eqn:fullG}-\eqref{eqn:fullW}].
With the initial assumptions
\begin{align}
\Sigma ^{\mathrm{EDMFT}}\big|_{C,\mathrm{loc}} &= \Sigma ^{GW}\big|_{C,\mathrm{loc}},
\\ \Pi^\mathrm{EDMFT}\big|_{C,\mathrm{loc}} &= \Pi^{GG}\big|_{C,\mathrm{loc}},
\end{align}
the selfenergy and screened interaction are computed according to Eqs.~\eqref{eqn:fullG}-\eqref{eqn:fullW}. 
If the self-consistency conditions (Eq.~\ref{eq:GWEDMFT-SCcond}) are not fulfilled the impurity selfenergy ($\Sigma ^{\mathrm{EDMFT}}\big|_{C,\mathrm{loc}}$) and polarization ($\Pi^\mathrm{EDMFT}\big|_{C,\mathrm{loc}}$) are computed using the \textsc{Alps} CT-Hyb impurity solver.\cite{Werner06ContinuousTime,Gull11Continuoustime}  
These values replace the initial guesses above, and yield a new Green's function and screened interaction.
The scheme is then iterated until the self-consistency conditions are fulfilled.
This cycle on TIERs~II and~I is implemented using the TRIQS framework \cite{Parcollet15Triqs}.
In the calculations presented in this paper, we did not have to employ any mixing to converge the results.
While multiple solutions may exist \cite{Stan15Unphysical}, there were no indications of additional unphysical solutions.

\subsection{Analytic continuations}
All analytic continuations were done using the maximum entropy method (MaxEnt)\cite{Bryan90Maximum,Gubernatis91Quantum} as implemented in Ref.~\onlinecite{MaxEnt}, except for the results in Fig.~\ref{fig:dimerUSw} 
which were obtained using a Pad\'e analytic continuation \cite{Vidberg77Solving}, since MaxEnt cannot handle noncausal spectral functions.

\section{Causality considerations}
\label{sec:causality}

\begin{figure}
  \includegraphics{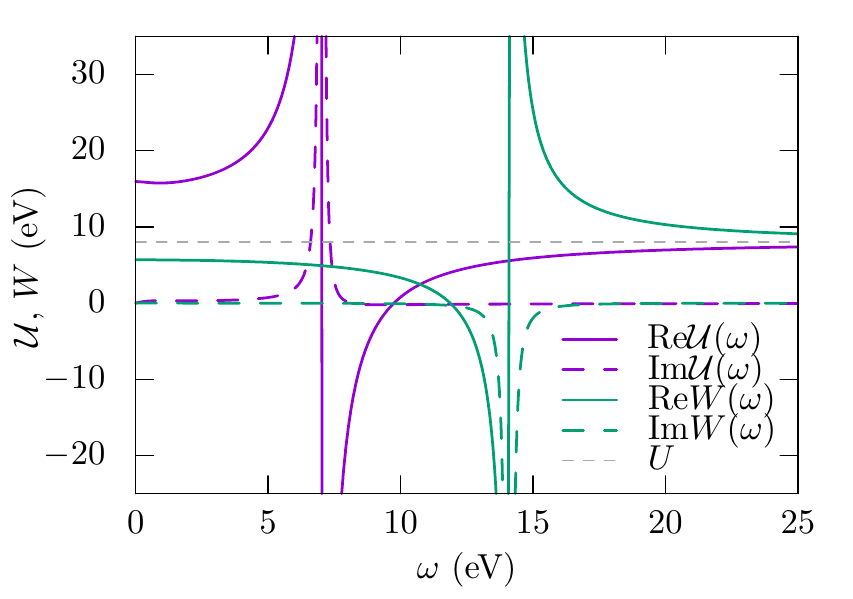}
  \caption{Non causal impurity interaction and the local fully screened interaction of the local dimer problem for $U=8$, $V=1$, $t=5$ and $\beta=5$.}
  \label{fig:dimerUSw}
\end{figure}

A recurring feature in $GW$+EDMFT based model and material calculations is that the impurity interaction~$\mathcal{U}(z)$, $z\in\mathbb{C}$ can become non-causal, that is, it may feature poles in the upper half of the complex plane instead of the lower half.
This leads to negative spectral weight of the bosonic modes $-\Im\mathcal{U}(\omega)$.
An example is shown in Fig.~\ref{fig:dimerUSw}, where $\Im\mathcal{U}(\omega)$ is positive in contrast to $\Im W(\omega)$, which has the expected negative peak.
This behavior can also been observed in Sec.~\ref{sec:SrMoO3} for SrMoO$_3$, but earlier examples include the simulations of SrVO$_3$~\cite{Boehnke16When} and model studies \cite{Ayral13Screening,Huang14Extended,vanLoon16Double,Stepanov16Selfconsistent,Ayral17Influence}, where this property can be seen at least indirectly.
From the Hilbert transform of $\mathcal{U}(\omega)$ it follows that 
\begin{align}
&\mathcal{U}(i\omega_{n+1})-\mathcal{U}(i\omega_n) \nonumber \\
&= -\frac{2}{\pi}\int_0^\infty d\omega\frac{\Im\left[\mathcal{U}(\omega)\right] \omega(\omega_{n+1}^2-\omega_{n}^2)}{(\omega_{n+1}^2 + \omega^2)(\omega_n^2 + \omega^2)} \nonumber \\
&\equiv -\frac{2}{\pi}\int_0^\infty d\omega\Im\mathcal{U}(\omega)f(\omega),
\label{eqn:causalitycond}
\end{align}
where $f(\omega)$ is a positive function on the interval $\omega \in [0,\infty)$.
Hence, if $\mathcal{U}(i\omega_{n+1})<\mathcal{U}(i\omega_n)$ for any Matsubara frequency ($i\omega_n$) the spectral weight must be noncausal in some frequency range. The converse is not necessarily true.

From the mean-field value of the (static) screened effective interaction for the $U$-$V$ model, $U_\mathrm{eff}=U-zV$ ($z$ being the coordination number),  
one can expect noncausal interactions to appear for peculiar choices of parameters, like an attractive nearest-neighbor interaction $V$.
Given however that the above-mentioned examples with noncausal impurity interactions are not in this regime, the question arises as to whether the noncausality is a consequence of the $GW$ approximation for the nonlocal components of the self-energy or a generic consequence of DMFT-type local approximations. 

To answer this question, we consider the simple case of a dimer problem, where the definition of the effective impurity model is merely a formal step, and where we can take \emph{all} nonlocal diagrams into account. 
Any formalism that maps to an auxiliary (impurity) problem does so by defining the self-consistency conditions. In the case of a derivation from a $\Psi$ functionalv \cite{Almbladh99Variational}, this is unambiguously\cite{Golez15Dynamics} given by
\begin{align}
  \label{eqn:selfconsistencyequationG}
  G_\mathrm{loc}(i\nu_n)=&G_\mathrm{imp}(i\nu_n),\\
  \label{eqn:selfconsistencyequationW}
  W_\mathrm{loc}(i\omega_n)=&W_\mathrm{imp}(i\omega_n)\;.
\end{align}
Dual boson uses the same conditions,\cite{Stepanov16Local} although originally a different procedure was proposed.\cite{Rubtsov12Dual}
In addition to these self-consistency conditions, a local formalism needs to define an approximation to the local $G$ and $W$ through lattice properties and impurity observables. 
As introduced before, $GW$+EDMFT takes the impurity self-energy and polarization and augments them by the nonlocal components within the $GW$ approximation.
Dual Boson introduces a dual expansion which is then cut at some perturbation order.

In this section we instead use the exact values for the local $G$ and $W$ of a dimer, and determine the 
impurity bath Green's function $\mathcal{G}$ and interaction $\mathcal{U}$ such that the solution of the impurity model reproduces these exact local $G$ and $W$.
Specifically, for the dimer with Hamiltonian
\begin{align}
H_\text{dimer}=&U(n_{1\uparrow}n_{1\downarrow}+n_{2\uparrow}n_{2\downarrow})+V n_1 n_2 \nonumber\\
& \hspace{15mm} -t \sum_\sigma(d^\dagger_{1\sigma} d_{2\sigma} + d^\dagger_{2\sigma} d_{1\sigma})
\end{align}
we compute 
$G_\mathrm{ex}$ and $W_\mathrm{ex}$ using exact diagonalization.\cite{pyED}
Then the functions $\mathcal{G}$ and $\mathcal{U}$ (i.e. the corresponding bath parameters) are determined numerically in such a way that the solution of the impurity model reproduces the local parts $G_\mathrm{ex,loc}$ and $W_\mathrm{ex,loc}$.
This is achieved by 
executing a self-consistency loop with the local $G$ and $W$ fixed to their exact values (here $i$ is the iteration)
\begin{align}
  \mathcal{G}_{i+1}=&(G_\mathrm{ex,loc}+\Sigma_{\mathrm{imp},i})^{-1},\\
  \mathcal{U}_{i+1}=&(W_\mathrm{ex,loc}+\Pi_{\mathrm{imp},i})^{-1}\;.
\end{align}
Figure~\ref{fig:dimerUSw} shows the impurity interaction on the real frequency axis for the dimer parameters $U=8$, $V=1$, $t=5$ and $\beta=5$.
The results were broadened slightly for better visibility.
The imaginary part is purely noncausal in this case, while for model and material calculations in $GW$+EDMFT we typically see a mixed behavior of causal or noncausal spectral weight at low frequencies and causal spectral weight at high frequencies.
Consistent with the strong antiscreening mode, the static impurity interaction is significantly \emph{increased} relative to the bare local interaction $U$.

Let us stress that the arguments presented here do \emph{not} depend on any particular non-local approximation. 
The \emph{exact} result for the dimer shows that given the self-consistency equations~\eqref{eqn:selfconsistencyequationG} and~\eqref{eqn:selfconsistencyequationW}, the auxiliary impurity problem can have a noncausal retarded interaction.  Hence, this is an intrinsic possibility of any local approximation.

\section{Results}
\label{sec:results}
\subsection{The cubic perovskites SrMoO$_3$ and SrVO$_3$}
\label{sec:SrMoO3}
\begin{figure}
  \includegraphics{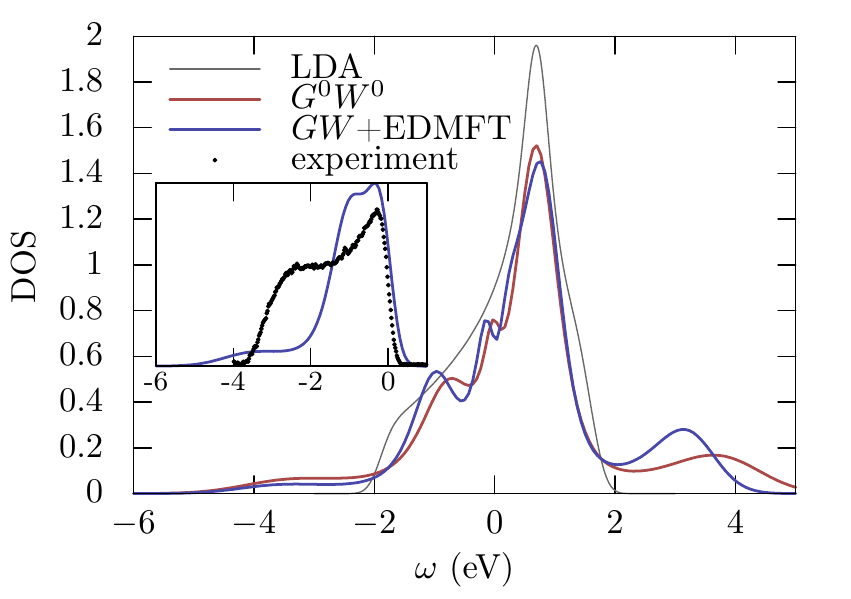}
  \caption{Spectral function of SrMoO$_3$. The experimental data for the inset is taken from Ref.~\onlinecite{Wadati14Photo}. The $GW$+EDMFT result in the inset has been adjusted to show the photoemission spectrum part only and a Gaussian filter has been applied to match the experimental resolution.}
  \label{fig:SrMoO3dos}
\end{figure}

\begin{figure*}
  \includegraphics{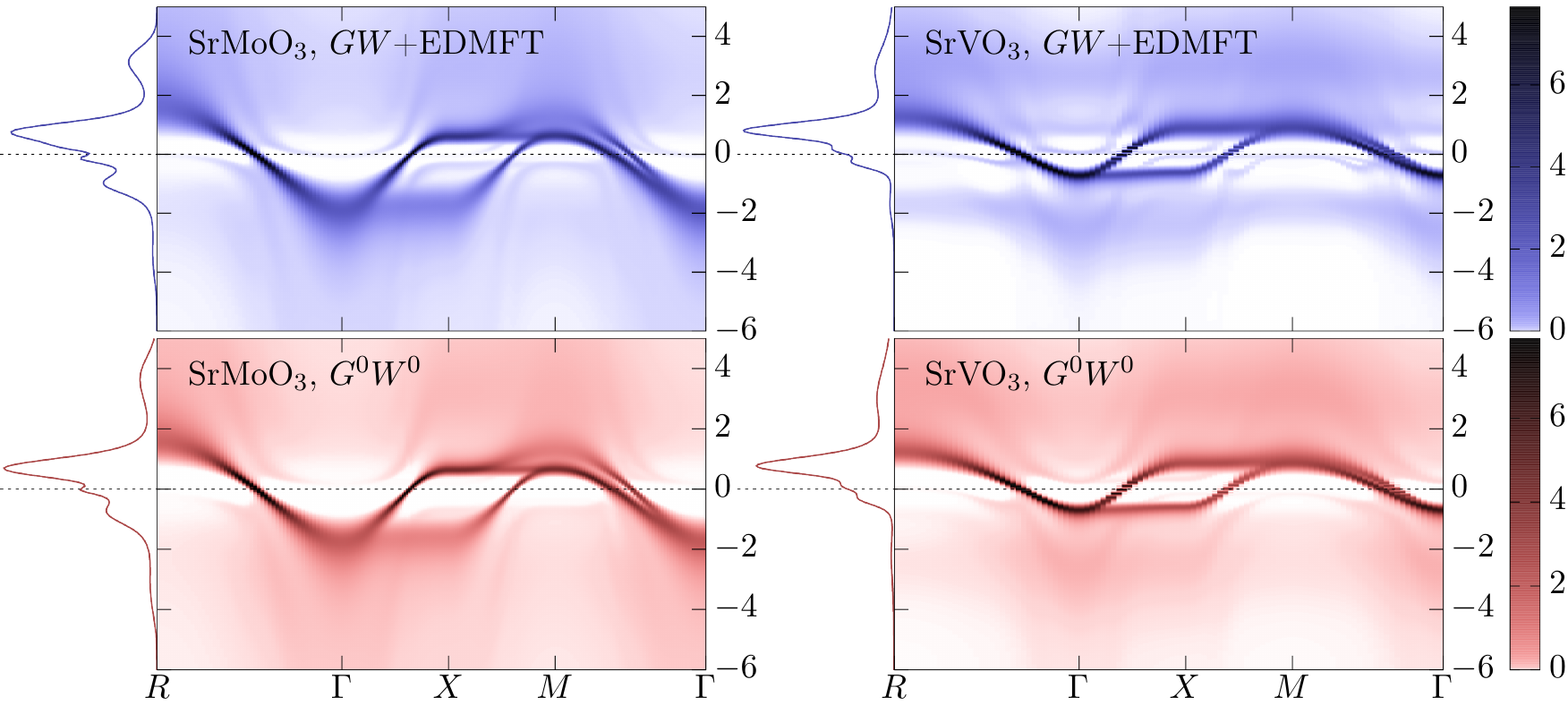}
  \caption{Spectral function of SrMoO$_3$ (left) and SrVO$_3$ (right). The SrVO$_3$ data are adapted from Ref.~\onlinecite{Boehnke16When}.}
  \label{fig:SrMoO3Akw}
\end{figure*}

\begin{figure}
  \includegraphics{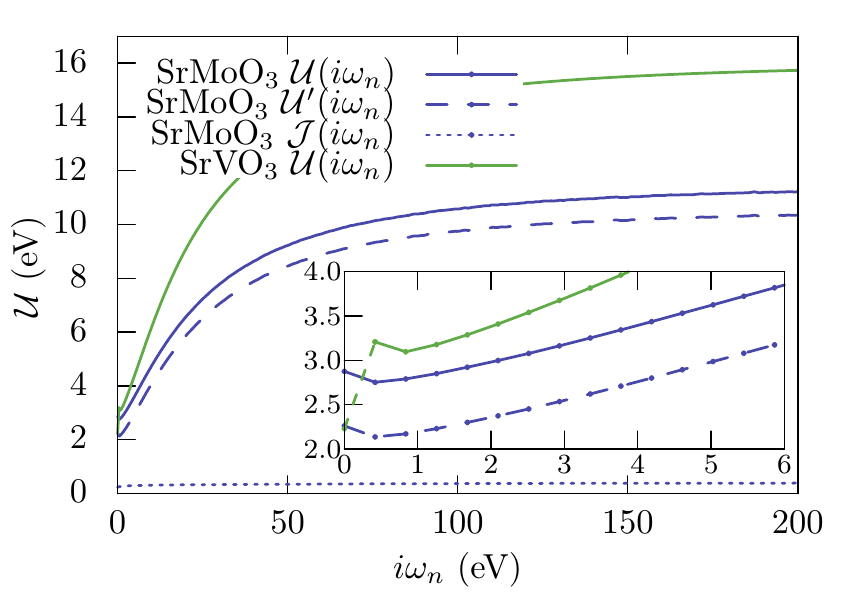}
  \caption{Different Kanamori-style components of the effective impurity interaction (see Eq.~\eqref{eqn:kanamoriU} for the naming convention) for SrMoO$_3$. For comparison we also show as a green line the result for SrVO$_3$, adapted from Ref.~\onlinecite{Boehnke16When} with the pole on the Matsubara axis (see text).}
  \label{fig:SrMoO3USijkl}
\end{figure}

\subsubsection{Spectral functions}

SrMoO$_3$ crystallizes in a cubic perovskite structure.\cite{Macquart2010Neutron} 
The conduction states originating from the Mo 4$d$ states are of $t_{2g}$ character and occupied by two electrons.
In LDA the $t_{2g}$ states form well-isolated bands around the Fermi energy with a bandwidth of roughly 3.8~eV.
SrMoO$_3$ is in many respects similar to the previously studied SrVO$_3$ \cite{Boehnke16When} but has twice the filling of the $t_{2g}$ shell
and also a slightly larger bandwidth. The larger bandwidth is expected  since  
the Mo 4$d$-states in SrMoO$_3$ 
 are less localized than the corresponding V 3$d$ states in SrVO$_3$.

The calculations in this section were performed for the inverse temperature $\beta=15\frac{1}{\mathrm{eV}}$ corresponding to the temperature $\approx774$K.

The photoemission spectra (PES) of SrMoO$_3$ measured in Ref.~\onlinecite{Wadati14Photo} (Inset in Fig.~\ref{fig:SrMoO3dos}) show a well defined quasiparticle peak together with a weak shoulder structure around $-2.5$~eV. 
Even though the specific heat coefficient is renormalized to approximately twice its LDA value 
the PES spectrum does not show any clear band narrowing compared to LDA.
The hump around $-2.5$~eV is not seen in bandstructure calculations and therefore presents a clear sign of electron correlations within the $t_{2g}$ band. However, in Ref.~\onlinecite{Wadati14Photo} it was shown that this structure cannot be reproduced in LDA+DMFT using any realistic values of the instantaneous local Hubbard interaction $U$.
Only with an unphysically large interaction of 5~eV a similar structure appeared in the LDA+DMFT spectra, but the band narrowing in this case was much too large. 
This led the authors to speculate that the shoulder structure was of plasmonic rather than Hubbard band character.\cite{Wadati14Photo}

The $GW$+EDMFT approach includes both the strong local correlations and the long-range screening effects within a single unified framework and can therefore capture both Mott physics and plasmonic excitations, as well as sub-plasmonic features originating from fluctuations within the correlated space, on an equal footing.
Furthermore, the multitier approach used in the present work does not include any adjustable parameters and therefore provides an ideal tool to distinguish plasmonic features from Hubbard bands.
When applied to SrVO$_3$ we already demonstrated in Ref.~\onlinecite{Boehnke16When} that the inclusion of long-range correlations changes the interpretation of the side-bands; the satellites in SrVO$_3$ should be interpreted as plasmonic satellites rather than Hubbard bands.
The situation in SrMoO$_3$ seems even more clearcut with no side-bands at all in plain DMFT.

In Fig.~\ref{fig:SrMoO3dos} we show the local spectral function for SrMoO$_3$ computed with the present multitier self-consistent $GW$+EDMFT scheme together with the experimental photoemission spectra from Ref.~\onlinecite{Wadati14Photo}.
The occupied part agrees well with the photoemission spectra and the shoulder structure around $-2.5$~eV is clearly visible.
We also predict a satellite feature in the unoccupied part of the spectrum, potentially visible in inverse photoemission experiments, centered at roughly 3~eV. 

The plasmonic signature for SrMoO$_3$ in the experimental data is more pronounced than what we find in our $GW$+EDMFT calculations.
The same is true for SrVO$_3$, where early measurements suggested\cite{Morikawa95Spectral} a very strong lower satellite
but later measurements with higher photon energies reveal a much reduced satellite intensity compared with the quasiparticle weight\cite{Sekiyama04Mutual}.
However, it is only very recently,\cite{Backes16Hubbard} that the experimental ratio in spectral weight between the lower satellite and the occupied part of the quasiparticle has been reexamined and found to be closer to 1:3, which is in reasonable agreement with the $GW$+EDMFT calculations in Ref.~\onlinecite{Boehnke16When} that suggest a ratio of~$\sim$1:4.
The difference to previous measurements has been attributed to oxygen vacancies in the system.
A similar situation seems to be present in other transition metal oxides like SrTiO$_3$\cite{Backes16Hubbard} and its interfaces.\cite{Dudy16InSitu,Sing17Influence}
Furthermore, extrinsic loss that can be large for plasmonic satellites\cite{Kas16Particle}, is not included in our calculations.
We propose that the difference in spectral weight between the photoemssion experiments and our \emph{ab initio} results in the inset of Fig.~\ref{fig:SrMoO3dos} may have a similar origin. 

In Fig.~\ref{fig:SrMoO3dos} we also show the spectral function obtained from the $G^0W^0$ approximation.
The $G^0W^0$ results agree very well with the $GW$+EDMFT results except for the position of the satellite in the unoccupied region that is overestimated within the single-shot $GW$ approximation.
The quasiparticle dispersion in the $\vec{k}$-resolved spectral function (Fig.~\ref{fig:SrMoO3Akw}) also agrees with the $GW$ results and shows a small band-narrowing compared to the LDA.
The dispersion of the satellites follows the dispersion of the quasiparticle bands. 
This behavior is similar to the behavior of plasmon satellites in the $G^0W^0$+cumulant expansion which
suggests that the satellites are of plasmonic origin \cite{Caruso15Band,Lischner15Satellite,Caruso15Spectral}.

\subsubsection{Effective interactions}

\begin{figure}
  \includegraphics{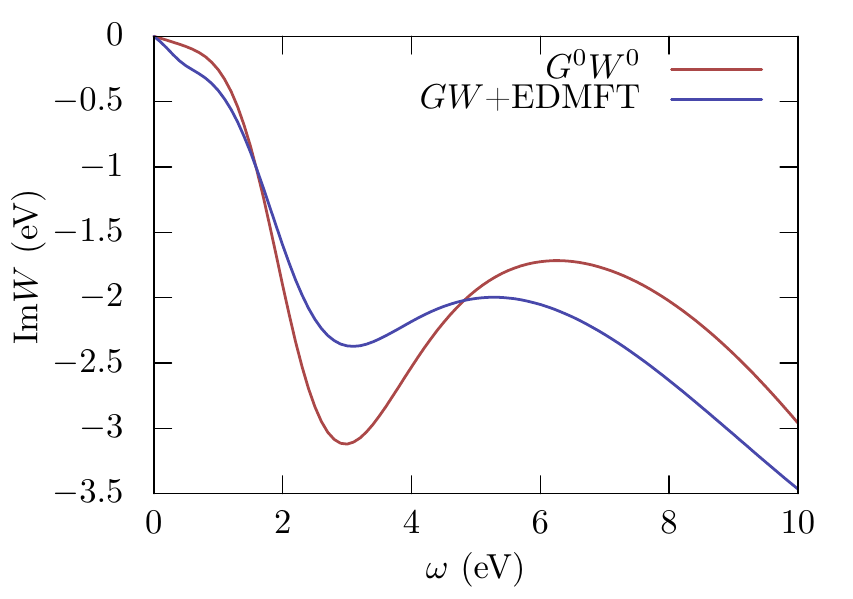}
  \caption{Local fully screened interaction of SrMoO$_3$.}
  \label{fig:SrMoO3Ww}
\end{figure}

To fully answer the question about the origin of the satellite features we look at the effective impurity interaction and
the fully screened interaction in Figs.~\ref{fig:SrMoO3USijkl} and \ref{fig:SrMoO3Ww}.
The static value of $\mathcal{U}$ is approximately 2.8 eV, which is clearly too small to explain the satellite features as Hubbard bands, since the separation of the satellites is approximately 5.5~eV.
On the other hand the positions of the satellites agree with the pole in $W$ in Fig.~\ref{fig:SrMoO3Ww}. 
This leads to the conclusion that the satellites are indeed of plasmonic orgin and hence SrMoO$_3$ and SrVO$_3$ are qualitatively similar, despite the qualitatively different predictions based on LDA+DMFT calculations.
It is noteworthy that for SrMoO$_3$ the plasma frequency is almost identical in $G^0W^0$ and $GW$+EDMFT while for SrVO$_3$ the plasma frequency is reduced in the full $GW$+EDMFT compared to one-shot $G^0W^0$.\cite{Boehnke16When}
Thus, RPA based on the LDA-bandstructure works better for SrMoO$_3$ which can be expected since this is a more extended system and hence for this compound $W^0$ is expected to yield a better plasma frequency.

In the following, we use SrMoO$_3$ as an example to discuss the frequency dependence and analytic properties of the effective impurity interaction $\mathcal{U}$.
Figure~\ref{fig:SrMoO3USijkl} shows the different components of $\mathcal{U}$ as defined in Eq.~\eqref{eqn:kanamoriU}. It is clear that the screenings
of $\mathcal{U}$ and $\mathcal{U}'$ are similar, while $J$ remains almost unscreened, which justifies the approximations in TIER I.

From Eq.~\eqref{eqn:causalitycond} it directly follows that if $\mathcal{U}(i\omega_n)$ is smaller than the static value for any Matsubara frequency $\omega_n$ then $\Im \mathcal{U}(\omega')$ has to be noncausal in some finite frequency range.
This is clearly the case for SrMoO$_3$ in Fig.~\ref{fig:SrMoO3USijkl}.
We further note that the effective impurity interaction for SrVO$_3$, shown in the same figure, does not only exhibit signs of a pole in the upper half plane,
for SrVO$_3$ the pole in $\mathcal{U}$ is located on the Matsubara axis between the first and second Matsubara frequencies, yielding a kink structure in $\mathcal{U}$.
This is not a problem per se, since poles in the upper half plane of $\mathcal{U}$ are to be expected (see Sec.~\ref{sec:causality}) and thus might end up on the Matsubara axis for certain parameters.
Yet, extra care has to be taken in 
this case to avoid numerical instabilities. 

\subsubsection{Filling dependence}

\begin{figure}
  \includegraphics{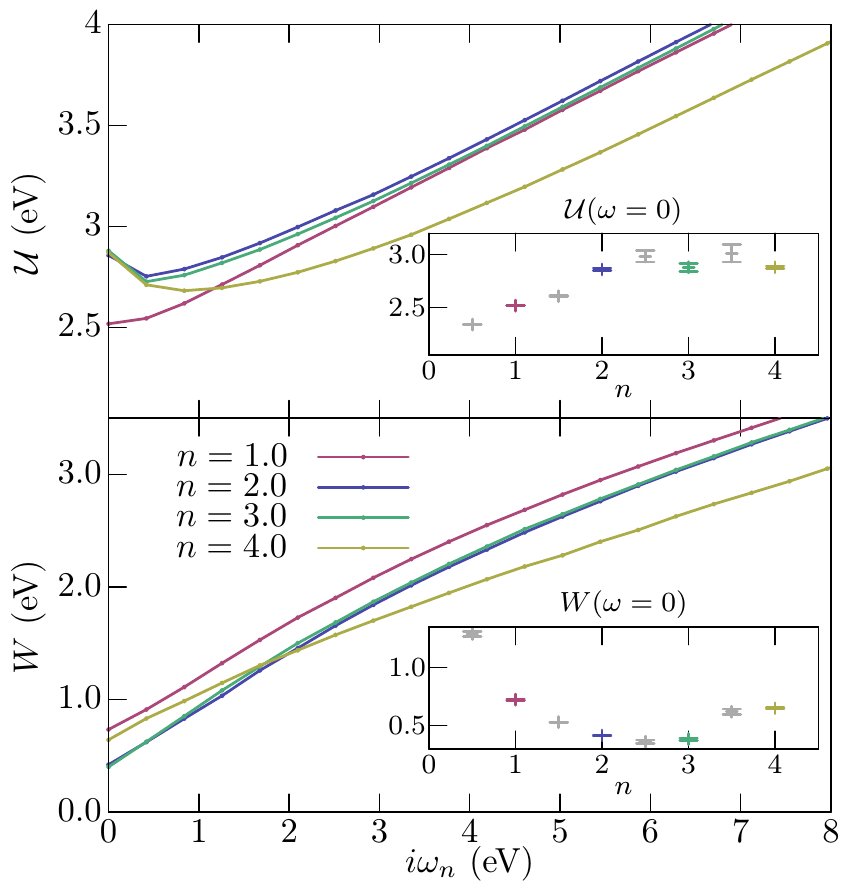}
  \caption{Effective impurity interaction $\mathcal{U}$ (upper panel) and local fully screened interaction $W$ (lower panel) for SrMoO$_3$ for different occupations. The inset shows the respective static values for different artificial occupations. The error bars are estimated from several successive iterations at self-consistency. $n=2.0$ corresponds to the physical case.}
  \label{fig:SrMoO3UWSn}
\end{figure}

To investigate the screening behavior and the noncausality further we have computed $\mathcal{U}(i\omega_n)$ for SrMoO$_3$ with different fillings of the $t_{2g}$ manifold; see  Fig.~\ref{fig:SrMoO3UWSn}. 
To reduce the computational time we held $\Sigma_r$ and $\Pi_r$ fixed at the experimental filling and only recalculated the model polarization and self-energies for the new fillings.
This is a reasonable approximation since all effects from within the low-energy subspace have been removed in $\Sigma_r$ and $\Pi_r$ and these quantities are therefore relatively insensitive
to small changes of the chemical potential.
For low fillings, $n\le 1$, $\mathcal{U}(i\omega_n)$ does not display any clear signs of noncausality. 
As the filling is increased towards the experimental filling ($n$=2) an antiscreening mode develops at low frequencies, which gives an increase in the static value and a negative slope of $\mathcal{U}(i\omega_n)$ between the first and second Matsubara frequencies.
The static value continues to increase and reaches its maximum near half-filling (see inset). 
To interpret the trend in $\mathcal{U}$ we also have to consider the fully screened interaction ($W$) for different fillings (Fig.~\ref{fig:SrMoO3UWSn}).
First of all one can note that even if the effective impurity interaction is noncausal for certain fillings, $W$ remains causal, as expected. 
The static value of $W$ follows the opposite trend to that of $\mathcal{U}$; i.e., it reaches its minimum at half filling, reflecting the increased screening as the number of free charge carriers is increased.

At self-consistency $W_{\mathrm{loc}}$ is obtained by screening $\mathcal{U}$ by the local polarization, $W_{\mathrm{loc}}=\mathcal{U} + \mathcal{U} \Pi_{\mathrm{loc}} W_{\mathrm{loc}}$.
Thus by comparing the two panels of Fig.~\ref{fig:SrMoO3UWSn} one can deduce that the local screening is strongest close to half filling.
It is also interesting to note that $\mathcal{U}$ becomes noncausal at low frequencies for the cases where the local screening is strong.

\subsection{Sodium as a model system}
\label{sec:Na}

\subsubsection{Results for different lattice constants}
Sodium provides an ideal playground for testing our approach.
Elemental Na has the electronic configuration [Ne]3s$^1$ and crystallizes in a bcc structure. 
In the solid the 3$s$-states hybridize with the unoccupied 3$p$ states to form a broad conduction band.
The conduction states are very delocalized and close to an electron gas model.
The main features of the occupied part of the experimental spectra is a well defined quasiparticle peak, a plasmonic satellite feature around $-7$~eV, which is repeated at approximately $-14$ eV \cite{79photoemission}.

The calculations in this section were performed for the inverse temperature $\beta=10\frac{1}{\mathrm{eV}}$ corresponding to $\approx1160$~K.

In the current work we will use sodium as a model system to scan different degrees of correlation. By successively increasing the lattice constant we increase the degree of local correlations
in a controlled manner.
This analysis is similar in spirit to the calculations on stretched diatomic molecules which are commonly used to benchmark quantum chemistry methods. 

\begin{figure}
  \includegraphics{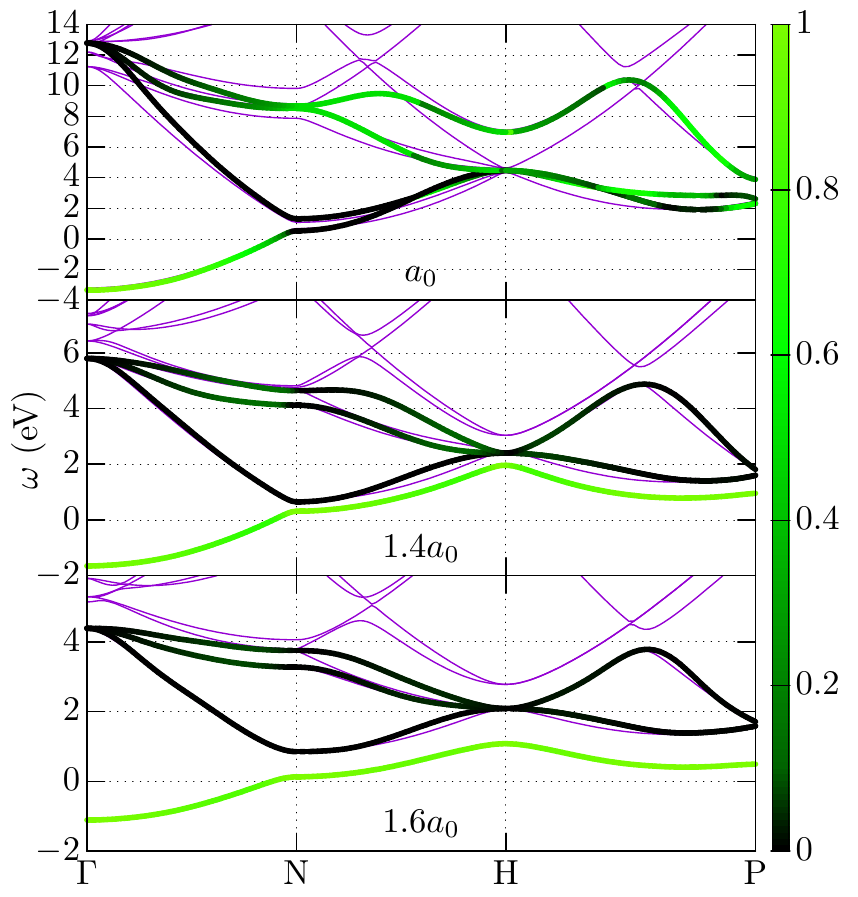}
  \caption{Interpolated band structure for Na with different lattice constants $a_0$. The color coding shows the ``$s$ character'' of the bands as defined by the $s$-like Wannier function.
  The solid purple lines show the original LDA bandstructure.}
  \label{fig:Naintband1}
\end{figure}

To faithfully reproduce the low-energy band structure we have to consider a 4-band model, consisting of the 3$s$ and 3$p$ states (see Fig.~\ref{fig:Naintband1}). 
For this material we will utilize the full strength of the multitier approach by choosing the $sp$ 4-band subspace for the self-consistent $GW$ calculation but only considering local EDMFT corrections for the $s$-state. Hence, the intermediate subspace is spanned by the full $s$ and $p$ Wannier functions but the correlated subspace is restricted to the $s$-like Wannier function.
We consider the experimental lattice constant $a_0$, as well as the artificially increased lattice constants $1.4a_0$ and $1.6a_0$.
\\

\begin{figure}
  \includegraphics{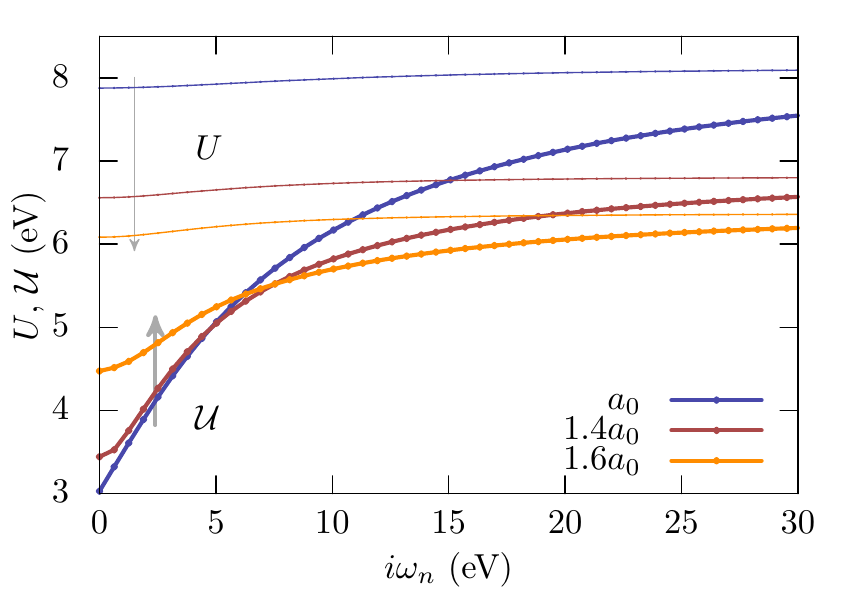}
  \caption{Comparisons between $U$ and $\mathcal{U}$ along the imaginary axis for Na with different lattice constants.}
  \label{fig:NaUs}
\end{figure}

\begin{figure*}
  \includegraphics{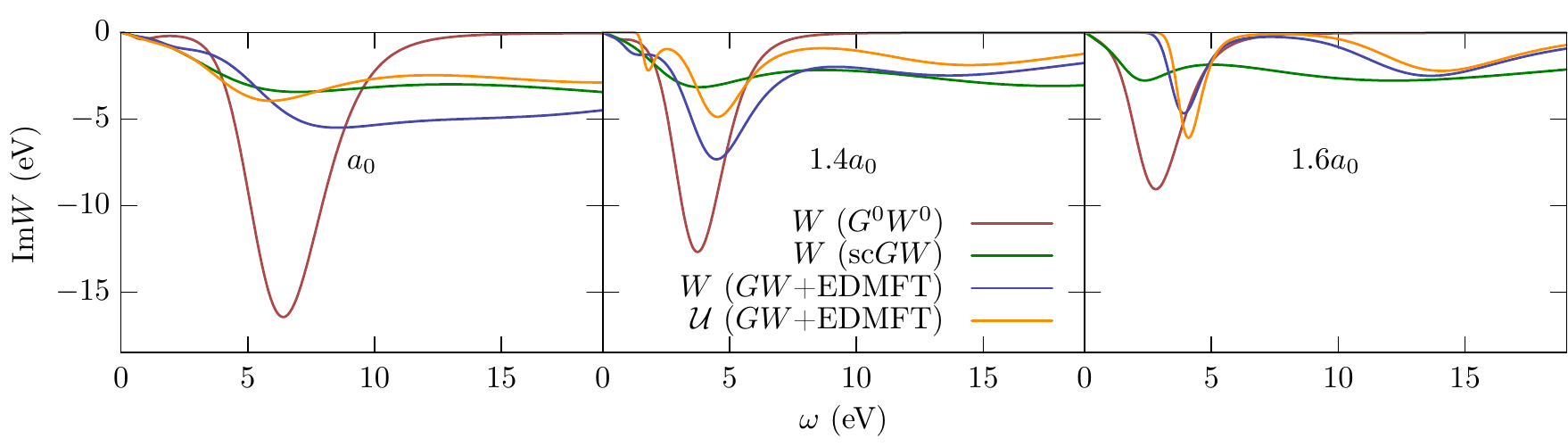}
  \caption{The $s$-character component of the local fully screened interactions for different lattice constants for Na.}
  \label{fig:NaW}
\end{figure*}

\begin{figure*}
  \includegraphics{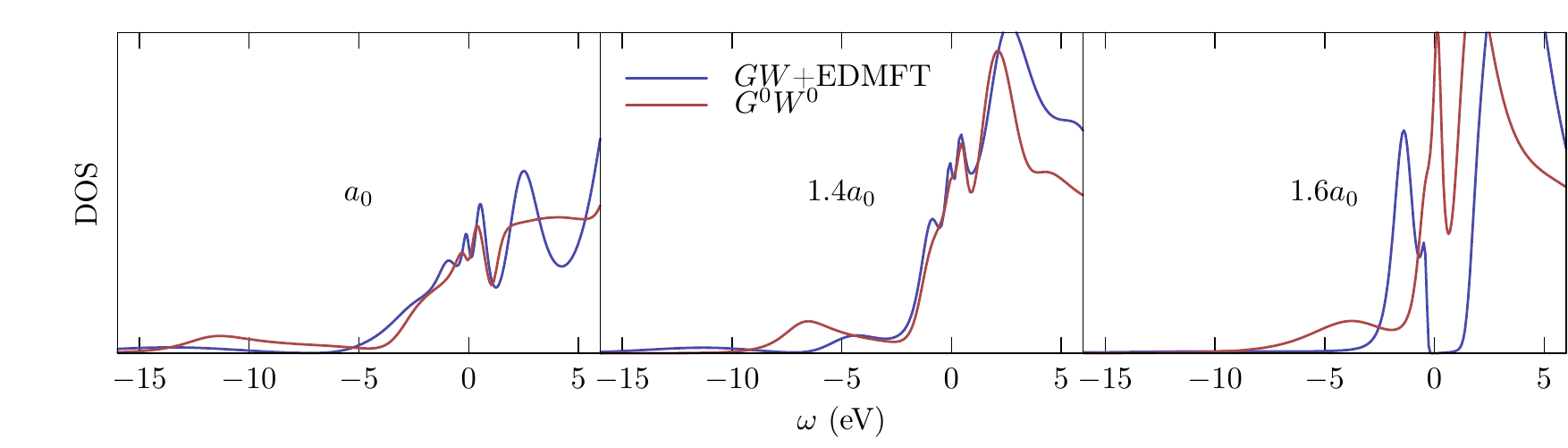}
  \caption{Local density of state for different lattice constants for Na.}
  \label{fig:Naspectra}
\end{figure*}

When the lattice constant is increased the
bandwidth of the conduction band is decreased, reflecting a decreasing hopping amplitude between the neighboring sites.
The model interaction, $U$ (Fig.~\ref{fig:NaUs}), 
is almost static for all lattice constants, implying that the model includes all important screening channels and hence also the dominant correlation contributions to the self-energy. 
Contrary to what one might expect, both $U$ and the bare interaction decreases as the lattice constant increases. A decreasing bare interaction implies that the localization of the Wannier basis states around the atomic positions is weaker 
for the larger lattice constants. 
Albeit counterintuitive, this phenomenon is well known and has been investigated for model systems in Ref. \onlinecite{Tomczak09Effective} and later for manganese monoxide under pressure in Ref. \onlinecite{Tomczak10Realistic}.
Since $U$ is only weakly screened the static value will follow the same trend as the bare interaction.

The effective impurity interaction $\mathcal{U}$ on the other hand follows the correct trend; i.e., it increases as the lattice constant is increased, reflecting the increasing importance of the local correlations.
That $\mathcal{U}$ and $U$ follow different trends illustrates the importance of the nonlocal $s$-$s$ screening as well as the $s$-$p$ screening channels for this system.

We will next discuss the fully screened interactions (Fig.~\ref{fig:NaW}) and the spectral functions (Fig.~\ref{fig:Naspectra}) for the different cases, comparing them both to self-consistent $GW$ and the one-shot $G^0W^0$. 
In the weakly correlated regime, where plasmonic physics dominates, the screened interaction already provides useful information about the spectral function:
\begin{enumerate}
 \item A peak in Im$W(\omega)$ at $\omega=\omega_\mathrm{p}$ will give a corresponding satellite feature in the spectral function at $E \pm n \omega'$, where $n$ is an integer and $E$ is the energy of the quasiparticle peak. The plus (minus) sign refers to unoccupied (occupied) states.
  \item For a given quasiparticle peak $G^0W^0$ will only give a single peak in the spectral function below and above the Fermi energy at too high energy.
  \item The renormalization of the quasiparticle peak will have a nontrivial dependence on the frequency and weight of the peak in Im$W$. 
  A peak with a large weight at low frequency will generally give the largest quasiparticle renormalization. 
  However the $\vec{k}$-dependence of the self-energy also influences the quasiparticle renormalization factor.
 \end{enumerate}
 
For elemental sodium (leftmost panels of Figs.~\ref{fig:NaW} and \ref{fig:Naspectra}) we know by comparing $G^0W^0$ calculations with the experimental spectra that the plasma frequency in $W^0$ is relatively good.
The discrepancies with the experimental spectra are mainly related to  self-energy corrections that can be accounted for using the cumulant expansion.
Also the quasiparticle renormalization is slightly underestimated in $G^0W^0$ \cite{Aryasetiawan96Multiple}.
Self-consistent $GW$ (sc$GW$ in Fig.~\ref{fig:NaW}) on the other hand severely worsens the result compared to experiment. The plasmon pole in $W$ is almost completely washed out, similar to what has been found for the electron gas.\cite{Holm98Fully} 
The local corrections from EDMFT in the full calculations ($GW$+EDMFT in Fig.~\ref{fig:NaW}) improves the sc$GW$ results 
but the strength of the pole is still severly underestimated and the position of the pole is at too high energy. This yields 
a weak plasmon between $-20$ to $-10$ eV in the spectral function ($a_0$ case in Fig.~\ref{fig:Naspectra}), in poor agreement with experiment.
Also the width of the quasiparticle peak is severely overestimated in the $GW$+EDMFT results.
The reason for the poor agreement with experiment is that the nonlocal screening is too big to be accounted for by 
only the first bubble diagram in the expansion. 
Thus, to get a good description of these kinds of very weakly correlated electron gas-like metals
it is necessary to include higher-order nonlocal screening beyond RPA. 

As the lattice constant is increased to 1.4$a_0$ the pole in $W^0$ is shifted to lower energies (Fig.~\ref{fig:NaW}). 
This is expected since the plasma frequency in the electron gas 
can be shown to be proportional to the square-root of the density and we effectively decrease the density by increasing the lattice constant.
sc$GW$ still gives a very wide and featureless plasmon similar to the original lattice constant.
However, $W$ for the full $GW$+EDMFT calculation develops a well defined plasmonic pole at slightly higher energy than the pole in $W^0$.
Hence, for this lattice constant we enter a regime where the $GW$+EDMFT approximation, which only takes into account the first bubble diagram in the nonlocal polarization, becomes physically reasonable.
In addition to the main peak in $W$ there is an additional shoulder structure around 15~eV. 
This structure, which is present also in sc$GW$ and the effective impurity interaction $\mathcal{U}$, is related to the lack of local corrections for the $s$-$p$ screening channel.
However, the structure is smaller in $GW$+EDMFT than in sc$GW$ which implies that the local corrections for the $s$-$s$ screening 
at least partially remove the unphysical high frequency structures in sc$GW$. Ideally though, all relevant low-energy screening channels
should be included in the correlated subspace and only less important screening channels should be treated in sc$GW$.

It is interesting to note that even though the peak in $W$ is at higher frequency in $GW$+EDMFT compared to $G^0W^0$, the plasmon 
satellite in the spectral function (Fig.~\ref{fig:Naspectra}) is closer to the quasiparticle peak.
Hence, the local vertex contributions to the selfenergy corrects the error in $G^0W^0$ and pulls the satellite closer to the quasiparticle.
There is also an additional satellite feature in the $GW$+EDMFT results around $-12$~eV. This feature is a combination of a repetition of the
main plasmon and a second plasmonic peak due to the shoulder structure around 15~eV in $W$.

Finally we will discuss the largest lattice constant $1.6a_0$ (rightmost panels of Figs.~\ref{fig:NaW} and \ref{fig:Naspectra}).
In this case the conduction band is well separated from the other bands and almost of pure $s$-character (see Fig.~\ref{fig:Naintband1}). 
This means that we can directly compare the bandwidth of the conduction band with the static value of the impurity interaction in Fig.~\ref{fig:NaUs}
to estimate the degree of local correlations. Since $\mathcal{U}$ is much larger than the bandwidth we get an insulating solution with an upper and lower Hubbard band separated by
approximately the static value of the interaction (Fig.~\ref{fig:NaW}). $G^0W^0$ clearly fails to capture the strong local correlations 
driving the metal to insulator transition and yields a metallic solution with a plasmonic satellite feature below the Fermi energy.
Due to the gap in the spectral function there are no excitations below $\approx 3$~eV, which means that Im$W$ is zero in this frequency range.
There is a peak in $W$ at around 4~eV corresponding to transitions between the lower and upper Hubbard band and an additional peak from the
$s$-$p$ screening at higher frequency. In the spectral function for the $1.6a_0$ case (Fig.~\ref{fig:Naspectra}) there is a weak 
(barely visible)
satellite feature
corresponding to the first peak in Im$W$.

\subsubsection{Effect of a local approximation in extended systems}
\begin{figure}
  \includegraphics{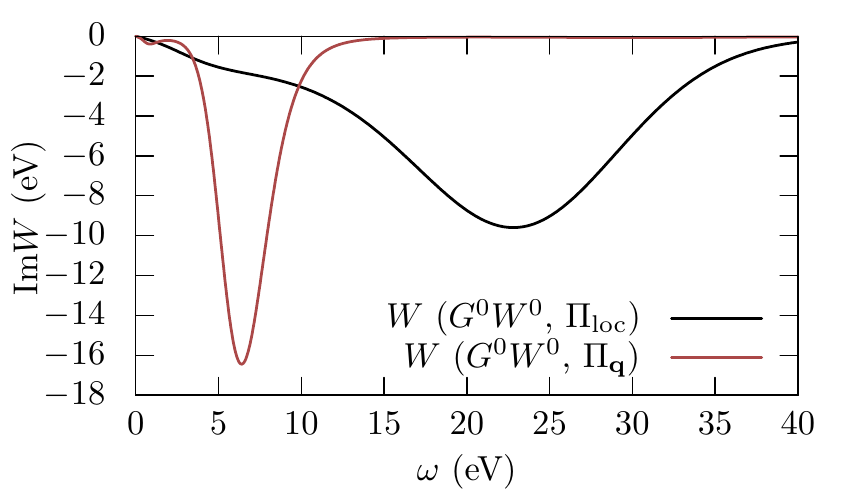}
  \caption{Comparison between the $s$-character component of the local fully screened interaction $W$ from a one-shot $GW$ calculation for elemental Na calculated with the complete $\vec{q}$-dependent polarization ($\Pi^{G^0G^0}_{\vec{q}}$) and the same quantity calculated with only the local projection of the polarization ($\Pi^{G^0G^0}_{\mathrm{loc}}$).}
  \label{fig:W0_piloc}
\end{figure}

\begin{figure}
  \includegraphics{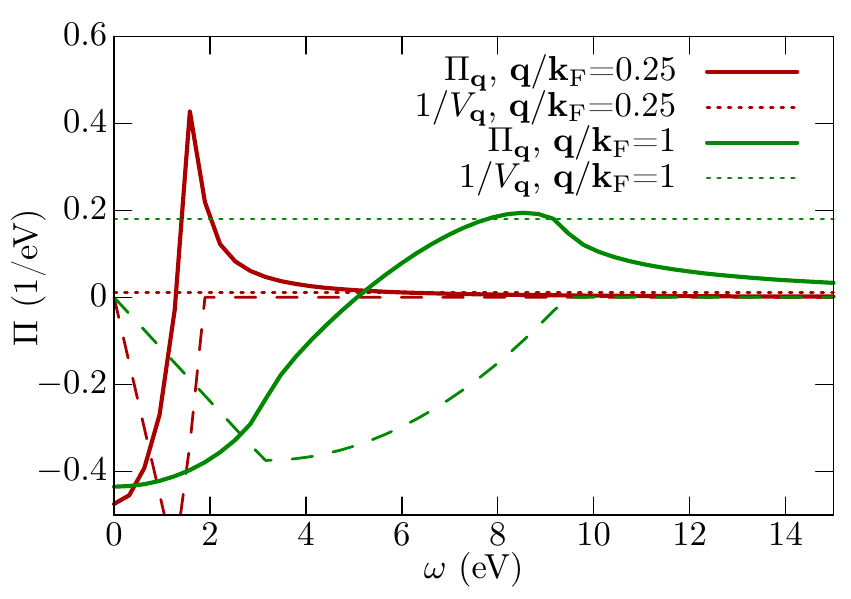}
  \caption{Analytic form of the polarization for the non-interacting electron gas for two different \vec{q} points. The real part is shown with solid thick lines and the corresponding imaginary part with thin dashed lines. The plasmon pole in $W$ occurs at the crossing between $1/V_\vec{q}$ and $\Pi_\vec{q}$. The Fermi energy was chosen to get a filling of $n=1$ electron.
  }
  \label{fig:pi_egas}
\end{figure}
In the $GW$+EDMFT results for elemental sodium above we found that the plasmon weight in $\Im W$ was reduced substantially and the position of the pole was pushed to higher energies. 
In sc$GW$ we found a similar reduction, but here the pole was pushed to lower energies instead.
To investigate these differences in behavior, which must originate from the local EDMFT contributions, we first consider the extreme case of approximating $\Pi_\vec{q}=\Pi_{\mathrm{loc}}$ in 
a simple one-shot $G_0W_0$-calculation for sodium and compare it to the regular $G_0W_0$ result with the 
full $\vec{q}$-dependent polarization $\Pi_{\vec{q}}$ (Fig. \ref{fig:W0_piloc}).
This clearly exhibits the same trend as the full $GW$+EDMFT results, albeit even more extreme: By only including a local polarization the plasmon peak in $W$ is shifted to much higher energies and
is broadened substantially.
Thus, in cases where the local polarization does not have more physical relevance than the non-local terms, the $GW$+EDMFT plasma frequency might end up getting overestimated.
Namely, for such a compound the nonlocal part is simply too large to be treated with only the first bubble diagram 
in the self-consistent expansion. Including all local diagrams but only the first-order nonlocal diagram leads to an underestimation of the $\vec{q}$-dependence
compared to the local contribution, which in turn shifts the position of the pole in $W$ to higher frequencies. However, as is evident from Fig.~\ref{fig:W0_piloc}, the
$GW$+EDMFT result should still be much better than what one would expect to get with an EDMFT-type approximation with only a local polarization.

\begin{figure}
  \includegraphics{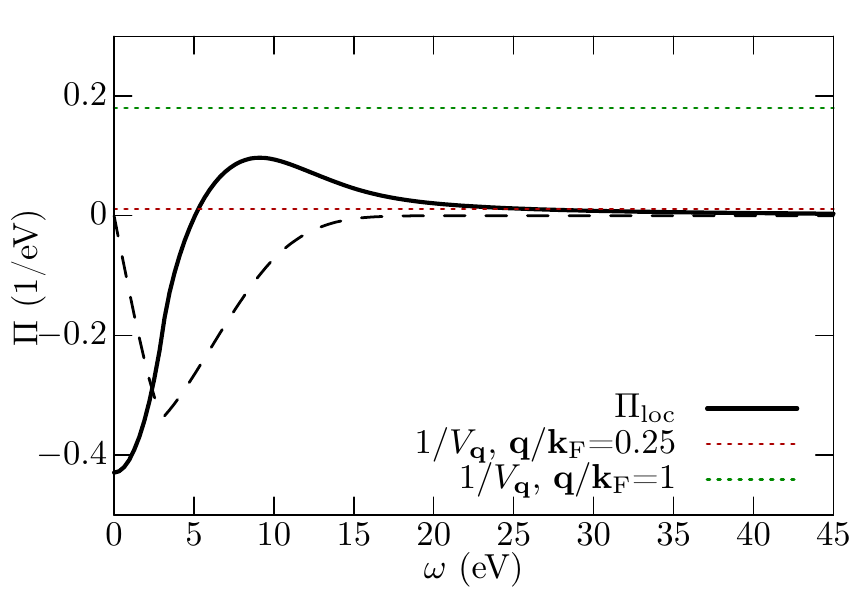}
  \caption{``Local'' polarization for the non-interacting electron gas. The real part is shown with solid thick lines and the corresponding imaginary part with thin dashed lines. The artificial unit cell was defined to be cubic with the same volume as the bcc unit cell for Na. 
  With these definitions the local polarization $\Pi_\mathrm{loc}$ was defined as the \vec{q} sum of $\Pi_\vec{q}$ in the first Brillouine zone.}
  \label{fig:piloc_egas}
\end{figure}

To understand this behavior in more detail we have to look at the specific form of $\Pi_{\vec{q}}$ and $\Pi_{\mathrm{loc}}$.
We do this for the non-interacting electron gas where the analytic form of the polarization is known explicitly (Figs. \ref{fig:pi_egas} and \ref{fig:piloc_egas})\cite{fetterwalecka}. 
\\

\begin{figure}
  \includegraphics{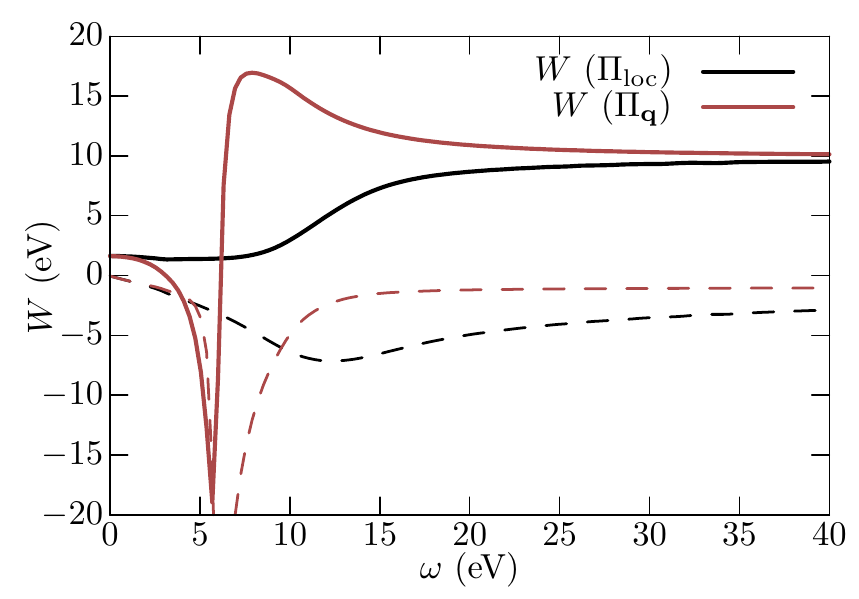}
  \caption{``Local'' projection of $W$ for the electron gas computed with the full $\mathbf{q}$-dependent polarization ($\Pi_{\mathbf{q}}$) as well as only the ``local'' polarization $\Pi_{\mathrm{loc}}$. 
  The real part is shown with solid thick lines and the corresponding imaginary part with thin dashed lines. The definition of the local projections are given in the caption of Fig.~\ref{fig:piloc_egas}.}
  \label{fig:W_egas}
\end{figure}

\begin{figure}
  \includegraphics{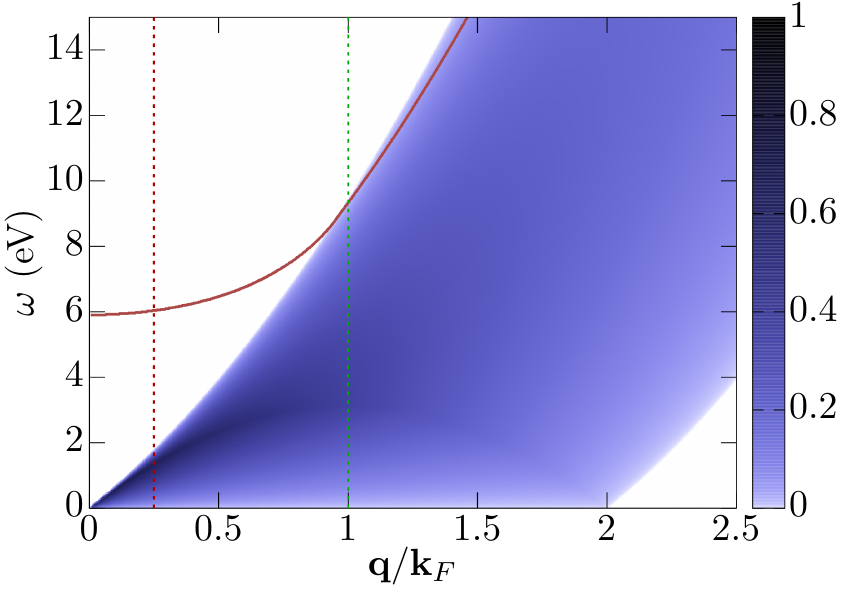}
  \caption{Imaginary part of the polarization for the noninteraction electron gas for different $\mathbf{q}$ and $\omega$ values.
  The solid red line shows the dispersion of the plasma frequency defined by $\mathrm{min}\{|\Pi(\mathbf{q},\omega)-1/V_{\mathbf{q}}|\}$. The two vertical lines correspond to the two cases considered in Fig.~\ref{fig:pi_egas}.}
  \label{fig:Landaudamp}
\end{figure}

The plasmon pole in $W_{\vec{q}}$ occurs at the zeros of the dielectric function 
\begin{align}
\epsilon_{\vec{q}}(\omega) = 1 - \Pi_{\vec{q}}(\omega)V_{\vec{q}}=0.
\label{Eq:zerocross}
\end{align}
From Fig.~\ref{fig:pi_egas} one can see that Eq.~\eqref{Eq:zerocross} is only fulfilled at the second crossing between the line
$1/V_{\vec{q}}$ and $\mathrm{Re}\Pi_{\vec{q}}(\omega)$, since the imaginary component of the polarization is big at the first crossing.
For the chosen parameters (see the caption of Fig. \ref{fig:pi_egas}) the position of the plasmon pole in $W_{\vec{q}}$ will have a small dispersion between approximately
6-10 eV which yields a sharp peak in $W_{\mathrm{loc}}$ in Fig.~\ref{fig:W_egas}. Furthermore, there will not be a well defined plasmon peak for all $\vec{q}$-values.
For $q/k_F \gtrsim 1$ the line $1/V_{\vec{q}}$ will not cross $\mathrm{Re}\Pi_{\vec{q}}(\omega)$, which results in a relatively broad and weak 
plasmonic feature for these $\vec{q}$-points. Another interesting observation is that the peak in $\mathrm{Im}\Pi_{\vec{q}}(\omega)$ is
sharper and shifted to lower frequencies for the $\vec{q}$-points close to the $\Gamma$-point compared to the large-$\vec{q}$ components as can also be understood from Fig.~\ref{fig:Landaudamp}.

To define a quantity corresponding to the local polarization for the electron gas we choose a cubic unit cell with the same volume as the bcc unit cell for sodium.
We then define $\Pi_{\mathrm{loc}}$ as the $\vec{q}$-sum of $\Pi_{\vec{q}}$ over the first Brillouin zone.
It should be noted that this definition of ``local polarization''  differs slightly from the local polarization in the full \emph{ab initio} calculations, 
where the local subspace is defined by MLWF:s and the screening from all bands are included. For the electron gas we use a plane-wave basis and only screening from the first (conduction) band is 
included in $\Pi_{\mathrm{loc}}$.
Hence, the following discussion should only be used to gain a qualitative understanding of Fig.~\ref{fig:W0_piloc} while some 
quantitative differences such as the exact position and weight of the plasmon pole might differ.

The local polarization (Fig. \ref{fig:piloc_egas}) contains a large contribution from the $\vec{q}$-points far away from the
$\Gamma$-point and therefore the peak in $\mathrm{Im}\Pi_{\mathrm{loc}}(\omega)$ is broad and peaked at a relatively high frequency.
This gives a correspondingly broad Kramers-Kronig feature in the real part at high energies.
For many $\vec{q}$-points there will not be a real crossing between $1/V_{\vec{q}}$ $\mathrm{Re}\Pi_{\mathrm{loc}}(\omega)$,
but if there is it will occur at energies larger than 10 eV. This yields a broad plasmon in $W$ at high frequency (Fig.~\ref{fig:W_egas}),
just as we observed for Na in Fig. \ref{fig:W0_piloc}.

In Fig.~\ref{fig:Landaudamp} we show the dispersion of the plasmon together with the imaginary part of the polarization for the noninteracting electron gas.
The two cases shown in
Fig.~\ref{fig:pi_egas} correspond to two points in Fig.~\ref{fig:Landaudamp}. 
The dispersion of the plasmon (solid curve in Fig.~\ref{fig:Landaudamp}) is defined as the
$(\mathbf{q},\omega)$ point where $|\Pi(\mathbf{q},\omega)-1/V_{\mathbf{q}}|$ takes its minimum value [which for the case $\Im\Pi_\mathbf{q}(\omega_p)=0$ is given by the crossing between
the line
$1/V_{\vec{q}}$ and $\mathrm{Re}\Pi_{\vec{q}}(\omega)$ as discussed above]. When $\Im \Pi_\vec{q}(\omega = \omega_p) \neq 0$ the strength of the plasmon in $W$ will be reduced and
the pole is broadened, corresponding to a finite lifetime of the plasmonic mode (e.g., Landau damping).

We can also use the non-interacting electron gas to gain a qualitative understanding of the effect of increasing the lattice constant for Na.
By increasing the lattice constant we effectively shrink the first Brillouin zone (1:st BZ). Therefore the maximum difference between
$\Pi_{\vec{q}}$ for different $\vec{q}\in$ 1:st BZ will decrease, that is $\Pi_{\vec{q}}$ becomes more local.
At some point the nonlocal components of $\Pi_{\vec{q}}$ are sufficiently small that a truncation at the first bubble diagram is appropriate. 
This is the point where $GW$+EDMFT becomes justified. At a much later point the nonlocal components of $\Pi_{\vec{q}}$ are
sufficiently small to be ignored completely, in which case an EDMFT treatment is sufficient. 

\section{Conclusions}
\label{sec:conclusions}
While the $GW$+EDMFT method was first proposed in 2003 the implementation of the fully self-consistent scheme for real materials was not realized until 2016 (Ref.~\onlinecite{Boehnke16When}). 
In this paper we provided a detailed description of the self-consistent multitier $GW$+EDMFT implementation used in Ref.~\onlinecite{Boehnke16When} and we tested the $GW$+EDMFT method 
for different systems with a focus on the effect of self-consistency.
We first applied the $GW$+EDMFT method to SrMoO$_3$, a cubic perovskite with 4$d$ valence electrons.
SrMoO$_3$ is in many respects similar to the 3$d$ cubic perovskite SrVO$_3$.
However, while SrVO$_3$ has previously been thought to be well described by LDA+DMFT, with a renormalized quasiparticle peak and Hubbard side-bands,\cite{Pavarini04Mott,Lechermann06Dynamical,Backes16Hubbard}
LDA+DMFT obviously fails to give a reasonable description of SrMoO$_3$. The interaction needed to produce the observed satellite features
yields a much too strong renormalization of the quasiparticle peak, which indicates that the satellites could be of plasmonic rather than Hubbard band character.\cite{Wadati14Photo} 
In this work we have shown that the parameter-free multitier $GW$+EDMFT scheme is able to describe the satellite features of SrMoO$_3$ and that the satellites are indeed of plasmonic origin.
These results can also be connected to our previous investigation of SrVO$_3$ and further support the main conclusion reached there,\cite{Boehnke16When} namely that $GW$+EDMFT is able to describe the satellite structures in terms of plasmonic fluctuations in a moderately correlated material.
We also used SrMoO$_3$ as a model system to provide insights into the effectiveness of screening and the causality of the impurity interaction by systematically changing the occupation of the $t_{2g}$ manifold.
Close to half-filling the local screening is strong, which yields a noncausal effective impurity interaction while all physical observables remain causal.

Finally we used sodium as a model system to investigate the performance of the method for different degrees of local correlations.
Starting from a weakly correlated metal with the experimental lattice constant, we successively increased the lattice constant and thus the degree of local correlations.
This allowed us to scan the metal to insulator transition in a controlled manner for a realistic material and to evaluate the performance of the multitier $GW$+EDMFT approach
in a wide parameter range.
We showed that the method performs well in the moderately to strongly correlated regime but underperforms in the very weakly correlated regime where the nonlocal screening is comparable to the local one. 
In the latter case it is no longer sufficient to treat the nonlocal correlations within the sc$GW$ approximation, i.e., nonlocal diagrams beyond RPA are needed. 
The $G^0W^0$ approximation on the other hand, works relatively well in this regime, which implies that the partial cancellation of diagrams in $G^0W^0$ is not restricted  to the local case. 
In the intermediate to strongly correlated regime the local vertex contributions to the self-energy remedy the well known problems with the one-shot $G^0W^0$ approximation, such as the overestimation of the satellite position in the spectral function compared to the pole in $\Im W$. 
$GW$+EDMFT also correctly captures the Mott-Hubbard metal to insulator transition.
The quasiparticle bandwidth is reduced compared to sc$GW$ and, contrary to the latter scheme, where the plasmon pole in $\Im W$ is washed out, the $GW$+EDMFT calculation yields well defined plasmonic peaks in $\Im W$.
The static model interaction, computed with the cRPA, exhibits the wrong trend for stretched sodium; it decreases as the lattice constant is increased.
In this work we showed that the additional screening channels included in $GW$+EDMFT solve this problem and provide the physically expected result, namely an impurity interaction which increases with increasing lattice constant.  

On the conceptual level, a noteworthy feature
in most previous self-consistent $GW$+EDMFT calculations, both for models \cite{Ayral13Screening,Huang14Extended,vanLoon16Double,Stepanov16Selfconsistent,Ayral17Influence}
and real materials,\cite{Boehnke16When} is that the effective interaction for the impurity problem $\mathcal{U}$ can become noncausal.
Using an exactly solvable dimer model, we showed that noncausal impurity interactions are not a specific feature of $GW$+EDMFT, and that any method that includes dynamic long-range polarizations is bound to yield noncausal effective impurity interactions in certain parameter regimes.
However, we also found that quantities that relate to physical observables, as opposed to auxiliary ones, remain causal.

The results of our study demonstrate that multitier self-consistent $GW$+EDMFT enables the \emph{ab initio} simulation of a broad range of materials in the intermediate to strongly correlated regime. The method is relatively cheap in terms of computational effort and memory requirements, and hence applicable to multiorbital systems. It does not contain any adjustable parameters, once the different subspaces, or tiers, have been fixed. In particular, the method provides a self-consistent calculation of the dynamically screened interaction parameters, and captures the effect of collective charge excitations. This distinguishes the method from standard LDA+DMFT simulations, and makes it the first true \emph{ab initio} simulation method for moderately to strongly correlated materials. As a fully Green's function based approach, multitier self-consistent $GW$+EDMFT is also a promising method for the study of nonequilibrium phenomena in correlated solids.

\begin{acknowledgments}
F.N. and F.A. acknowledge financial support from the Swedish Research Council (VR).
L.B. is supported by the Swiss National Science Foundation through NCCR MARVEL.
P.W. acknowledges support from ERC Consolidator Grant No. 724103. 
The computations were performed on resources provided by the Swedish National Infrastructure for Computing (SNIC) at LUNARC and at the CSCS Dora cluster provided by MARVEL.

LB would like to thank Denis Gole\v{z} and Thomas Ayral for valuable discussions and Hugo Strand for providing his exact diagonalization code.\cite{pyED}

F.N. and L.B. contributed equally to this work.
\end{acknowledgments}

% \bibliography{all}

\begin{thebibliography}{94}%
\makeatletter
\providecommand \@ifxundefined [1]{%
 \@ifx{#1\undefined}
}%
\providecommand \@ifnum [1]{%
 \ifnum #1\expandafter \@firstoftwo
 \else \expandafter \@secondoftwo
 \fi
}%
\providecommand \@ifx [1]{%
 \ifx #1\expandafter \@firstoftwo
 \else \expandafter \@secondoftwo
 \fi
}%
\providecommand \natexlab [1]{#1}%
\providecommand \enquote  [1]{``#1''}%
\providecommand \bibnamefont  [1]{#1}%
\providecommand \bibfnamefont [1]{#1}%
\providecommand \citenamefont [1]{#1}%
\providecommand \href@noop [0]{\@secondoftwo}%
\providecommand \href [0]{\begingroup \@sanitize@url \@href}%
\providecommand \@href[1]{\@@startlink{#1}\@@href}%
\providecommand \@@href[1]{\endgroup#1\@@endlink}%
\providecommand \@sanitize@url [0]{\catcode `\\12\catcode `\$12\catcode
  `\&12\catcode `\#12\catcode `\^12\catcode `\_12\catcode `\%12\relax}%
\providecommand \@@startlink[1]{}%
\providecommand \@@endlink[0]{}%
\providecommand \url  [0]{\begingroup\@sanitize@url \@url }%
\providecommand \@url [1]{\endgroup\@href {#1}{\urlprefix }}%
\providecommand \urlprefix  [0]{URL }%
\providecommand \Eprint [0]{\href }%
\providecommand \doibase [0]{http://dx.doi.org/}%
\providecommand \selectlanguage [0]{\@gobble}%
\providecommand \bibinfo  [0]{\@secondoftwo}%
\providecommand \bibfield  [0]{\@secondoftwo}%
\providecommand \translation [1]{[#1]}%
\providecommand \BibitemOpen [0]{}%
\providecommand \bibitemStop [0]{}%
\providecommand \bibitemNoStop [0]{.\EOS\space}%
\providecommand \EOS [0]{\spacefactor3000\relax}%
\providecommand \BibitemShut  [1]{\csname bibitem#1\endcsname}%
\let\auto@bib@innerbib\@empty
%</preamble>
\bibitem [{\citenamefont {Hedin}(1965)}]{Hedin65New}%
  \BibitemOpen
  \bibfield  {author} {\bibinfo {author} {\bibfnamefont {L.}~\bibnamefont
  {Hedin}},\ }\href {\doibase 10.1103/PhysRev.139.A796} {\bibfield  {journal}
  {\bibinfo  {journal} {Phys. Rev.}\ }\textbf {\bibinfo {volume} {139}},\
  \bibinfo {pages} {A796} (\bibinfo {year} {1965})}\BibitemShut {NoStop}%
\bibitem [{\citenamefont {Kohn}\ and\ \citenamefont
  {Sham}(1965)}]{Kohn65SelfConsistent}%
  \BibitemOpen
  \bibfield  {author} {\bibinfo {author} {\bibfnamefont {W.}~\bibnamefont
  {Kohn}}\ and\ \bibinfo {author} {\bibfnamefont {L.~J.}\ \bibnamefont
  {Sham}},\ }\href {\doibase 10.1103/PhysRev.140.A1133} {\bibfield  {journal}
  {\bibinfo  {journal} {Phys. Rev.}\ }\textbf {\bibinfo {volume} {140}},\
  \bibinfo {pages} {A1133} (\bibinfo {year} {1965})}\BibitemShut {NoStop}%
\bibitem [{\citenamefont {Aryasetiawan}\ and\ \citenamefont
  {Gunnarsson}(1998)}]{Aryasetiawan98GW}%
  \BibitemOpen
  \bibfield  {author} {\bibinfo {author} {\bibfnamefont {F.}~\bibnamefont
  {Aryasetiawan}}\ and\ \bibinfo {author} {\bibfnamefont {O.}~\bibnamefont
  {Gunnarsson}},\ }\href {\doibase 10.1088/0034-4885/61/3/002} {\bibfield
  {journal} {\bibinfo  {journal} {{Reports on Progress in Physics}}\ }\textbf
  {\bibinfo {volume} {61}},\ \bibinfo {pages} {237} (\bibinfo {year}
  {1998})}\BibitemShut {NoStop}%
\bibitem [{\citenamefont {van Schilfgaarde}\ \emph {et~al.}(2006)\citenamefont
  {van Schilfgaarde}, \citenamefont {Kotani},\ and\ \citenamefont
  {Faleev}}]{Schilfgaarde06Quasiparticle}%
  \BibitemOpen
  \bibfield  {author} {\bibinfo {author} {\bibfnamefont {M.}~\bibnamefont {van
  Schilfgaarde}}, \bibinfo {author} {\bibfnamefont {T.}~\bibnamefont {Kotani}},
  \ and\ \bibinfo {author} {\bibfnamefont {S.}~\bibnamefont {Faleev}},\ }\href
  {\doibase 10.1103/PhysRevLett.96.226402} {\bibfield  {journal} {\bibinfo
  {journal} {Phys. Rev. Lett.}\ }\textbf {\bibinfo {volume} {96}},\ \bibinfo
  {pages} {226402} (\bibinfo {year} {2006})}\BibitemShut {NoStop}%
\bibitem [{\citenamefont {Lyo}\ and\ \citenamefont
  {Plummer}(1988)}]{Lyo88Quasiparticle}%
  \BibitemOpen
  \bibfield  {author} {\bibinfo {author} {\bibfnamefont {I.-W.}\ \bibnamefont
  {Lyo}}\ and\ \bibinfo {author} {\bibfnamefont {E.~W.}\ \bibnamefont
  {Plummer}},\ }\href {\doibase 10.1103/PhysRevLett.60.1558} {\bibfield
  {journal} {\bibinfo  {journal} {Phys. Rev. Lett.}\ }\textbf {\bibinfo
  {volume} {60}},\ \bibinfo {pages} {1558} (\bibinfo {year}
  {1988})}\BibitemShut {NoStop}%
\bibitem [{\citenamefont {Bruneval}\ \emph {et~al.}(2006)\citenamefont
  {Bruneval}, \citenamefont {Vast},\ and\ \citenamefont
  {Reining}}]{Bruneval06Effect}%
  \BibitemOpen
  \bibfield  {author} {\bibinfo {author} {\bibfnamefont {F.}~\bibnamefont
  {Bruneval}}, \bibinfo {author} {\bibfnamefont {N.}~\bibnamefont {Vast}}, \
  and\ \bibinfo {author} {\bibfnamefont {L.}~\bibnamefont {Reining}},\ }\href
  {\doibase 10.1103/PhysRevB.74.045102} {\bibfield  {journal} {\bibinfo
  {journal} {Phys. Rev. B}\ }\textbf {\bibinfo {volume} {74}},\ \bibinfo
  {pages} {045102} (\bibinfo {year} {2006})}\BibitemShut {NoStop}%
\bibitem [{\citenamefont {Georges}\ \emph {et~al.}(1996)\citenamefont
  {Georges}, \citenamefont {Kotliar}, \citenamefont {Krauth},\ and\
  \citenamefont {Rozenberg}}]{Georges96Dynamical}%
  \BibitemOpen
  \bibfield  {author} {\bibinfo {author} {\bibfnamefont {A.}~\bibnamefont
  {Georges}}, \bibinfo {author} {\bibfnamefont {G.}~\bibnamefont {Kotliar}},
  \bibinfo {author} {\bibfnamefont {W.}~\bibnamefont {Krauth}}, \ and\ \bibinfo
  {author} {\bibfnamefont {M.~J.}\ \bibnamefont {Rozenberg}},\ }\href {\doibase
  10.1103/RevModPhys.68.13} {\bibfield  {journal} {\bibinfo  {journal} {{Rev.
  Mod. Phys.}}\ }\textbf {\bibinfo {volume} {68}},\ \bibinfo {pages} {13}
  (\bibinfo {year} {1996})}\BibitemShut {NoStop}%
\bibitem [{\citenamefont {Karolak}\ \emph {et~al.}(2010)\citenamefont
  {Karolak}, \citenamefont {Ulm}, \citenamefont {Wehling}, \citenamefont
  {Mazurenko}, \citenamefont {Poteryaev},\ and\ \citenamefont
  {Lichtenstein}}]{Karolak10Double}%
  \BibitemOpen
  \bibfield  {author} {\bibinfo {author} {\bibfnamefont {M.}~\bibnamefont
  {Karolak}}, \bibinfo {author} {\bibfnamefont {G.}~\bibnamefont {Ulm}},
  \bibinfo {author} {\bibfnamefont {T.}~\bibnamefont {Wehling}}, \bibinfo
  {author} {\bibfnamefont {V.}~\bibnamefont {Mazurenko}}, \bibinfo {author}
  {\bibfnamefont {A.}~\bibnamefont {Poteryaev}}, \ and\ \bibinfo {author}
  {\bibfnamefont {A.}~\bibnamefont {Lichtenstein}},\ }\href {\doibase
  https://doi.org/10.1016/j.elspec.2010.05.021} {\bibfield  {journal} {\bibinfo
   {journal} {Journal of Electron Spectroscopy and Related Phenomena}\ }\textbf
  {\bibinfo {volume} {181}},\ \bibinfo {pages} {11 } (\bibinfo {year}
  {2010})},\ \bibinfo {note} {proceedings of International Workshop on Strong
  Correlations and Angle-Resolved Photoemission Spectroscopy 2009}\BibitemShut
  {NoStop}%
\bibitem [{\citenamefont {Aryasetiawan}\ \emph {et~al.}(2004)\citenamefont
  {Aryasetiawan}, \citenamefont {Imada}, \citenamefont {Georges}, \citenamefont
  {Kotliar}, \citenamefont {Biermann},\ and\ \citenamefont
  {Lichtenstein}}]{Aryasetiawan04Frequencydependent}%
  \BibitemOpen
  \bibfield  {author} {\bibinfo {author} {\bibfnamefont {F.}~\bibnamefont
  {Aryasetiawan}}, \bibinfo {author} {\bibfnamefont {M.}~\bibnamefont {Imada}},
  \bibinfo {author} {\bibfnamefont {A.}~\bibnamefont {Georges}}, \bibinfo
  {author} {\bibfnamefont {G.}~\bibnamefont {Kotliar}}, \bibinfo {author}
  {\bibfnamefont {S.}~\bibnamefont {Biermann}}, \ and\ \bibinfo {author}
  {\bibfnamefont {A.~I.}\ \bibnamefont {Lichtenstein}},\ }\href {\doibase
  10.1103/PhysRevB.70.195104} {\bibfield  {journal} {\bibinfo  {journal}
  {{Phys. Rev. B}}\ }\textbf {\bibinfo {volume} {70}},\ \bibinfo {pages}
  {195104} (\bibinfo {year} {2004})}\BibitemShut {NoStop}%
\bibitem [{\citenamefont {Boehnke}\ \emph {et~al.}(2016)\citenamefont
  {Boehnke}, \citenamefont {Nilsson}, \citenamefont {Aryasetiawan},\ and\
  \citenamefont {Werner}}]{Boehnke16When}%
  \BibitemOpen
  \bibfield  {author} {\bibinfo {author} {\bibfnamefont {L.}~\bibnamefont
  {Boehnke}}, \bibinfo {author} {\bibfnamefont {F.}~\bibnamefont {Nilsson}},
  \bibinfo {author} {\bibfnamefont {F.}~\bibnamefont {Aryasetiawan}}, \ and\
  \bibinfo {author} {\bibfnamefont {P.}~\bibnamefont {Werner}},\ }\href
  {\doibase 10.1103/PhysRevB.94.201106} {\bibfield  {journal} {\bibinfo
  {journal} {Phys. Rev. B}\ }\textbf {\bibinfo {volume} {94}},\ \bibinfo
  {pages} {201106} (\bibinfo {year} {2016})}\BibitemShut {NoStop}%
\bibitem [{\citenamefont {Pavarini}\ \emph {et~al.}(2004)\citenamefont
  {Pavarini}, \citenamefont {Biermann}, \citenamefont {Poteryaev},
  \citenamefont {Lichtenstein}, \citenamefont {Georges},\ and\ \citenamefont
  {Andersen}}]{Pavarini04Mott}%
  \BibitemOpen
  \bibfield  {author} {\bibinfo {author} {\bibfnamefont {E.}~\bibnamefont
  {Pavarini}}, \bibinfo {author} {\bibfnamefont {S.}~\bibnamefont {Biermann}},
  \bibinfo {author} {\bibfnamefont {A.}~\bibnamefont {Poteryaev}}, \bibinfo
  {author} {\bibfnamefont {A.~I.}\ \bibnamefont {Lichtenstein}}, \bibinfo
  {author} {\bibfnamefont {A.}~\bibnamefont {Georges}}, \ and\ \bibinfo
  {author} {\bibfnamefont {O.~K.}\ \bibnamefont {Andersen}},\ }\href {\doibase
  10.1103/PhysRevLett.92.176403} {\bibfield  {journal} {\bibinfo  {journal}
  {{Phys. Rev. Lett.}}\ }\textbf {\bibinfo {volume} {92}},\ \bibinfo {pages}
  {176403} (\bibinfo {year} {2004})}\BibitemShut {NoStop}%
\bibitem [{\citenamefont {Lechermann}\ \emph {et~al.}(2006)\citenamefont
  {Lechermann}, \citenamefont {Georges}, \citenamefont {Poteryaev},
  \citenamefont {Biermann}, \citenamefont {Posternak}, \citenamefont
  {Yamasaki},\ and\ \citenamefont {Andersen}}]{Lechermann06Dynamical}%
  \BibitemOpen
  \bibfield  {author} {\bibinfo {author} {\bibfnamefont {F.}~\bibnamefont
  {Lechermann}}, \bibinfo {author} {\bibfnamefont {A.}~\bibnamefont {Georges}},
  \bibinfo {author} {\bibfnamefont {A.}~\bibnamefont {Poteryaev}}, \bibinfo
  {author} {\bibfnamefont {S.}~\bibnamefont {Biermann}}, \bibinfo {author}
  {\bibfnamefont {M.}~\bibnamefont {Posternak}}, \bibinfo {author}
  {\bibfnamefont {A.}~\bibnamefont {Yamasaki}}, \ and\ \bibinfo {author}
  {\bibfnamefont {O.~K.}\ \bibnamefont {Andersen}},\ }\href {\doibase
  10.1103/PhysRevB.74.125120} {\bibfield  {journal} {\bibinfo  {journal}
  {{Phys. Rev. B}}\ }\textbf {\bibinfo {volume} {74}},\ \bibinfo {pages}
  {125120} (\bibinfo {year} {2006})}\BibitemShut {NoStop}%
\bibitem [{\citenamefont {Si}\ and\ \citenamefont
  {Smith}(1996)}]{Si96Kosterlitz}%
  \BibitemOpen
  \bibfield  {author} {\bibinfo {author} {\bibfnamefont {Q.}~\bibnamefont
  {Si}}\ and\ \bibinfo {author} {\bibfnamefont {J.~L.}\ \bibnamefont {Smith}},\
  }\href {\doibase 10.1103/PhysRevLett.77.3391} {\bibfield  {journal} {\bibinfo
   {journal} {Phys. Rev. Lett.}\ }\textbf {\bibinfo {volume} {77}},\ \bibinfo
  {pages} {3391} (\bibinfo {year} {1996})}\BibitemShut {NoStop}%
\bibitem [{\citenamefont {Smith}\ and\ \citenamefont
  {Si}(2000)}]{Smith00Spatial}%
  \BibitemOpen
  \bibfield  {author} {\bibinfo {author} {\bibfnamefont {J.~L.}\ \bibnamefont
  {Smith}}\ and\ \bibinfo {author} {\bibfnamefont {Q.}~\bibnamefont {Si}},\
  }\href {\doibase 10.1103/PhysRevB.61.5184} {\bibfield  {journal} {\bibinfo
  {journal} {Phys. Rev. B}\ }\textbf {\bibinfo {volume} {61}},\ \bibinfo
  {pages} {5184} (\bibinfo {year} {2000})}\BibitemShut {NoStop}%
\bibitem [{\citenamefont {Chitra}\ and\ \citenamefont
  {Kotliar}(2001)}]{Chitra01Effective}%
  \BibitemOpen
  \bibfield  {author} {\bibinfo {author} {\bibfnamefont {R.}~\bibnamefont
  {Chitra}}\ and\ \bibinfo {author} {\bibfnamefont {G.}~\bibnamefont
  {Kotliar}},\ }\href {\doibase 10.1103/PhysRevB.63.115110} {\bibfield
  {journal} {\bibinfo  {journal} {Phys. Rev. B}\ }\textbf {\bibinfo {volume}
  {63}},\ \bibinfo {pages} {115110} (\bibinfo {year} {2001})}\BibitemShut
  {NoStop}%
\bibitem [{\citenamefont {Sun}\ and\ \citenamefont
  {Kotliar}(2002)}]{Sun02Extended}%
  \BibitemOpen
  \bibfield  {author} {\bibinfo {author} {\bibfnamefont {P.}~\bibnamefont
  {Sun}}\ and\ \bibinfo {author} {\bibfnamefont {G.}~\bibnamefont {Kotliar}},\
  }\href {\doibase 10.1103/PhysRevB.66.085120} {\bibfield  {journal} {\bibinfo
  {journal} {{Phys. Rev. B}}\ }\textbf {\bibinfo {volume} {66}},\ \bibinfo
  {pages} {085120} (\bibinfo {year} {2002})}\BibitemShut {NoStop}%
\bibitem [{\citenamefont {Biermann}\ \emph {et~al.}(2003)\citenamefont
  {Biermann}, \citenamefont {Aryasetiawan},\ and\ \citenamefont
  {Georges}}]{Biermann03FirstPrinciples}%
  \BibitemOpen
  \bibfield  {author} {\bibinfo {author} {\bibfnamefont {S.}~\bibnamefont
  {Biermann}}, \bibinfo {author} {\bibfnamefont {F.}~\bibnamefont
  {Aryasetiawan}}, \ and\ \bibinfo {author} {\bibfnamefont {A.}~\bibnamefont
  {Georges}},\ }\href {\doibase 10.1103/PhysRevLett.90.086402} {\bibfield
  {journal} {\bibinfo  {journal} {{Physical Review Letters}}\ }\textbf
  {\bibinfo {volume} {90}},\ \bibinfo {pages} {086402} (\bibinfo {year}
  {2003})}\BibitemShut {NoStop}%
\bibitem [{Note1()}]{Note1}%
  \BibitemOpen
  \bibinfo {note} {This formalism has been introduced in \cite
  {Biermann03FirstPrinciples} as $GW$+DMFT. Here, we use the term $GW$+EDMFT to
  distinguish this approach from recent implementations that discard the
  screening of the impurity interaction}\BibitemShut {NoStop}%
\bibitem [{\citenamefont {Gole\ifmmode~\check{z}\else \v{z}\fi{}}\ \emph
  {et~al.}(2017)\citenamefont {Gole\ifmmode~\check{z}\else \v{z}\fi{}},
  \citenamefont {Boehnke}, \citenamefont {Strand}, \citenamefont {Eckstein},\
  and\ \citenamefont {Werner}}]{Golez17Nonequilibrium}%
  \BibitemOpen
  \bibfield  {author} {\bibinfo {author} {\bibfnamefont {D.}~\bibnamefont
  {Gole\ifmmode~\check{z}\else \v{z}\fi{}}}, \bibinfo {author} {\bibfnamefont
  {L.}~\bibnamefont {Boehnke}}, \bibinfo {author} {\bibfnamefont {H.~U.~R.}\
  \bibnamefont {Strand}}, \bibinfo {author} {\bibfnamefont {M.}~\bibnamefont
  {Eckstein}}, \ and\ \bibinfo {author} {\bibfnamefont {P.}~\bibnamefont
  {Werner}},\ }\href {\doibase 10.1103/PhysRevLett.118.246402} {\bibfield
  {journal} {\bibinfo  {journal} {Phys. Rev. Lett.}\ }\textbf {\bibinfo
  {volume} {118}},\ \bibinfo {pages} {246402} (\bibinfo {year}
  {2017})}\BibitemShut {NoStop}%
\bibitem [{\citenamefont {Werner}\ and\ \citenamefont
  {Casula}(2016)}]{IOPreview}%
  \BibitemOpen
  \bibfield  {author} {\bibinfo {author} {\bibfnamefont {P.}~\bibnamefont
  {Werner}}\ and\ \bibinfo {author} {\bibfnamefont {M.}~\bibnamefont
  {Casula}},\ }\href {http://stacks.iop.org/0953-8984/28/i=38/a=383001}
  {\bibfield  {journal} {\bibinfo  {journal} {Journal of Physics: Condensed
  Matter}\ }\textbf {\bibinfo {volume} {28}},\ \bibinfo {pages} {383001}
  (\bibinfo {year} {2016})}\BibitemShut {NoStop}%
\bibitem [{\citenamefont {Sakuma}\ \emph {et~al.}(2013)\citenamefont {Sakuma},
  \citenamefont {Werner},\ and\ \citenamefont
  {Aryasetiawan}}]{Sakuma13Electronic}%
  \BibitemOpen
  \bibfield  {author} {\bibinfo {author} {\bibfnamefont {R.}~\bibnamefont
  {Sakuma}}, \bibinfo {author} {\bibfnamefont {P.}~\bibnamefont {Werner}}, \
  and\ \bibinfo {author} {\bibfnamefont {F.}~\bibnamefont {Aryasetiawan}},\
  }\href {\doibase 10.1103/PhysRevB.88.235110} {\bibfield  {journal} {\bibinfo
  {journal} {{Phys. Rev. B}}\ }\textbf {\bibinfo {volume} {88}},\ \bibinfo
  {pages} {235110} (\bibinfo {year} {2013})}\BibitemShut {NoStop}%
\bibitem [{\citenamefont {Tomczak}\ \emph {et~al.}(2014)\citenamefont
  {Tomczak}, \citenamefont {Casula}, \citenamefont {Miyake},\ and\
  \citenamefont {Biermann}}]{Tomczak14Asymmetry}%
  \BibitemOpen
  \bibfield  {author} {\bibinfo {author} {\bibfnamefont {J.~M.}\ \bibnamefont
  {Tomczak}}, \bibinfo {author} {\bibfnamefont {M.}~\bibnamefont {Casula}},
  \bibinfo {author} {\bibfnamefont {T.}~\bibnamefont {Miyake}}, \ and\ \bibinfo
  {author} {\bibfnamefont {S.}~\bibnamefont {Biermann}},\ }\href {\doibase
  10.1103/PhysRevB.90.165138} {\bibfield  {journal} {\bibinfo  {journal}
  {{Phys. Rev. B}}\ }\textbf {\bibinfo {volume} {90}},\ \bibinfo {pages}
  {165138} (\bibinfo {year} {2014})}\BibitemShut {NoStop}%
\bibitem [{\citenamefont {Choi}\ \emph {et~al.}(2016)\citenamefont {Choi},
  \citenamefont {Kutepov}, \citenamefont {Haule}, \citenamefont {van
  Schilfgaarde},\ and\ \citenamefont {Kotliar}}]{Choi-First}%
  \BibitemOpen
  \bibfield  {author} {\bibinfo {author} {\bibfnamefont {S.}~\bibnamefont
  {Choi}}, \bibinfo {author} {\bibfnamefont {A.}~\bibnamefont {Kutepov}},
  \bibinfo {author} {\bibfnamefont {K.}~\bibnamefont {Haule}}, \bibinfo
  {author} {\bibfnamefont {M.}~\bibnamefont {van Schilfgaarde}}, \ and\
  \bibinfo {author} {\bibfnamefont {G.}~\bibnamefont {Kotliar}},\ }\href
  {\doibase 10.1038/npjquantmats.2016.1} {\bibfield  {journal} {\bibinfo
  {journal} {Npj Quantum Materials}\ }\textbf {\bibinfo {volume} {1}},\
  \bibinfo {pages} {16001} (\bibinfo {year} {2016})}\BibitemShut {NoStop}%
\bibitem [{\citenamefont {Karlsson}(2005)}]{Karlsson05Selfconsistent}%
  \BibitemOpen
  \bibfield  {author} {\bibinfo {author} {\bibfnamefont {K.}~\bibnamefont
  {Karlsson}},\ }\href@noop {} {\bibfield  {journal} {\bibinfo  {journal}
  {Journal of Physics: Condensed Matter}\ }\textbf {\bibinfo {volume} {17}},\
  \bibinfo {pages} {7573} (\bibinfo {year} {2005})}\BibitemShut {NoStop}%
\bibitem [{\citenamefont {Ayral}\ \emph {et~al.}(2013)\citenamefont {Ayral},
  \citenamefont {Biermann},\ and\ \citenamefont {Werner}}]{Ayral13Screening}%
  \BibitemOpen
  \bibfield  {author} {\bibinfo {author} {\bibfnamefont {T.}~\bibnamefont
  {Ayral}}, \bibinfo {author} {\bibfnamefont {S.}~\bibnamefont {Biermann}}, \
  and\ \bibinfo {author} {\bibfnamefont {P.}~\bibnamefont {Werner}},\ }\href
  {\doibase 10.1103/PhysRevB.87.125149} {\bibfield  {journal} {\bibinfo
  {journal} {{Phys. Rev. B}}\ }\textbf {\bibinfo {volume} {87}},\ \bibinfo
  {pages} {125149} (\bibinfo {year} {2013})}\BibitemShut {NoStop}%
\bibitem [{\citenamefont {Huang}\ \emph {et~al.}(2014)\citenamefont {Huang},
  \citenamefont {Ayral}, \citenamefont {Biermann},\ and\ \citenamefont
  {Werner}}]{Huang14Extended}%
  \BibitemOpen
  \bibfield  {author} {\bibinfo {author} {\bibfnamefont {L.}~\bibnamefont
  {Huang}}, \bibinfo {author} {\bibfnamefont {T.}~\bibnamefont {Ayral}},
  \bibinfo {author} {\bibfnamefont {S.}~\bibnamefont {Biermann}}, \ and\
  \bibinfo {author} {\bibfnamefont {P.}~\bibnamefont {Werner}},\ }\href
  {\doibase 10.1103/PhysRevB.90.195114} {\bibfield  {journal} {\bibinfo
  {journal} {Phys. Rev. B}\ }\textbf {\bibinfo {volume} {90}},\ \bibinfo
  {pages} {195114} (\bibinfo {year} {2014})}\BibitemShut {NoStop}%
\bibitem [{\citenamefont {van Loon}\ \emph {et~al.}(2016)\citenamefont {van
  Loon}, \citenamefont {Krien}, \citenamefont {Hafermann}, \citenamefont
  {Stepanov}, \citenamefont {Lichtenstein},\ and\ \citenamefont
  {Katsnelson}}]{vanLoon16Double}%
  \BibitemOpen
  \bibfield  {author} {\bibinfo {author} {\bibfnamefont {E.~G. C.~P.}\
  \bibnamefont {van Loon}}, \bibinfo {author} {\bibfnamefont {F.}~\bibnamefont
  {Krien}}, \bibinfo {author} {\bibfnamefont {H.}~\bibnamefont {Hafermann}},
  \bibinfo {author} {\bibfnamefont {E.~A.}\ \bibnamefont {Stepanov}}, \bibinfo
  {author} {\bibfnamefont {A.~I.}\ \bibnamefont {Lichtenstein}}, \ and\
  \bibinfo {author} {\bibfnamefont {M.~I.}\ \bibnamefont {Katsnelson}},\ }\href
  {\doibase 10.1103/PhysRevB.93.155162} {\bibfield  {journal} {\bibinfo
  {journal} {Phys. Rev. B}\ }\textbf {\bibinfo {volume} {93}},\ \bibinfo
  {pages} {155162} (\bibinfo {year} {2016})}\BibitemShut {NoStop}%
\bibitem [{\citenamefont {Stepanov}\ \emph
  {et~al.}(2016{\natexlab{a}})\citenamefont {Stepanov}, \citenamefont {van
  Loon}, \citenamefont {Katanin}, \citenamefont {Lichtenstein}, \citenamefont
  {Katsnelson},\ and\ \citenamefont {Rubtsov}}]{Stepanov16Selfconsistent}%
  \BibitemOpen
  \bibfield  {author} {\bibinfo {author} {\bibfnamefont {E.~A.}\ \bibnamefont
  {Stepanov}}, \bibinfo {author} {\bibfnamefont {E.~G. C.~P.}\ \bibnamefont
  {van Loon}}, \bibinfo {author} {\bibfnamefont {A.~A.}\ \bibnamefont
  {Katanin}}, \bibinfo {author} {\bibfnamefont {A.~I.}\ \bibnamefont
  {Lichtenstein}}, \bibinfo {author} {\bibfnamefont {M.~I.}\ \bibnamefont
  {Katsnelson}}, \ and\ \bibinfo {author} {\bibfnamefont {A.~N.}\ \bibnamefont
  {Rubtsov}},\ }\href {\doibase 10.1103/PhysRevB.93.045107} {\bibfield
  {journal} {\bibinfo  {journal} {Phys. Rev. B}\ }\textbf {\bibinfo {volume}
  {93}},\ \bibinfo {pages} {045107} (\bibinfo {year}
  {2016}{\natexlab{a}})}\BibitemShut {NoStop}%
\bibitem [{\citenamefont {Ayral}\ \emph {et~al.}(2017)\citenamefont {Ayral},
  \citenamefont {Biermann}, \citenamefont {Werner},\ and\ \citenamefont
  {Boehnke}}]{Ayral17Influence}%
  \BibitemOpen
  \bibfield  {author} {\bibinfo {author} {\bibfnamefont {T.}~\bibnamefont
  {Ayral}}, \bibinfo {author} {\bibfnamefont {S.}~\bibnamefont {Biermann}},
  \bibinfo {author} {\bibfnamefont {P.}~\bibnamefont {Werner}}, \ and\ \bibinfo
  {author} {\bibfnamefont {L.}~\bibnamefont {Boehnke}},\ }\href {\doibase
  10.1103/PhysRevB.95.245130} {\bibfield  {journal} {\bibinfo  {journal} {Phys.
  Rev. B}\ }\textbf {\bibinfo {volume} {95}},\ \bibinfo {pages} {245130}
  (\bibinfo {year} {2017})}\BibitemShut {NoStop}%
\bibitem [{\citenamefont {Lee}\ and\ \citenamefont
  {Haule}(2017)}]{Lee17diatomic}%
  \BibitemOpen
  \bibfield  {author} {\bibinfo {author} {\bibfnamefont {J.}~\bibnamefont
  {Lee}}\ and\ \bibinfo {author} {\bibfnamefont {K.}~\bibnamefont {Haule}},\
  }\href {\doibase 10.1103/PhysRevB.95.155104} {\bibfield  {journal} {\bibinfo
  {journal} {Phys. Rev. B}\ }\textbf {\bibinfo {volume} {95}},\ \bibinfo
  {pages} {155104} (\bibinfo {year} {2017})}\BibitemShut {NoStop}%
\bibitem [{\citenamefont {Wadati}\ \emph {et~al.}(2014)\citenamefont {Wadati},
  \citenamefont {Mravlje}, \citenamefont {Yoshimatsu}, \citenamefont
  {Kumigashira}, \citenamefont {Oshima}, \citenamefont {Sugiyama},
  \citenamefont {Ikenaga}, \citenamefont {Fujimori}, \citenamefont {Georges},
  \citenamefont {Radetinac}, \citenamefont {Takahashi}, \citenamefont
  {Kawasaki},\ and\ \citenamefont {Tokura}}]{Wadati14Photo}%
  \BibitemOpen
  \bibfield  {author} {\bibinfo {author} {\bibfnamefont {H.}~\bibnamefont
  {Wadati}}, \bibinfo {author} {\bibfnamefont {J.}~\bibnamefont {Mravlje}},
  \bibinfo {author} {\bibfnamefont {K.}~\bibnamefont {Yoshimatsu}}, \bibinfo
  {author} {\bibfnamefont {H.}~\bibnamefont {Kumigashira}}, \bibinfo {author}
  {\bibfnamefont {M.}~\bibnamefont {Oshima}}, \bibinfo {author} {\bibfnamefont
  {T.}~\bibnamefont {Sugiyama}}, \bibinfo {author} {\bibfnamefont
  {E.}~\bibnamefont {Ikenaga}}, \bibinfo {author} {\bibfnamefont
  {A.}~\bibnamefont {Fujimori}}, \bibinfo {author} {\bibfnamefont
  {A.}~\bibnamefont {Georges}}, \bibinfo {author} {\bibfnamefont
  {A.}~\bibnamefont {Radetinac}}, \bibinfo {author} {\bibfnamefont {K.~S.}\
  \bibnamefont {Takahashi}}, \bibinfo {author} {\bibfnamefont {M.}~\bibnamefont
  {Kawasaki}}, \ and\ \bibinfo {author} {\bibfnamefont {Y.}~\bibnamefont
  {Tokura}},\ }\href {\doibase 10.1103/PhysRevB.90.205131} {\bibfield
  {journal} {\bibinfo  {journal} {Phys. Rev. B}\ }\textbf {\bibinfo {volume}
  {90}},\ \bibinfo {pages} {205131} (\bibinfo {year} {2014})}\BibitemShut
  {NoStop}%
\bibitem [{\citenamefont {Marzari}\ and\ \citenamefont
  {Vanderbilt}(1997)}]{Marzari97Maximally}%
  \BibitemOpen
  \bibfield  {author} {\bibinfo {author} {\bibfnamefont {N.}~\bibnamefont
  {Marzari}}\ and\ \bibinfo {author} {\bibfnamefont {D.}~\bibnamefont
  {Vanderbilt}},\ }\href {\doibase 10.1103/PhysRevB.56.12847} {\bibfield
  {journal} {\bibinfo  {journal} {{Phys. Rev. B}}\ }\textbf {\bibinfo {volume}
  {56}},\ \bibinfo {pages} {12847} (\bibinfo {year} {1997})}\BibitemShut
  {NoStop}%
\bibitem [{\citenamefont {Mostofi}\ \emph {et~al.}(2008)\citenamefont
  {Mostofi}, \citenamefont {Yates}, \citenamefont {Lee}, \citenamefont {Souza},
  \citenamefont {Vanderbilt},\ and\ \citenamefont
  {Marzari}}]{Mostofi08wannier90}%
  \BibitemOpen
  \bibfield  {author} {\bibinfo {author} {\bibfnamefont {A.~A.}\ \bibnamefont
  {Mostofi}}, \bibinfo {author} {\bibfnamefont {J.~R.}\ \bibnamefont {Yates}},
  \bibinfo {author} {\bibfnamefont {Y.-S.}\ \bibnamefont {Lee}}, \bibinfo
  {author} {\bibfnamefont {I.}~\bibnamefont {Souza}}, \bibinfo {author}
  {\bibfnamefont {D.}~\bibnamefont {Vanderbilt}}, \ and\ \bibinfo {author}
  {\bibfnamefont {N.}~\bibnamefont {Marzari}},\ }\href {\doibase
  {http://dx.doi.org/10.1016/j.cpc.2007.11.016}} {\bibfield  {journal}
  {\bibinfo  {journal} {{Computer Physics Communications }}\ }\textbf {\bibinfo
  {volume} {178}},\ \bibinfo {pages} {685} (\bibinfo {year}
  {2008})}\BibitemShut {NoStop}%
\bibitem [{\citenamefont {H\"ugel}\ \emph {et~al.}(2016)\citenamefont
  {H\"ugel}, \citenamefont {Werner}, \citenamefont {Pollet},\ and\
  \citenamefont {Strand}}]{Hugel16Bosonic}%
  \BibitemOpen
  \bibfield  {author} {\bibinfo {author} {\bibfnamefont {D.}~\bibnamefont
  {H\"ugel}}, \bibinfo {author} {\bibfnamefont {P.}~\bibnamefont {Werner}},
  \bibinfo {author} {\bibfnamefont {L.}~\bibnamefont {Pollet}}, \ and\ \bibinfo
  {author} {\bibfnamefont {H.~U.~R.}\ \bibnamefont {Strand}},\ }\href {\doibase
  10.1103/PhysRevB.94.195119} {\bibfield  {journal} {\bibinfo  {journal} {Phys.
  Rev. B}\ }\textbf {\bibinfo {volume} {94}},\ \bibinfo {pages} {195119}
  (\bibinfo {year} {2016})}\BibitemShut {NoStop}%
\bibitem [{\citenamefont {Kanamori}(1963)}]{Kanamori63Electron}%
  \BibitemOpen
  \bibfield  {author} {\bibinfo {author} {\bibfnamefont {J.}~\bibnamefont
  {Kanamori}},\ }\href {\doibase 10.1143/PTP.30.275} {\bibfield  {journal}
  {\bibinfo  {journal} {Progress of Theoretical Physics}\ }\textbf {\bibinfo
  {volume} {30}},\ \bibinfo {pages} {275} (\bibinfo {year} {1963})}\BibitemShut
  {NoStop}%
\bibitem [{\citenamefont {ALMBLADH}\ \emph {et~al.}(1999)\citenamefont
  {ALMBLADH}, \citenamefont {BARTH},\ and\ \citenamefont
  {LEEUWEN}}]{Almbladh99Variational}%
  \BibitemOpen
  \bibfield  {author} {\bibinfo {author} {\bibfnamefont {C.-O.}\ \bibnamefont
  {ALMBLADH}}, \bibinfo {author} {\bibfnamefont {U.~V.}\ \bibnamefont {BARTH}},
  \ and\ \bibinfo {author} {\bibfnamefont {R.~V.}\ \bibnamefont {LEEUWEN}},\
  }\href {\doibase 10.1142/S0217979299000436} {\bibfield  {journal} {\bibinfo
  {journal} {International Journal of Modern Physics B}\ }\textbf {\bibinfo
  {volume} {13}},\ \bibinfo {pages} {535} (\bibinfo {year} {1999})}\BibitemShut
  {NoStop}%
\bibitem [{\citenamefont {{Rohringer}}\ \emph {et~al.}(2017)\citenamefont
  {{Rohringer}}, \citenamefont {{Hafermann}}, \citenamefont {{Toschi}},
  \citenamefont {{Katanin}}, \citenamefont {{Antipov}}, \citenamefont
  {{Katsnelson}}, \citenamefont {{Lichtenstein}}, \citenamefont {{Rubtsov}},\
  and\ \citenamefont {{Held}}}]{Rohringer17Diagrammatic}%
  \BibitemOpen
  \bibfield  {author} {\bibinfo {author} {\bibfnamefont {G.}~\bibnamefont
  {{Rohringer}}}, \bibinfo {author} {\bibfnamefont {H.}~\bibnamefont
  {{Hafermann}}}, \bibinfo {author} {\bibfnamefont {A.}~\bibnamefont
  {{Toschi}}}, \bibinfo {author} {\bibfnamefont {A.~A.}\ \bibnamefont
  {{Katanin}}}, \bibinfo {author} {\bibfnamefont {A.~E.}\ \bibnamefont
  {{Antipov}}}, \bibinfo {author} {\bibfnamefont {M.~I.}\ \bibnamefont
  {{Katsnelson}}}, \bibinfo {author} {\bibfnamefont {A.~I.}\ \bibnamefont
  {{Lichtenstein}}}, \bibinfo {author} {\bibfnamefont {A.~N.}\ \bibnamefont
  {{Rubtsov}}}, \ and\ \bibinfo {author} {\bibfnamefont {K.}~\bibnamefont
  {{Held}}},\ }\href@noop {} {\bibfield  {journal} {\bibinfo  {journal} {ArXiv
  e-prints}\ } (\bibinfo {year} {2017})},\ \Eprint
  {http://arxiv.org/abs/1705.00024} {arXiv:1705.00024 [cond-mat.str-el]}
  \BibitemShut {NoStop}%
\bibitem [{\citenamefont {Toschi}\ \emph {et~al.}(2007)\citenamefont {Toschi},
  \citenamefont {Katanin},\ and\ \citenamefont {Held}}]{Toschi07Dynamical}%
  \BibitemOpen
  \bibfield  {author} {\bibinfo {author} {\bibfnamefont {A.}~\bibnamefont
  {Toschi}}, \bibinfo {author} {\bibfnamefont {A.~A.}\ \bibnamefont {Katanin}},
  \ and\ \bibinfo {author} {\bibfnamefont {K.}~\bibnamefont {Held}},\ }\href
  {\doibase 10.1103/PhysRevB.75.045118} {\bibfield  {journal} {\bibinfo
  {journal} {Phys. Rev. B}\ }\textbf {\bibinfo {volume} {75}},\ \bibinfo
  {pages} {045118} (\bibinfo {year} {2007})}\BibitemShut {NoStop}%
\bibitem [{\citenamefont {Ayral}\ and\ \citenamefont
  {Parcollet}(2015)}]{Ayral15Mott}%
  \BibitemOpen
  \bibfield  {author} {\bibinfo {author} {\bibfnamefont {T.}~\bibnamefont
  {Ayral}}\ and\ \bibinfo {author} {\bibfnamefont {O.}~\bibnamefont
  {Parcollet}},\ }\href {\doibase 10.1103/PhysRevB.92.115109} {\bibfield
  {journal} {\bibinfo  {journal} {Phys. Rev. B}\ }\textbf {\bibinfo {volume}
  {92}},\ \bibinfo {pages} {115109} (\bibinfo {year} {2015})}\BibitemShut
  {NoStop}%
\bibitem [{\citenamefont {Ayral}\ and\ \citenamefont
  {Parcollet}(2016{\natexlab{a}})}]{Ayral16Motta}%
  \BibitemOpen
  \bibfield  {author} {\bibinfo {author} {\bibfnamefont {T.}~\bibnamefont
  {Ayral}}\ and\ \bibinfo {author} {\bibfnamefont {O.}~\bibnamefont
  {Parcollet}},\ }\href {\doibase 10.1103/PhysRevB.93.235124} {\bibfield
  {journal} {\bibinfo  {journal} {Phys. Rev. B}\ }\textbf {\bibinfo {volume}
  {93}},\ \bibinfo {pages} {235124} (\bibinfo {year}
  {2016}{\natexlab{a}})}\BibitemShut {NoStop}%
\bibitem [{\citenamefont {Holm}\ and\ \citenamefont {von
  Barth}(1998)}]{Holm98Fully}%
  \BibitemOpen
  \bibfield  {author} {\bibinfo {author} {\bibfnamefont {B.}~\bibnamefont
  {Holm}}\ and\ \bibinfo {author} {\bibfnamefont {U.}~\bibnamefont {von
  Barth}},\ }\href {\doibase 10.1103/PhysRevB.57.2108} {\bibfield  {journal}
  {\bibinfo  {journal} {Phys. Rev. B}\ }\textbf {\bibinfo {volume} {57}},\
  \bibinfo {pages} {2108} (\bibinfo {year} {1998})}\BibitemShut {NoStop}%
\bibitem [{\citenamefont {Garc\'{\i}a-Gonz\'alez}\ and\ \citenamefont
  {Godby}(2001)}]{GarciaSelf-Consistent}%
  \BibitemOpen
  \bibfield  {author} {\bibinfo {author} {\bibfnamefont {P.}~\bibnamefont
  {Garc\'{\i}a-Gonz\'alez}}\ and\ \bibinfo {author} {\bibfnamefont {R.~W.}\
  \bibnamefont {Godby}},\ }\href {\doibase 10.1103/PhysRevB.63.075112}
  {\bibfield  {journal} {\bibinfo  {journal} {Phys. Rev. B}\ }\textbf {\bibinfo
  {volume} {63}},\ \bibinfo {pages} {075112} (\bibinfo {year}
  {2001})}\BibitemShut {NoStop}%
\bibitem [{\citenamefont {Koval}\ \emph {et~al.}(2014)\citenamefont {Koval},
  \citenamefont {Foerster},\ and\ \citenamefont
  {S\'anchez-Portal}}]{Koval14Fully}%
  \BibitemOpen
  \bibfield  {author} {\bibinfo {author} {\bibfnamefont {P.}~\bibnamefont
  {Koval}}, \bibinfo {author} {\bibfnamefont {D.}~\bibnamefont {Foerster}}, \
  and\ \bibinfo {author} {\bibfnamefont {D.}~\bibnamefont {S\'anchez-Portal}},\
  }\href {\doibase 10.1103/PhysRevB.89.155417} {\bibfield  {journal} {\bibinfo
  {journal} {Phys. Rev. B}\ }\textbf {\bibinfo {volume} {89}},\ \bibinfo
  {pages} {155417} (\bibinfo {year} {2014})}\BibitemShut {NoStop}%
\bibitem [{\citenamefont {Gatti}\ and\ \citenamefont
  {Guzzo}(2013)}]{Gatti13Dynamical}%
  \BibitemOpen
  \bibfield  {author} {\bibinfo {author} {\bibfnamefont {M.}~\bibnamefont
  {Gatti}}\ and\ \bibinfo {author} {\bibfnamefont {M.}~\bibnamefont {Guzzo}},\
  }\href {\doibase 10.1103/PhysRevB.87.155147} {\bibfield  {journal} {\bibinfo
  {journal} {{Phys. Rev. B}}\ }\textbf {\bibinfo {volume} {87}},\ \bibinfo
  {pages} {155147} (\bibinfo {year} {2013})}\BibitemShut {NoStop}%
\bibitem [{\citenamefont {Kotliar}\ \emph {et~al.}(2006)\citenamefont
  {Kotliar}, \citenamefont {Savrasov}, \citenamefont {Haule}, \citenamefont
  {Oudovenko}, \citenamefont {Parcollet},\ and\ \citenamefont
  {Marianetti}}]{Kotliar06Electronic}%
  \BibitemOpen
  \bibfield  {author} {\bibinfo {author} {\bibfnamefont {G.}~\bibnamefont
  {Kotliar}}, \bibinfo {author} {\bibfnamefont {S.~Y.}\ \bibnamefont
  {Savrasov}}, \bibinfo {author} {\bibfnamefont {K.}~\bibnamefont {Haule}},
  \bibinfo {author} {\bibfnamefont {V.~S.}\ \bibnamefont {Oudovenko}}, \bibinfo
  {author} {\bibfnamefont {O.}~\bibnamefont {Parcollet}}, \ and\ \bibinfo
  {author} {\bibfnamefont {C.~A.}\ \bibnamefont {Marianetti}},\ }\href
  {\doibase 10.1103/RevModPhys.78.865} {\bibfield  {journal} {\bibinfo
  {journal} {Rev. Mod. Phys.}\ }\textbf {\bibinfo {volume} {78}},\ \bibinfo
  {pages} {865} (\bibinfo {year} {2006})}\BibitemShut {NoStop}%
\bibitem [{\citenamefont {Werner}\ \emph {et~al.}(2012)\citenamefont {Werner},
  \citenamefont {Casula}, \citenamefont {Miyake}, \citenamefont {Aryasetiawan},
  \citenamefont {Millis},\ and\ \citenamefont {Biermann}}]{Werner12Satellites}%
  \BibitemOpen
  \bibfield  {author} {\bibinfo {author} {\bibfnamefont {P.}~\bibnamefont
  {Werner}}, \bibinfo {author} {\bibfnamefont {M.}~\bibnamefont {Casula}},
  \bibinfo {author} {\bibfnamefont {T.}~\bibnamefont {Miyake}}, \bibinfo
  {author} {\bibfnamefont {F.}~\bibnamefont {Aryasetiawan}}, \bibinfo {author}
  {\bibfnamefont {A.~J.}\ \bibnamefont {Millis}}, \ and\ \bibinfo {author}
  {\bibfnamefont {S.}~\bibnamefont {Biermann}},\ }\href {\doibase
  10.1038/nphys2250} {\bibfield  {journal} {\bibinfo  {journal} {Nat Phys}\
  }\textbf {\bibinfo {volume} {8}},\ \bibinfo {pages} {331} (\bibinfo {year}
  {2012})}\BibitemShut {NoStop}%
\bibitem [{\citenamefont {Werner}\ \emph {et~al.}(2015)\citenamefont {Werner},
  \citenamefont {Sakuma}, \citenamefont {Nilsson},\ and\ \citenamefont
  {Aryasetiawan}}]{Werner15Dynamical}%
  \BibitemOpen
  \bibfield  {author} {\bibinfo {author} {\bibfnamefont {P.}~\bibnamefont
  {Werner}}, \bibinfo {author} {\bibfnamefont {R.}~\bibnamefont {Sakuma}},
  \bibinfo {author} {\bibfnamefont {F.}~\bibnamefont {Nilsson}}, \ and\
  \bibinfo {author} {\bibfnamefont {F.}~\bibnamefont {Aryasetiawan}},\ }\href
  {\doibase 10.1103/PhysRevB.91.125142} {\bibfield  {journal} {\bibinfo
  {journal} {{Phys. Rev. B}}\ }\textbf {\bibinfo {volume} {91}},\ \bibinfo
  {pages} {125142} (\bibinfo {year} {2015})}\BibitemShut {NoStop}%
\bibitem [{\citenamefont {van Roekeghem}\ \emph {et~al.}(2014)\citenamefont
  {van Roekeghem}, \citenamefont {Ayral}, \citenamefont {Tomczak},
  \citenamefont {Casula}, \citenamefont {Xu}, \citenamefont {Ding},
  \citenamefont {Ferrero}, \citenamefont {Parcollet}, \citenamefont {Jiang},\
  and\ \citenamefont {Biermann}}]{vanRoekeghem14Dynamical}%
  \BibitemOpen
  \bibfield  {author} {\bibinfo {author} {\bibfnamefont {A.}~\bibnamefont {van
  Roekeghem}}, \bibinfo {author} {\bibfnamefont {T.}~\bibnamefont {Ayral}},
  \bibinfo {author} {\bibfnamefont {J.~M.}\ \bibnamefont {Tomczak}}, \bibinfo
  {author} {\bibfnamefont {M.}~\bibnamefont {Casula}}, \bibinfo {author}
  {\bibfnamefont {N.}~\bibnamefont {Xu}}, \bibinfo {author} {\bibfnamefont
  {H.}~\bibnamefont {Ding}}, \bibinfo {author} {\bibfnamefont {M.}~\bibnamefont
  {Ferrero}}, \bibinfo {author} {\bibfnamefont {O.}~\bibnamefont {Parcollet}},
  \bibinfo {author} {\bibfnamefont {H.}~\bibnamefont {Jiang}}, \ and\ \bibinfo
  {author} {\bibfnamefont {S.}~\bibnamefont {Biermann}},\ }\href {\doibase
  10.1103/PhysRevLett.113.266403} {\bibfield  {journal} {\bibinfo  {journal}
  {Phys. Rev. Lett.}\ }\textbf {\bibinfo {volume} {113}},\ \bibinfo {pages}
  {266403} (\bibinfo {year} {2014})}\BibitemShut {NoStop}%
\bibitem [{\citenamefont {Casula}\ \emph {et~al.}(2012)\citenamefont {Casula},
  \citenamefont {Rubtsov},\ and\ \citenamefont {Biermann}}]{Casula12Dynamical}%
  \BibitemOpen
  \bibfield  {author} {\bibinfo {author} {\bibfnamefont {M.}~\bibnamefont
  {Casula}}, \bibinfo {author} {\bibfnamefont {A.}~\bibnamefont {Rubtsov}}, \
  and\ \bibinfo {author} {\bibfnamefont {S.}~\bibnamefont {Biermann}},\ }\href
  {\doibase 10.1103/PhysRevB.85.035115} {\bibfield  {journal} {\bibinfo
  {journal} {{Phys. Rev. B}}\ }\textbf {\bibinfo {volume} {85}},\ \bibinfo
  {pages} {035115} (\bibinfo {year} {2012})}\BibitemShut {NoStop}%
\bibitem [{\citenamefont {Toschi}\ \emph {et~al.}(2011)\citenamefont {Toschi},
  \citenamefont {Rohringer}, \citenamefont {Katanin},\ and\ \citenamefont
  {Held}}]{Toschi11Abinitio}%
  \BibitemOpen
  \bibfield  {author} {\bibinfo {author} {\bibfnamefont {A.}~\bibnamefont
  {Toschi}}, \bibinfo {author} {\bibfnamefont {G.}~\bibnamefont {Rohringer}},
  \bibinfo {author} {\bibfnamefont {A.}~\bibnamefont {Katanin}}, \ and\
  \bibinfo {author} {\bibfnamefont {K.}~\bibnamefont {Held}},\ }\href {\doibase
  10.1002/andp.201100036} {\bibfield  {journal} {\bibinfo  {journal} {Annalen
  der Physik}\ }\textbf {\bibinfo {volume} {523}},\ \bibinfo {pages} {698}
  (\bibinfo {year} {2011})}\BibitemShut {NoStop}%
\bibitem [{\citenamefont {Galler}\ \emph {et~al.}(2017)\citenamefont {Galler},
  \citenamefont {Thunstr\"om}, \citenamefont {Gunacker}, \citenamefont
  {Tomczak},\ and\ \citenamefont {Held}}]{Galler17Abinitio}%
  \BibitemOpen
  \bibfield  {author} {\bibinfo {author} {\bibfnamefont {A.}~\bibnamefont
  {Galler}}, \bibinfo {author} {\bibfnamefont {P.}~\bibnamefont {Thunstr\"om}},
  \bibinfo {author} {\bibfnamefont {P.}~\bibnamefont {Gunacker}}, \bibinfo
  {author} {\bibfnamefont {J.~M.}\ \bibnamefont {Tomczak}}, \ and\ \bibinfo
  {author} {\bibfnamefont {K.}~\bibnamefont {Held}},\ }\href {\doibase
  10.1103/PhysRevB.95.115107} {\bibfield  {journal} {\bibinfo  {journal} {Phys.
  Rev. B}\ }\textbf {\bibinfo {volume} {95}},\ \bibinfo {pages} {115107}
  (\bibinfo {year} {2017})}\BibitemShut {NoStop}%
\bibitem [{\citenamefont {Rubtsov}\ \emph {et~al.}(2008)\citenamefont
  {Rubtsov}, \citenamefont {Katsnelson},\ and\ \citenamefont
  {Lichtenstein}}]{Rubtsov08Dual}%
  \BibitemOpen
  \bibfield  {author} {\bibinfo {author} {\bibfnamefont {A.~N.}\ \bibnamefont
  {Rubtsov}}, \bibinfo {author} {\bibfnamefont {M.~I.}\ \bibnamefont
  {Katsnelson}}, \ and\ \bibinfo {author} {\bibfnamefont {A.~I.}\ \bibnamefont
  {Lichtenstein}},\ }\href {\doibase 10.1103/PhysRevB.77.033101} {\bibfield
  {journal} {\bibinfo  {journal} {Phys. Rev. B}\ }\textbf {\bibinfo {volume}
  {77}},\ \bibinfo {pages} {033101} (\bibinfo {year} {2008})}\BibitemShut
  {NoStop}%
\bibitem [{\citenamefont {Rubtsov}\ \emph {et~al.}(2012)\citenamefont
  {Rubtsov}, \citenamefont {Katsnelson},\ and\ \citenamefont
  {Lichtenstein}}]{Rubtsov12Dual}%
  \BibitemOpen
  \bibfield  {author} {\bibinfo {author} {\bibfnamefont {A.~N.}\ \bibnamefont
  {Rubtsov}}, \bibinfo {author} {\bibfnamefont {M.~I.}\ \bibnamefont
  {Katsnelson}}, \ and\ \bibinfo {author} {\bibfnamefont {A.~I.}\ \bibnamefont
  {Lichtenstein}},\ }\href {\doibase 10.1016/j.aop.2012.01.002} {\bibfield
  {journal} {\bibinfo  {journal} {{Annals of Physics, Volume 327, Issue 5, p.
  1320-1335.}}\ }\textbf {\bibinfo {volume} {327}},\ \bibinfo {pages} {1320}
  (\bibinfo {year} {2012})}\BibitemShut {NoStop}%
\bibitem [{\citenamefont {Stepanov}\ \emph
  {et~al.}(2016{\natexlab{b}})\citenamefont {Stepanov}, \citenamefont {Huber},
  \citenamefont {van Loon}, \citenamefont {Lichtenstein},\ and\ \citenamefont
  {Katsnelson}}]{Stepanov16Local}%
  \BibitemOpen
  \bibfield  {author} {\bibinfo {author} {\bibfnamefont {E.~A.}\ \bibnamefont
  {Stepanov}}, \bibinfo {author} {\bibfnamefont {A.}~\bibnamefont {Huber}},
  \bibinfo {author} {\bibfnamefont {E.~G. C.~P.}\ \bibnamefont {van Loon}},
  \bibinfo {author} {\bibfnamefont {A.~I.}\ \bibnamefont {Lichtenstein}}, \
  and\ \bibinfo {author} {\bibfnamefont {M.~I.}\ \bibnamefont {Katsnelson}},\
  }\href {\doibase 10.1103/PhysRevB.94.205110} {\bibfield  {journal} {\bibinfo
  {journal} {Phys. Rev. B}\ }\textbf {\bibinfo {volume} {94}},\ \bibinfo
  {pages} {205110} (\bibinfo {year} {2016}{\natexlab{b}})}\BibitemShut
  {NoStop}%
\bibitem [{\citenamefont {Ayral}\ and\ \citenamefont
  {Parcollet}(2016{\natexlab{b}})}]{Ayral16Mottb}%
  \BibitemOpen
  \bibfield  {author} {\bibinfo {author} {\bibfnamefont {T.}~\bibnamefont
  {Ayral}}\ and\ \bibinfo {author} {\bibfnamefont {O.}~\bibnamefont
  {Parcollet}},\ }\href {\doibase 10.1103/PhysRevB.94.075159} {\bibfield
  {journal} {\bibinfo  {journal} {Phys. Rev. B}\ }\textbf {\bibinfo {volume}
  {94}},\ \bibinfo {pages} {075159} (\bibinfo {year}
  {2016}{\natexlab{b}})}\BibitemShut {NoStop}%
\bibitem [{\citenamefont {Friedrich}\ \emph {et~al.}(2010)\citenamefont
  {Friedrich}, \citenamefont {Bl{\"{u}}gel},\ and\ \citenamefont
  {Schindlmayr}}]{Friedrich10Efficient}%
  \BibitemOpen
  \bibfield  {author} {\bibinfo {author} {\bibfnamefont {C.}~\bibnamefont
  {Friedrich}}, \bibinfo {author} {\bibfnamefont {S.}~\bibnamefont
  {Bl{\"{u}}gel}}, \ and\ \bibinfo {author} {\bibfnamefont {A.}~\bibnamefont
  {Schindlmayr}},\ }\href {\doibase 10.1103/PhysRevB.81.125102} {\bibfield
  {journal} {\bibinfo  {journal} {{Phys. Rev. B}}\ }\textbf {\bibinfo {volume}
  {81}},\ \bibinfo {pages} {125102} (\bibinfo {year} {2010})}\BibitemShut
  {NoStop}%
\bibitem [{\citenamefont {{The FLEUR group}}()}]{fleur}%
  \BibitemOpen
  \bibfield  {author} {\bibinfo {author} {\bibnamefont {{The FLEUR group}}},\
  }\href@noop {} {\enquote {\bibinfo {title} {{The FLEUR project}},}\ }\bibinfo
  {howpublished} {\url{http://www.flapw.de}}\BibitemShut {NoStop}%
\bibitem [{\citenamefont {Freimuth}\ \emph {et~al.}(2008)\citenamefont
  {Freimuth}, \citenamefont {Mokrousov}, \citenamefont {Wortmann},
  \citenamefont {Heinze},\ and\ \citenamefont
  {Bl{\"{u}}gel}}]{Freimuth08Maximally}%
  \BibitemOpen
  \bibfield  {author} {\bibinfo {author} {\bibfnamefont {F.}~\bibnamefont
  {Freimuth}}, \bibinfo {author} {\bibfnamefont {Y.}~\bibnamefont {Mokrousov}},
  \bibinfo {author} {\bibfnamefont {D.}~\bibnamefont {Wortmann}}, \bibinfo
  {author} {\bibfnamefont {S.}~\bibnamefont {Heinze}}, \ and\ \bibinfo {author}
  {\bibfnamefont {S.}~\bibnamefont {Bl{\"{u}}gel}},\ }\href {\doibase
  10.1103/PhysRevB.78.035120} {\bibfield  {journal} {\bibinfo  {journal}
  {{Phys. Rev. B}}\ }\textbf {\bibinfo {volume} {78}},\ \bibinfo {pages}
  {035120} (\bibinfo {year} {2008})}\BibitemShut {NoStop}%
\bibitem [{\citenamefont {Sakuma}(2013)}]{Sakuma13Symmetryadapted}%
  \BibitemOpen
  \bibfield  {author} {\bibinfo {author} {\bibfnamefont {R.}~\bibnamefont
  {Sakuma}},\ }\href {\doibase 10.1103/PhysRevB.87.235109} {\bibfield
  {journal} {\bibinfo  {journal} {{Phys. Rev. B}}\ }\textbf {\bibinfo {volume}
  {87}},\ \bibinfo {pages} {235109} (\bibinfo {year} {2013})}\BibitemShut
  {NoStop}%
\bibitem [{\citenamefont {Werner}\ \emph {et~al.}(2006)\citenamefont {Werner},
  \citenamefont {Comanac}, \citenamefont {de' Medici}, \citenamefont {Troyer},\
  and\ \citenamefont {Millis}}]{Werner06ContinuousTime}%
  \BibitemOpen
  \bibfield  {author} {\bibinfo {author} {\bibfnamefont {P.}~\bibnamefont
  {Werner}}, \bibinfo {author} {\bibfnamefont {A.}~\bibnamefont {Comanac}},
  \bibinfo {author} {\bibfnamefont {L.}~\bibnamefont {de' Medici}}, \bibinfo
  {author} {\bibfnamefont {M.}~\bibnamefont {Troyer}}, \ and\ \bibinfo {author}
  {\bibfnamefont {A.~J.}\ \bibnamefont {Millis}},\ }\href {\doibase
  10.1103/PhysRevLett.97.076405} {\bibfield  {journal} {\bibinfo  {journal}
  {Phys. Rev. Lett.}\ }\textbf {\bibinfo {volume} {97}},\ \bibinfo {pages}
  {076405} (\bibinfo {year} {2006})}\BibitemShut {NoStop}%
\bibitem [{\citenamefont {Werner}\ and\ \citenamefont
  {Millis}(2007)}]{Werner07Efficient}%
  \BibitemOpen
  \bibfield  {author} {\bibinfo {author} {\bibfnamefont {P.}~\bibnamefont
  {Werner}}\ and\ \bibinfo {author} {\bibfnamefont {A.~J.}\ \bibnamefont
  {Millis}},\ }\href {\doibase 10.1103/PhysRevLett.99.146404} {\bibfield
  {journal} {\bibinfo  {journal} {{Phys. Rev. Lett.}}\ }\textbf {\bibinfo
  {volume} {99}},\ \bibinfo {pages} {146404} (\bibinfo {year}
  {2007})}\BibitemShut {NoStop}%
\bibitem [{\citenamefont {Gull}\ \emph {et~al.}(2011)\citenamefont {Gull},
  \citenamefont {Millis}, \citenamefont {Lichtenstein}, \citenamefont
  {Rubtsov}, \citenamefont {Troyer},\ and\ \citenamefont
  {Werner}}]{Gull11Continuoustime}%
  \BibitemOpen
  \bibfield  {author} {\bibinfo {author} {\bibfnamefont {E.}~\bibnamefont
  {Gull}}, \bibinfo {author} {\bibfnamefont {A.~J.}\ \bibnamefont {Millis}},
  \bibinfo {author} {\bibfnamefont {A.~I.}\ \bibnamefont {Lichtenstein}},
  \bibinfo {author} {\bibfnamefont {A.~N.}\ \bibnamefont {Rubtsov}}, \bibinfo
  {author} {\bibfnamefont {M.}~\bibnamefont {Troyer}}, \ and\ \bibinfo {author}
  {\bibfnamefont {P.}~\bibnamefont {Werner}},\ }\href {\doibase
  10.1103/RevModPhys.83.349} {\bibfield  {journal} {\bibinfo  {journal}
  {{Review of Modern Physics, vol. 83, Issue 2, pp. 349-404}}\ }\textbf
  {\bibinfo {volume} {83}},\ \bibinfo {pages} {349} (\bibinfo {year}
  {2011})}\BibitemShut {NoStop}%
\bibitem [{\citenamefont {Bauer}\ \emph {et~al.}(2011)\citenamefont {Bauer},
  \citenamefont {Carr}, \citenamefont {Evertz}, \citenamefont {Feiguin},
  \citenamefont {Freire}, \citenamefont {Fuchs}, \citenamefont {Gamper},
  \citenamefont {Gukelberger}, \citenamefont {Gull}, \citenamefont {Guertler},
  \citenamefont {Hehn}, \citenamefont {Igarashi}, \citenamefont {Isakov},
  \citenamefont {Koop}, \citenamefont {Ma}, \citenamefont {Mates},
  \citenamefont {Matsuo}, \citenamefont {Parcollet}, \citenamefont {Pawowski},
  \citenamefont {Picon}, \citenamefont {Pollet}, \citenamefont {Santos},
  \citenamefont {Scarola}, \citenamefont {Schollwck}, \citenamefont {Silva},
  \citenamefont {Surer}, \citenamefont {Todo}, \citenamefont {Trebst},
  \citenamefont {Troyer}, \citenamefont {Wall}, \citenamefont {Werner},\ and\
  \citenamefont {Wessel}}]{Bauer11Alps}%
  \BibitemOpen
  \bibfield  {author} {\bibinfo {author} {\bibfnamefont {B.}~\bibnamefont
  {Bauer}}, \bibinfo {author} {\bibfnamefont {L.~D.}\ \bibnamefont {Carr}},
  \bibinfo {author} {\bibfnamefont {H.~G.}\ \bibnamefont {Evertz}}, \bibinfo
  {author} {\bibfnamefont {A.}~\bibnamefont {Feiguin}}, \bibinfo {author}
  {\bibfnamefont {J.}~\bibnamefont {Freire}}, \bibinfo {author} {\bibfnamefont
  {S.}~\bibnamefont {Fuchs}}, \bibinfo {author} {\bibfnamefont
  {L.}~\bibnamefont {Gamper}}, \bibinfo {author} {\bibfnamefont
  {J.}~\bibnamefont {Gukelberger}}, \bibinfo {author} {\bibfnamefont
  {E.}~\bibnamefont {Gull}}, \bibinfo {author} {\bibfnamefont {S.}~\bibnamefont
  {Guertler}}, \bibinfo {author} {\bibfnamefont {A.}~\bibnamefont {Hehn}},
  \bibinfo {author} {\bibfnamefont {R.}~\bibnamefont {Igarashi}}, \bibinfo
  {author} {\bibfnamefont {S.~V.}\ \bibnamefont {Isakov}}, \bibinfo {author}
  {\bibfnamefont {D.}~\bibnamefont {Koop}}, \bibinfo {author} {\bibfnamefont
  {P.~N.}\ \bibnamefont {Ma}}, \bibinfo {author} {\bibfnamefont
  {P.}~\bibnamefont {Mates}}, \bibinfo {author} {\bibfnamefont
  {H.}~\bibnamefont {Matsuo}}, \bibinfo {author} {\bibfnamefont
  {O.}~\bibnamefont {Parcollet}}, \bibinfo {author} {\bibfnamefont
  {G.}~\bibnamefont {Pawowski}}, \bibinfo {author} {\bibfnamefont {J.~D.}\
  \bibnamefont {Picon}}, \bibinfo {author} {\bibfnamefont {L.}~\bibnamefont
  {Pollet}}, \bibinfo {author} {\bibfnamefont {E.}~\bibnamefont {Santos}},
  \bibinfo {author} {\bibfnamefont {V.~W.}\ \bibnamefont {Scarola}}, \bibinfo
  {author} {\bibfnamefont {U.}~\bibnamefont {Schollwck}}, \bibinfo {author}
  {\bibfnamefont {C.}~\bibnamefont {Silva}}, \bibinfo {author} {\bibfnamefont
  {B.}~\bibnamefont {Surer}}, \bibinfo {author} {\bibfnamefont
  {S.}~\bibnamefont {Todo}}, \bibinfo {author} {\bibfnamefont {S.}~\bibnamefont
  {Trebst}}, \bibinfo {author} {\bibfnamefont {M.}~\bibnamefont {Troyer}},
  \bibinfo {author} {\bibfnamefont {M.~L.}\ \bibnamefont {Wall}}, \bibinfo
  {author} {\bibfnamefont {P.}~\bibnamefont {Werner}}, \ and\ \bibinfo {author}
  {\bibfnamefont {S.}~\bibnamefont {Wessel}},\ }\href
  {http://stacks.iop.org/1742-5468/2011/i=05/a=P05001} {\bibfield  {journal}
  {\bibinfo  {journal} {Journal of Statistical Mechanics: Theory and
  Experiment}\ }\textbf {\bibinfo {volume} {2011}},\ \bibinfo {pages} {P05001}
  (\bibinfo {year} {2011})}\BibitemShut {NoStop}%
\bibitem [{\citenamefont {{ALPS collaboration}}()}]{Alps}%
  \BibitemOpen
  \bibfield  {author} {\bibinfo {author} {\bibnamefont {{ALPS
  collaboration}}},\ }\href@noop {} {\enquote {\bibinfo {title} {{ALPS
  project}},}\ }\bibinfo {howpublished}
  {\url{http://alps.comp-phys.org}}\BibitemShut {NoStop}%
\bibitem [{\citenamefont {Hafermann}\ \emph {et~al.}(2013)\citenamefont
  {Hafermann}, \citenamefont {Werner},\ and\ \citenamefont
  {Gull}}]{Hafermann13Efficient}%
  \BibitemOpen
  \bibfield  {author} {\bibinfo {author} {\bibfnamefont {H.}~\bibnamefont
  {Hafermann}}, \bibinfo {author} {\bibfnamefont {P.}~\bibnamefont {Werner}}, \
  and\ \bibinfo {author} {\bibfnamefont {E.}~\bibnamefont {Gull}},\ }\href
  {\doibase http://dx.doi.org/10.1016/j.cpc.2012.12.013} {\bibfield  {journal}
  {\bibinfo  {journal} {Comput. Phys. Commun.}\ }\textbf {\bibinfo {volume}
  {184}},\ \bibinfo {pages} {1280 } (\bibinfo {year} {2013})}\BibitemShut
  {NoStop}%
\bibitem [{\citenamefont {Werner}\ and\ \citenamefont
  {Millis}(2010)}]{Werner10Dynamical}%
  \BibitemOpen
  \bibfield  {author} {\bibinfo {author} {\bibfnamefont {P.}~\bibnamefont
  {Werner}}\ and\ \bibinfo {author} {\bibfnamefont {A.~J.}\ \bibnamefont
  {Millis}},\ }\href {\doibase 10.1103/PhysRevLett.104.146401} {\bibfield
  {journal} {\bibinfo  {journal} {{Phys. Rev. Lett.}}\ }\textbf {\bibinfo
  {volume} {104}},\ \bibinfo {pages} {146401} (\bibinfo {year}
  {2010})}\BibitemShut {NoStop}%
\bibitem [{\citenamefont {Boehnke}\ \emph {et~al.}(2011)\citenamefont
  {Boehnke}, \citenamefont {Hafermann}, \citenamefont {Ferrero}, \citenamefont
  {Lechermann},\ and\ \citenamefont {Parcollet}}]{Boehnke11Orthogonal}%
  \BibitemOpen
  \bibfield  {author} {\bibinfo {author} {\bibfnamefont {L.}~\bibnamefont
  {Boehnke}}, \bibinfo {author} {\bibfnamefont {H.}~\bibnamefont {Hafermann}},
  \bibinfo {author} {\bibfnamefont {M.}~\bibnamefont {Ferrero}}, \bibinfo
  {author} {\bibfnamefont {F.}~\bibnamefont {Lechermann}}, \ and\ \bibinfo
  {author} {\bibfnamefont {O.}~\bibnamefont {Parcollet}},\ }\href {\doibase
  10.1103/PhysRevB.84.075145} {\bibfield  {journal} {\bibinfo  {journal} {Phys.
  Rev. B}\ }\textbf {\bibinfo {volume} {84}},\ \bibinfo {pages} {075145}
  (\bibinfo {year} {2011})}\BibitemShut {NoStop}%
\bibitem [{\citenamefont {Steiner}\ \emph {et~al.}(2015)\citenamefont
  {Steiner}, \citenamefont {Nomura},\ and\ \citenamefont
  {Werner}}]{Steiner15Double}%
  \BibitemOpen
  \bibfield  {author} {\bibinfo {author} {\bibfnamefont {K.}~\bibnamefont
  {Steiner}}, \bibinfo {author} {\bibfnamefont {Y.}~\bibnamefont {Nomura}}, \
  and\ \bibinfo {author} {\bibfnamefont {P.}~\bibnamefont {Werner}},\ }\href
  {\doibase 10.1103/PhysRevB.92.115123} {\bibfield  {journal} {\bibinfo
  {journal} {Phys. Rev. B}\ }\textbf {\bibinfo {volume} {92}},\ \bibinfo
  {pages} {115123} (\bibinfo {year} {2015})}\BibitemShut {NoStop}%
\bibitem [{\citenamefont {Grewe}\ and\ \citenamefont
  {Keiter}(1981)}]{Grewe81Diagrammatic}%
  \BibitemOpen
  \bibfield  {author} {\bibinfo {author} {\bibfnamefont {N.}~\bibnamefont
  {Grewe}}\ and\ \bibinfo {author} {\bibfnamefont {H.}~\bibnamefont {Keiter}},\
  }\href {\doibase 10.1103/PhysRevB.24.4420} {\bibfield  {journal} {\bibinfo
  {journal} {Phys. Rev. B}\ }\textbf {\bibinfo {volume} {24}},\ \bibinfo
  {pages} {4420} (\bibinfo {year} {1981})}\BibitemShut {NoStop}%
\bibitem [{\citenamefont {Haule}\ \emph {et~al.}(2002)\citenamefont {Haule},
  \citenamefont {Rosch}, \citenamefont {Kroha},\ and\ \citenamefont
  {W\"olfle}}]{Haule02Pseudogap}%
  \BibitemOpen
  \bibfield  {author} {\bibinfo {author} {\bibfnamefont {K.}~\bibnamefont
  {Haule}}, \bibinfo {author} {\bibfnamefont {A.}~\bibnamefont {Rosch}},
  \bibinfo {author} {\bibfnamefont {J.}~\bibnamefont {Kroha}}, \ and\ \bibinfo
  {author} {\bibfnamefont {P.}~\bibnamefont {W\"olfle}},\ }\href {\doibase
  10.1103/PhysRevLett.89.236402} {\bibfield  {journal} {\bibinfo  {journal}
  {Phys. Rev. Lett.}\ }\textbf {\bibinfo {volume} {89}},\ \bibinfo {pages}
  {236402} (\bibinfo {year} {2002})}\BibitemShut {NoStop}%
\bibitem [{\citenamefont {Gole\ifmmode~\check{z}\else \v{z}\fi{}}\ \emph
  {et~al.}(2015)\citenamefont {Gole\ifmmode~\check{z}\else \v{z}\fi{}},
  \citenamefont {Eckstein},\ and\ \citenamefont {Werner}}]{Golez15Dynamics}%
  \BibitemOpen
  \bibfield  {author} {\bibinfo {author} {\bibfnamefont {D.}~\bibnamefont
  {Gole\ifmmode~\check{z}\else \v{z}\fi{}}}, \bibinfo {author} {\bibfnamefont
  {M.}~\bibnamefont {Eckstein}}, \ and\ \bibinfo {author} {\bibfnamefont
  {P.}~\bibnamefont {Werner}},\ }\href {\doibase 10.1103/PhysRevB.92.195123}
  {\bibfield  {journal} {\bibinfo  {journal} {Phys. Rev. B}\ }\textbf {\bibinfo
  {volume} {92}},\ \bibinfo {pages} {195123} (\bibinfo {year}
  {2015})}\BibitemShut {NoStop}%
\bibitem [{\citenamefont {Werner}\ and\ \citenamefont
  {Eckstein}(2013)}]{Werner08Phonon}%
  \BibitemOpen
  \bibfield  {author} {\bibinfo {author} {\bibfnamefont {P.}~\bibnamefont
  {Werner}}\ and\ \bibinfo {author} {\bibfnamefont {M.}~\bibnamefont
  {Eckstein}},\ }\href {\doibase 10.1103/PhysRevB.88.165108} {\bibfield
  {journal} {\bibinfo  {journal} {Phys. Rev. B}\ }\textbf {\bibinfo {volume}
  {88}},\ \bibinfo {pages} {165108} (\bibinfo {year} {2013})}\BibitemShut
  {NoStop}%
\bibitem [{\citenamefont {Parcollet}\ \emph {et~al.}(2015)\citenamefont
  {Parcollet}, \citenamefont {Ferrero}, \citenamefont {Ayral}, \citenamefont
  {Hafermann}, \citenamefont {Krivenko}, \citenamefont {Messio},\ and\
  \citenamefont {Seth}}]{Parcollet15Triqs}%
  \BibitemOpen
  \bibfield  {author} {\bibinfo {author} {\bibfnamefont {O.}~\bibnamefont
  {Parcollet}}, \bibinfo {author} {\bibfnamefont {M.}~\bibnamefont {Ferrero}},
  \bibinfo {author} {\bibfnamefont {T.}~\bibnamefont {Ayral}}, \bibinfo
  {author} {\bibfnamefont {H.}~\bibnamefont {Hafermann}}, \bibinfo {author}
  {\bibfnamefont {I.}~\bibnamefont {Krivenko}}, \bibinfo {author}
  {\bibfnamefont {L.}~\bibnamefont {Messio}}, \ and\ \bibinfo {author}
  {\bibfnamefont {P.}~\bibnamefont {Seth}},\ }\href {\doibase
  http://dx.doi.org/10.1016/j.cpc.2015.04.023} {\bibfield  {journal} {\bibinfo
  {journal} {Computer Physics Communications}\ }\textbf {\bibinfo {volume}
  {196}},\ \bibinfo {pages} {398 } (\bibinfo {year} {2015})}\BibitemShut
  {NoStop}%
\bibitem [{\citenamefont {Stan}\ \emph {et~al.}(2015)\citenamefont {Stan},
  \citenamefont {Romaniello}, \citenamefont {Rigamonti}, \citenamefont
  {Reining},\ and\ \citenamefont {Berger}}]{Stan15Unphysical}%
  \BibitemOpen
  \bibfield  {author} {\bibinfo {author} {\bibfnamefont {A.}~\bibnamefont
  {Stan}}, \bibinfo {author} {\bibfnamefont {P.}~\bibnamefont {Romaniello}},
  \bibinfo {author} {\bibfnamefont {S.}~\bibnamefont {Rigamonti}}, \bibinfo
  {author} {\bibfnamefont {L.}~\bibnamefont {Reining}}, \ and\ \bibinfo
  {author} {\bibfnamefont {J.~A.}\ \bibnamefont {Berger}},\ }\href
  {http://stacks.iop.org/1367-2630/17/i=9/a=093045} {\bibfield  {journal}
  {\bibinfo  {journal} {New Journal of Physics}\ }\textbf {\bibinfo {volume}
  {17}},\ \bibinfo {pages} {093045} (\bibinfo {year} {2015})}\BibitemShut
  {NoStop}%
\bibitem [{\citenamefont {Bryan}(1990)}]{Bryan90Maximum}%
  \BibitemOpen
  \bibfield  {author} {\bibinfo {author} {\bibfnamefont {R.~K.}\ \bibnamefont
  {Bryan}},\ }\href {\doibase 10.1007/BF02427376} {\bibfield  {journal}
  {\bibinfo  {journal} {Eur. Biophys. J.}\ }\textbf {\bibinfo {volume} {18}},\
  \bibinfo {pages} {165} (\bibinfo {year} {1990})}\BibitemShut {NoStop}%
\bibitem [{\citenamefont {Gubernatis}\ \emph {et~al.}(1991)\citenamefont
  {Gubernatis}, \citenamefont {Jarrell}, \citenamefont {Silver},\ and\
  \citenamefont {Sivia}}]{Gubernatis91Quantum}%
  \BibitemOpen
  \bibfield  {author} {\bibinfo {author} {\bibfnamefont {J.~E.}\ \bibnamefont
  {Gubernatis}}, \bibinfo {author} {\bibfnamefont {M.}~\bibnamefont {Jarrell}},
  \bibinfo {author} {\bibfnamefont {R.~N.}\ \bibnamefont {Silver}}, \ and\
  \bibinfo {author} {\bibfnamefont {D.~S.}\ \bibnamefont {Sivia}},\ }\href
  {\doibase 10.1103/PhysRevB.44.6011} {\bibfield  {journal} {\bibinfo
  {journal} {Phys. Rev. B}\ }\textbf {\bibinfo {volume} {44}},\ \bibinfo
  {pages} {6011} (\bibinfo {year} {1991})}\BibitemShut {NoStop}%
\bibitem [{\citenamefont {Boehnke}()}]{MaxEnt}%
  \BibitemOpen
  \bibfield  {author} {\bibinfo {author} {\bibfnamefont {L.}~\bibnamefont
  {Boehnke}},\ }\href@noop {} {\enquote {\bibinfo {title} {{Bryan MaxEnt
  analytical continuation}},}\ }\bibinfo {howpublished}
  {\url{http://www.bitbucket.com/lewinboehnke/maxent}}\BibitemShut {NoStop}%
\bibitem [{\citenamefont {Vidberg}\ and\ \citenamefont
  {Serene}(1977)}]{Vidberg77Solving}%
  \BibitemOpen
  \bibfield  {author} {\bibinfo {author} {\bibfnamefont {H.~J.}\ \bibnamefont
  {Vidberg}}\ and\ \bibinfo {author} {\bibfnamefont {J.~W.}\ \bibnamefont
  {Serene}},\ }\href {\doibase 10.1007/BF00655090} {\bibfield  {journal}
  {\bibinfo  {journal} {Journal of Low Temperature Physics}\ }\textbf {\bibinfo
  {volume} {29}},\ \bibinfo {pages} {179} (\bibinfo {year} {1977})}\BibitemShut
  {NoStop}%
\bibitem [{\citenamefont {Strand}()}]{pyED}%
  \BibitemOpen
  \bibfield  {author} {\bibinfo {author} {\bibfnamefont {H.}~\bibnamefont
  {Strand}},\ }\href@noop {} {\enquote {\bibinfo {title} {{PYED}},}\ }\bibinfo
  {howpublished} {\url{https://github.com/hugostrand/pyed}}\BibitemShut
  {NoStop}%
\bibitem [{\citenamefont {Macquart}\ \emph {et~al.}(2010)\citenamefont
  {Macquart}, \citenamefont {Kennedy},\ and\ \citenamefont
  {Avdeev}}]{Macquart2010Neutron}%
  \BibitemOpen
  \bibfield  {author} {\bibinfo {author} {\bibfnamefont {R.~B.}\ \bibnamefont
  {Macquart}}, \bibinfo {author} {\bibfnamefont {B.~J.}\ \bibnamefont
  {Kennedy}}, \ and\ \bibinfo {author} {\bibfnamefont {M.}~\bibnamefont
  {Avdeev}},\ }\href {\doibase https://doi.org/10.1016/j.jssc.2009.11.005}
  {\bibfield  {journal} {\bibinfo  {journal} {Journal of Solid State
  Chemistry}\ }\textbf {\bibinfo {volume} {183}},\ \bibinfo {pages} {250 }
  (\bibinfo {year} {2010})}\BibitemShut {NoStop}%
\bibitem [{\citenamefont {Morikawa}\ \emph {et~al.}(1995)\citenamefont
  {Morikawa}, \citenamefont {Mizokawa}, \citenamefont {Kobayashi},
  \citenamefont {Fujimori}, \citenamefont {Eisaki}, \citenamefont {Uchida},
  \citenamefont {Iga},\ and\ \citenamefont {Nishihara}}]{Morikawa95Spectral}%
  \BibitemOpen
  \bibfield  {author} {\bibinfo {author} {\bibfnamefont {K.}~\bibnamefont
  {Morikawa}}, \bibinfo {author} {\bibfnamefont {T.}~\bibnamefont {Mizokawa}},
  \bibinfo {author} {\bibfnamefont {K.}~\bibnamefont {Kobayashi}}, \bibinfo
  {author} {\bibfnamefont {A.}~\bibnamefont {Fujimori}}, \bibinfo {author}
  {\bibfnamefont {H.}~\bibnamefont {Eisaki}}, \bibinfo {author} {\bibfnamefont
  {S.}~\bibnamefont {Uchida}}, \bibinfo {author} {\bibfnamefont
  {F.}~\bibnamefont {Iga}}, \ and\ \bibinfo {author} {\bibfnamefont
  {Y.}~\bibnamefont {Nishihara}},\ }\href {\doibase 10.1103/PhysRevB.52.13711}
  {\bibfield  {journal} {\bibinfo  {journal} {{Phys. Rev. B}}\ }\textbf
  {\bibinfo {volume} {52}},\ \bibinfo {pages} {13711} (\bibinfo {year}
  {1995})}\BibitemShut {NoStop}%
\bibitem [{\citenamefont {Sekiyama}\ \emph {et~al.}(2004)\citenamefont
  {Sekiyama}, \citenamefont {Fujiwara}, \citenamefont {Imada}, \citenamefont
  {Suga}, \citenamefont {Eisaki}, \citenamefont {Uchida}, \citenamefont
  {Takegahara}, \citenamefont {Harima}, \citenamefont {Saitoh}, \citenamefont
  {Nekrasov}, \citenamefont {Keller}, \citenamefont {Kondakov}, \citenamefont
  {Kozhevnikov}, \citenamefont {Pruschke}, \citenamefont {Held}, \citenamefont
  {Vollhardt},\ and\ \citenamefont {Anisimov}}]{Sekiyama04Mutual}%
  \BibitemOpen
  \bibfield  {author} {\bibinfo {author} {\bibfnamefont {A.}~\bibnamefont
  {Sekiyama}}, \bibinfo {author} {\bibfnamefont {H.}~\bibnamefont {Fujiwara}},
  \bibinfo {author} {\bibfnamefont {S.}~\bibnamefont {Imada}}, \bibinfo
  {author} {\bibfnamefont {S.}~\bibnamefont {Suga}}, \bibinfo {author}
  {\bibfnamefont {H.}~\bibnamefont {Eisaki}}, \bibinfo {author} {\bibfnamefont
  {S.~I.}\ \bibnamefont {Uchida}}, \bibinfo {author} {\bibfnamefont
  {K.}~\bibnamefont {Takegahara}}, \bibinfo {author} {\bibfnamefont
  {H.}~\bibnamefont {Harima}}, \bibinfo {author} {\bibfnamefont
  {Y.}~\bibnamefont {Saitoh}}, \bibinfo {author} {\bibfnamefont {I.~A.}\
  \bibnamefont {Nekrasov}}, \bibinfo {author} {\bibfnamefont {G.}~\bibnamefont
  {Keller}}, \bibinfo {author} {\bibfnamefont {D.~E.}\ \bibnamefont
  {Kondakov}}, \bibinfo {author} {\bibfnamefont {A.~V.}\ \bibnamefont
  {Kozhevnikov}}, \bibinfo {author} {\bibfnamefont {T.}~\bibnamefont
  {Pruschke}}, \bibinfo {author} {\bibfnamefont {K.}~\bibnamefont {Held}},
  \bibinfo {author} {\bibfnamefont {D.}~\bibnamefont {Vollhardt}}, \ and\
  \bibinfo {author} {\bibfnamefont {V.~I.}\ \bibnamefont {Anisimov}},\ }\href
  {\doibase 10.1103/PhysRevLett.93.156402} {\bibfield  {journal} {\bibinfo
  {journal} {Phys. Rev. Lett.}\ }\textbf {\bibinfo {volume} {93}},\ \bibinfo
  {pages} {156402} (\bibinfo {year} {2004})}\BibitemShut {NoStop}%
\bibitem [{\citenamefont {Backes}\ \emph {et~al.}(2016)\citenamefont {Backes},
  \citenamefont {R\"odel}, \citenamefont {Fortuna}, \citenamefont
  {Frantzeskakis}, \citenamefont {Le~F\`evre}, \citenamefont {Bertran},
  \citenamefont {Kobayashi}, \citenamefont {Yukawa}, \citenamefont
  {Mitsuhashi}, \citenamefont {Kitamura}, \citenamefont {Horiba}, \citenamefont
  {Kumigashira}, \citenamefont {Saint-Martin}, \citenamefont {Fouchet},
  \citenamefont {Berini}, \citenamefont {Dumont}, \citenamefont {Kim},
  \citenamefont {Lechermann}, \citenamefont {Jeschke}, \citenamefont
  {Rozenberg}, \citenamefont {Valent\'{\i}},\ and\ \citenamefont
  {Santander-Syro}}]{Backes16Hubbard}%
  \BibitemOpen
  \bibfield  {author} {\bibinfo {author} {\bibfnamefont {S.}~\bibnamefont
  {Backes}}, \bibinfo {author} {\bibfnamefont {T.~C.}\ \bibnamefont {R\"odel}},
  \bibinfo {author} {\bibfnamefont {F.}~\bibnamefont {Fortuna}}, \bibinfo
  {author} {\bibfnamefont {E.}~\bibnamefont {Frantzeskakis}}, \bibinfo {author}
  {\bibfnamefont {P.}~\bibnamefont {Le~F\`evre}}, \bibinfo {author}
  {\bibfnamefont {F.}~\bibnamefont {Bertran}}, \bibinfo {author} {\bibfnamefont
  {M.}~\bibnamefont {Kobayashi}}, \bibinfo {author} {\bibfnamefont
  {R.}~\bibnamefont {Yukawa}}, \bibinfo {author} {\bibfnamefont
  {T.}~\bibnamefont {Mitsuhashi}}, \bibinfo {author} {\bibfnamefont
  {M.}~\bibnamefont {Kitamura}}, \bibinfo {author} {\bibfnamefont
  {K.}~\bibnamefont {Horiba}}, \bibinfo {author} {\bibfnamefont
  {H.}~\bibnamefont {Kumigashira}}, \bibinfo {author} {\bibfnamefont
  {R.}~\bibnamefont {Saint-Martin}}, \bibinfo {author} {\bibfnamefont
  {A.}~\bibnamefont {Fouchet}}, \bibinfo {author} {\bibfnamefont
  {B.}~\bibnamefont {Berini}}, \bibinfo {author} {\bibfnamefont
  {Y.}~\bibnamefont {Dumont}}, \bibinfo {author} {\bibfnamefont {A.~J.}\
  \bibnamefont {Kim}}, \bibinfo {author} {\bibfnamefont {F.}~\bibnamefont
  {Lechermann}}, \bibinfo {author} {\bibfnamefont {H.~O.}\ \bibnamefont
  {Jeschke}}, \bibinfo {author} {\bibfnamefont {M.~J.}\ \bibnamefont
  {Rozenberg}}, \bibinfo {author} {\bibfnamefont {R.}~\bibnamefont
  {Valent\'{\i}}}, \ and\ \bibinfo {author} {\bibfnamefont {A.~F.}\
  \bibnamefont {Santander-Syro}},\ }\href {\doibase 10.1103/PhysRevB.94.241110}
  {\bibfield  {journal} {\bibinfo  {journal} {Phys. Rev. B}\ }\textbf {\bibinfo
  {volume} {94}},\ \bibinfo {pages} {241110} (\bibinfo {year}
  {2016})}\BibitemShut {NoStop}%
\bibitem [{\citenamefont {Dudy}\ \emph {et~al.}(2016)\citenamefont {Dudy},
  \citenamefont {Sing}, \citenamefont {Scheiderer}, \citenamefont {Denlinger},
  \citenamefont {Schütz}, \citenamefont {Gabel}, \citenamefont {Buchwald},
  \citenamefont {Schlueter}, \citenamefont {Lee},\ and\ \citenamefont
  {Claessen}}]{Dudy16InSitu}%
  \BibitemOpen
  \bibfield  {author} {\bibinfo {author} {\bibfnamefont {L.}~\bibnamefont
  {Dudy}}, \bibinfo {author} {\bibfnamefont {M.}~\bibnamefont {Sing}}, \bibinfo
  {author} {\bibfnamefont {P.}~\bibnamefont {Scheiderer}}, \bibinfo {author}
  {\bibfnamefont {J.~D.}\ \bibnamefont {Denlinger}}, \bibinfo {author}
  {\bibfnamefont {P.}~\bibnamefont {Schütz}}, \bibinfo {author} {\bibfnamefont
  {J.}~\bibnamefont {Gabel}}, \bibinfo {author} {\bibfnamefont
  {M.}~\bibnamefont {Buchwald}}, \bibinfo {author} {\bibfnamefont
  {C.}~\bibnamefont {Schlueter}}, \bibinfo {author} {\bibfnamefont {T.-L.}\
  \bibnamefont {Lee}}, \ and\ \bibinfo {author} {\bibfnamefont
  {R.}~\bibnamefont {Claessen}},\ }\href {\doibase 10.1002/adma.201600046}
  {\bibfield  {journal} {\bibinfo  {journal} {Advanced Materials}\ }\textbf
  {\bibinfo {volume} {28}},\ \bibinfo {pages} {7443} (\bibinfo {year}
  {2016})}\BibitemShut {NoStop}%
\bibitem [{\citenamefont {Sing}\ \emph {et~al.}(2017)\citenamefont {Sing},
  \citenamefont {Jeschke}, \citenamefont {Lechermann}, \citenamefont
  {Valent{\'i}},\ and\ \citenamefont {Claessen}}]{Sing17Influence}%
  \BibitemOpen
  \bibfield  {author} {\bibinfo {author} {\bibfnamefont {M.}~\bibnamefont
  {Sing}}, \bibinfo {author} {\bibfnamefont {H.~O.}\ \bibnamefont {Jeschke}},
  \bibinfo {author} {\bibfnamefont {F.}~\bibnamefont {Lechermann}}, \bibinfo
  {author} {\bibfnamefont {R.}~\bibnamefont {Valent{\'i}}}, \ and\ \bibinfo
  {author} {\bibfnamefont {R.}~\bibnamefont {Claessen}},\ }\href {\doibase
  10.1140/epjst/e2017-70059-7} {\bibfield  {journal} {\bibinfo  {journal} {The
  European Physical Journal Special Topics}\ }\textbf {\bibinfo {volume}
  {226}},\ \bibinfo {pages} {2457} (\bibinfo {year} {2017})}\BibitemShut
  {NoStop}%
\bibitem [{\citenamefont {Kas}\ \emph {et~al.}(2016)\citenamefont {Kas},
  \citenamefont {Rehr},\ and\ \citenamefont {Curtis}}]{Kas16Particle}%
  \BibitemOpen
  \bibfield  {author} {\bibinfo {author} {\bibfnamefont {J.~J.}\ \bibnamefont
  {Kas}}, \bibinfo {author} {\bibfnamefont {J.~J.}\ \bibnamefont {Rehr}}, \
  and\ \bibinfo {author} {\bibfnamefont {J.~B.}\ \bibnamefont {Curtis}},\
  }\href {\doibase 10.1103/PhysRevB.94.035156} {\bibfield  {journal} {\bibinfo
  {journal} {Phys. Rev. B}\ }\textbf {\bibinfo {volume} {94}},\ \bibinfo
  {pages} {035156} (\bibinfo {year} {2016})}\BibitemShut {NoStop}%
\bibitem [{\citenamefont {Caruso}\ \emph {et~al.}(2015)\citenamefont {Caruso},
  \citenamefont {Lambert},\ and\ \citenamefont {Giustino}}]{Caruso15Band}%
  \BibitemOpen
  \bibfield  {author} {\bibinfo {author} {\bibfnamefont {F.}~\bibnamefont
  {Caruso}}, \bibinfo {author} {\bibfnamefont {H.}~\bibnamefont {Lambert}}, \
  and\ \bibinfo {author} {\bibfnamefont {F.}~\bibnamefont {Giustino}},\ }\href
  {\doibase 10.1103/PhysRevLett.114.146404} {\bibfield  {journal} {\bibinfo
  {journal} {Phys. Rev. Lett.}\ }\textbf {\bibinfo {volume} {114}},\ \bibinfo
  {pages} {146404} (\bibinfo {year} {2015})}\BibitemShut {NoStop}%
\bibitem [{\citenamefont {Lischner}\ \emph {et~al.}(2015)\citenamefont
  {Lischner}, \citenamefont {P\'alsson}, \citenamefont {Vigil-Fowler},
  \citenamefont {Nemsak}, \citenamefont {Avila}, \citenamefont {Asensio},
  \citenamefont {Fadley},\ and\ \citenamefont {Louie}}]{Lischner15Satellite}%
  \BibitemOpen
  \bibfield  {author} {\bibinfo {author} {\bibfnamefont {J.}~\bibnamefont
  {Lischner}}, \bibinfo {author} {\bibfnamefont {G.~K.}\ \bibnamefont
  {P\'alsson}}, \bibinfo {author} {\bibfnamefont {D.}~\bibnamefont
  {Vigil-Fowler}}, \bibinfo {author} {\bibfnamefont {S.}~\bibnamefont
  {Nemsak}}, \bibinfo {author} {\bibfnamefont {J.}~\bibnamefont {Avila}},
  \bibinfo {author} {\bibfnamefont {M.~C.}\ \bibnamefont {Asensio}}, \bibinfo
  {author} {\bibfnamefont {C.~S.}\ \bibnamefont {Fadley}}, \ and\ \bibinfo
  {author} {\bibfnamefont {S.~G.}\ \bibnamefont {Louie}},\ }\href {\doibase
  10.1103/PhysRevB.91.205113} {\bibfield  {journal} {\bibinfo  {journal} {Phys.
  Rev. B}\ }\textbf {\bibinfo {volume} {91}},\ \bibinfo {pages} {205113}
  (\bibinfo {year} {2015})}\BibitemShut {NoStop}%
\bibitem [{\citenamefont {Caruso}\ and\ \citenamefont
  {Giustino}(2015)}]{Caruso15Spectral}%
  \BibitemOpen
  \bibfield  {author} {\bibinfo {author} {\bibfnamefont {F.}~\bibnamefont
  {Caruso}}\ and\ \bibinfo {author} {\bibfnamefont {F.}~\bibnamefont
  {Giustino}},\ }\href {\doibase 10.1103/PhysRevB.92.045123} {\bibfield
  {journal} {\bibinfo  {journal} {Phys. Rev. B}\ }\textbf {\bibinfo {volume}
  {92}},\ \bibinfo {pages} {045123} (\bibinfo {year} {2015})}\BibitemShut
  {NoStop}%
\bibitem [{\citenamefont {Steiner}\ \emph {et~al.}(1979)\citenamefont
  {Steiner}, \citenamefont {H{\"o}chst},\ and\ \citenamefont
  {H{\"u}fner}}]{79photoemission}%
  \BibitemOpen
  \bibfield  {author} {\bibinfo {author} {\bibfnamefont {P.}~\bibnamefont
  {Steiner}}, \bibinfo {author} {\bibfnamefont {H.}~\bibnamefont {H{\"o}chst}},
  \ and\ \bibinfo {author} {\bibfnamefont {S.}~\bibnamefont {H{\"u}fner}},\
  }\enquote {\bibinfo {title} {Simple metals},}\ in\ \href {\doibase
  10.1007/3-540-09202-1_7} {\emph {\bibinfo {booktitle} {Photoemission in
  Solids II: Case Studies}}},\ \bibinfo {editor} {edited by\ \bibinfo {editor}
  {\bibfnamefont {L.}~\bibnamefont {Ley}}\ and\ \bibinfo {editor}
  {\bibfnamefont {M.}~\bibnamefont {Cardona}}}\ (\bibinfo  {publisher}
  {Springer Berlin Heidelberg},\ \bibinfo {address} {Berlin, Heidelberg},\
  \bibinfo {year} {1979})\ pp.\ \bibinfo {pages} {349--372}\BibitemShut
  {NoStop}%
\bibitem [{\citenamefont {Tomczak}\ \emph {et~al.}(2009)\citenamefont
  {Tomczak}, \citenamefont {Miyake}, \citenamefont {Sakuma},\ and\
  \citenamefont {Aryasetiawan}}]{Tomczak09Effective}%
  \BibitemOpen
  \bibfield  {author} {\bibinfo {author} {\bibfnamefont {J.~M.}\ \bibnamefont
  {Tomczak}}, \bibinfo {author} {\bibfnamefont {T.}~\bibnamefont {Miyake}},
  \bibinfo {author} {\bibfnamefont {R.}~\bibnamefont {Sakuma}}, \ and\ \bibinfo
  {author} {\bibfnamefont {F.}~\bibnamefont {Aryasetiawan}},\ }\href {\doibase
  10.1103/PhysRevB.79.235133} {\bibfield  {journal} {\bibinfo  {journal} {Phys.
  Rev. B}\ }\textbf {\bibinfo {volume} {79}},\ \bibinfo {pages} {235133}
  (\bibinfo {year} {2009})}\BibitemShut {NoStop}%
\bibitem [{\citenamefont {Tomczak}\ \emph {et~al.}(2010)\citenamefont
  {Tomczak}, \citenamefont {Miyake},\ and\ \citenamefont
  {Aryasetiawan}}]{Tomczak10Realistic}%
  \BibitemOpen
  \bibfield  {author} {\bibinfo {author} {\bibfnamefont {J.~M.}\ \bibnamefont
  {Tomczak}}, \bibinfo {author} {\bibfnamefont {T.}~\bibnamefont {Miyake}}, \
  and\ \bibinfo {author} {\bibfnamefont {F.}~\bibnamefont {Aryasetiawan}},\
  }\href {\doibase 10.1103/PhysRevB.81.115116} {\bibfield  {journal} {\bibinfo
  {journal} {Phys. Rev. B}\ }\textbf {\bibinfo {volume} {81}},\ \bibinfo
  {pages} {115116} (\bibinfo {year} {2010})}\BibitemShut {NoStop}%
\bibitem [{\citenamefont {Aryasetiawan}\ \emph {et~al.}(1996)\citenamefont
  {Aryasetiawan}, \citenamefont {Hedin},\ and\ \citenamefont
  {Karlsson}}]{Aryasetiawan96Multiple}%
  \BibitemOpen
  \bibfield  {author} {\bibinfo {author} {\bibfnamefont {F.}~\bibnamefont
  {Aryasetiawan}}, \bibinfo {author} {\bibfnamefont {L.}~\bibnamefont {Hedin}},
  \ and\ \bibinfo {author} {\bibfnamefont {K.}~\bibnamefont {Karlsson}},\
  }\href {\doibase 10.1103/PhysRevLett.77.2268} {\bibfield  {journal} {\bibinfo
   {journal} {Phys. Rev. Lett.}\ }\textbf {\bibinfo {volume} {77}},\ \bibinfo
  {pages} {2268} (\bibinfo {year} {1996})}\BibitemShut {NoStop}%
\bibitem [{\citenamefont {Fetter}\ and\ \citenamefont
  {Walecka}(2003)}]{fetterwalecka}%
  \BibitemOpen
  \bibfield  {author} {\bibinfo {author} {\bibfnamefont {A.}~\bibnamefont
  {Fetter}}\ and\ \bibinfo {author} {\bibfnamefont {J.}~\bibnamefont
  {Walecka}},\ }\href@noop {} {\emph {\bibinfo {title} {Quantum Theory of
  Many-Particle Systems}}}\ (\bibinfo  {publisher} {Dover, Mineola, New York},\
  \bibinfo {year} {2003})\BibitemShut {NoStop}%
\end{thebibliography}

\end{document}